\documentclass[iop, twocolumn, tighten]{emulateapj}
 
\usepackage{graphicx} 
\usepackage{dcolumn} 
\usepackage{times} 
\usepackage{amsmath} 
\usepackage{enumerate} 
\usepackage{natbib} 
\usepackage{url} 
\usepackage[utf8]{inputenc}
\usepackage{color}
\usepackage{sistyle} 
\usepackage[english]{babel} 
\usepackage[colorlinks=true,urlcolor=cyan,citecolor=blue,linkcolor=blue]{hyperref}
 
\newcommand{\beq}{\begin{equation}}
\newcommand{\eeq}{\end{equation}}
\DeclareMathOperator \dm {d}

\def\EAH{{\it Einstein@Home}}
\def\simlt{\mathrel{\hbox{\rlap{\hbox{\lower4pt\hbox{$\sim$}}}\hbox{$<$}}}}
\def\simgt{\mathrel{\hbox{\rlap{\hbox{\lower4pt\hbox{$\sim$}}}\hbox{$>$}}}}

\newcommand{\bwswitch}{0}

\shorttitle{Einstein@Home and PSR~J2007+2722}
\shortauthors{Allen et al.}

\journalinfo{{\it Published 29 July 2013 in the Astrophysical Journal.} Reference: B. Allen et al., ApJ 773, 91
(2013). \href{http://dx.doi.org/10.1088/0004-637X/773/2/91}{doi:10.1088/0004-637X/773/2/91}}
\submitted{}

\begin{document}

\title{The Einstein@Home search for radio pulsars and PSR~J2007+2722
  discovery} \author{
B.~Allen\altaffilmark{1,2, 3},
B.~Knispel\altaffilmark{1,2},  
J.~M.~Cordes\altaffilmark{4},
J.~S.~Deneva\altaffilmark{5},
J.~W.~T.~Hessels\altaffilmark{6,7},
D.~Anderson\altaffilmark{8},
C.~Aulbert\altaffilmark{1,2},
O.~Bock\altaffilmark{1,2},
A.~Brazier\altaffilmark{4,9},
S.~Chatterjee\altaffilmark{4},
P.~B.~Demorest\altaffilmark{10},
H.~B.~Eggenstein\altaffilmark{1,2},
H.~Fehrmann\altaffilmark{1,2},
E.~V.~Gotthelf\altaffilmark{11},
D.~Hammer\altaffilmark{3},
V.~M.~Kaspi\altaffilmark{12},
M.~Kramer\altaffilmark{13},
A.~G.~Lyne\altaffilmark{14}, 
B.~Machenschalk\altaffilmark{1,2},
M.~A.~McLaughlin\altaffilmark{15},
C.~Messenger\altaffilmark{1,2},
H.~J.~Pletsch\altaffilmark{1,2},
S.~M.~Ransom\altaffilmark{10},
I.~H.~Stairs\altaffilmark{16}, 
B.~W.~Stappers\altaffilmark{14},
N.~D.~R.~Bhat\altaffilmark{17,18}, 
S.~Bogdanov\altaffilmark{11},
F.~Camilo\altaffilmark{5,11},
D.~J.~Champion\altaffilmark{13},
F.~Crawford\altaffilmark{19},
G.~Desvignes\altaffilmark{20},
P.~C.~C.~Freire\altaffilmark{13},
G.~Heald\altaffilmark{6},
F.~A.~Jenet\altaffilmark{21}, 
P.~Lazarus\altaffilmark{13}, 
K.~J.~Lee\altaffilmark{13},
J.~van~Leeuwen\altaffilmark{6,7}, 
R.~Lynch\altaffilmark{12}, 
M.~A.~Papa\altaffilmark{1,2,3},
R.~Prix\altaffilmark{1,2},
R.~Rosen\altaffilmark{22}, 
P.~Scholz\altaffilmark{12},
X.~Siemens\altaffilmark{3}, 
K.~Stovall\altaffilmark{21}, 
A.~Venkataraman\altaffilmark{5},
W.~Zhu\altaffilmark{16}
}

\altaffiltext{1}{Max-Planck-Institut f\"ur Gravitationsphysik, D-30167 Hannover, Germany; \href{mailto:bruce.allen@aei.mpg.de}{bruce.allen@aei.mpg.de}}
\altaffiltext{2}{Leibniz Universit{\"a}t Hannover, D-30167 Hannover, Germany}
\altaffiltext{3}{Department of Physics, University of Wisconsin - Milwaukee, Milwaukee WI 53211, USA}
\altaffiltext{4}{Department of Astronomy, Cornell University, Ithaca, NY 14853, USA} 
\altaffiltext{5}{Arecibo Observatory, HC3 Box 53995, Arecibo, PR 00612, USA}
\altaffiltext{6}{ASTRON, the Netherlands Institute for Radio Astronomy, Postbus 2, 7990 AA, Dwingeloo, The Netherlands} 
\altaffiltext{7}{Astronomical Institute ``Anton Pannekoek'', University of Amsterdam, Science Park 904, 1098 XH Amsterdam, The Netherlands}
\altaffiltext{8}{Space Sciences Laboratory, University~of California -- Berkeley, USA} 
\altaffiltext{9}{NAIC, Cornell University, Ithaca, NY 14853, USA} 
\altaffiltext{10}{NRAO (National Radio Astronomy Observatory), Charlottesville, VA 22903, USA} 
\altaffiltext{11}{Columbia Astrophysics Laboratory, Columbia University, New York, NY 10027, USA} 
\altaffiltext{12}{Department~of Physics, McGill University, Montreal, QC H3A2T8, Canada}
\altaffiltext{13}{Max-Planck-Institut f\"ur Radioastronomie, D-53121 Bonn, Germany} 
\altaffiltext{14}{Jodrell Bank Centre for Astrophys., School of Physics and Astronomy, University of Manchester, Manch., M13 9PL, UK} 
\altaffiltext{15}{Department of Physics, West Virginia University, Morgantown, WV 26506, USA} 
\altaffiltext{16}{Department~of Physics and Astronomy, University~of British Columbia, Vancouver, BC V6T 1Z1, Canada} 
\altaffiltext{17}{Center for Astrophysics and Supercomputing, Swinburne University,  Hawthorn, Victoria 3122, Australia} 
\altaffiltext{18}{International Centre for Radio Astronomy Research, Curtin University, Bentley, WA 6102, Australia} 
\altaffiltext{19}{Department of Physics and Astronomy, Franklin and Marshall College, Lancaster, PA 17604-3003, USA} 
\altaffiltext{20}{Department of Astronomy and Radio Astronomy Laboratory, University of California, Berkeley, CA 94720, USA} 
\altaffiltext{21}{Center for Gravitational Wave Astronomy, University of Texas - Brownsville, TX 78520, USA} 
\altaffiltext{22}{NRAO, P.O. Box 2, Green Bank WV 24944, USA}

\begin{abstract}\noindent
  Einstein@Home aggregates the computer power of hundreds of thousands
  of volunteers from 193 countries, to search for new neutron stars
  using data from electromagnetic and gravitational-wave detectors.
  This paper presents a detailed description of the search for new
  radio pulsars using Pulsar ALFA survey data from the Arecibo
  Observatory. The enormous computing power allows this search to
  cover a new region of parameter space; it can detect pulsars in
  binary systems with orbital periods as short as 11\,minutes.  We also
  describe the first Einstein@Home discovery, the 40.8~Hz isolated
  pulsar PSR~J2007+2722, and provide a full timing model.
  PSR~J2007+2722's pulse profile is remarkably wide with emission over
  almost the entire spin period.  This neutron star is most likely a
  disrupted recycled pulsar, about as old as its characteristic
  spin-down age of 404~Myr.  However there is a small chance that it
  was born recently, with a low magnetic field. If so, upper limits on
  the X-ray flux suggest but can not prove that PSR~J2007+2722 is at
  least $\sim 100$~kyr old.  In the future, we expect that the massive
  computing power provided by volunteers should enable many additional
  radio pulsar discoveries.
\end{abstract}
 

\section{Introduction }
\label{s:introduction}

\EAH{} is an on-going volunteer distributed computing project
\citep{Anderson:2006:DRS:1188455.1188586}, launched in early 2005.
More than 330\,000 members of the general public have
``signed up'' their laptop and desktop computers. When otherwise idle,
these computers download observational data from the \EAH{} servers,
search the data for weak astrophysical signals, and return the results
of the analysis. The collective computing power is on par with the
largest supercomputers in the world.

The goal of \EAH{} is to discover neutron stars, using data from an
international network of gravitational-wave (GW) detectors
\citep{lrr-2009-2}, from radio telescopes \citep{ls98,LorimerKramer},
and from the Large Area Telescope \citep[LAT;][]{2009ApJ...697.1071A}
gamma-ray detector onboard the \textit{Fermi} Satellite. Because the
expected signals are weak, and the source
parameters\footnote{Depending upon the type of search, these unknown
  parameters might include the sky position, spin frequency, spin-down
  rate, orbital parameters, etc.} are unknown, the sensitivity of the
GW searches \citep{bc1,bc2} the radio pulsar searches
\citep{springerlink:10.1007/978-1-84628-757-2}, and the gamma-ray
searches \citep{PletschAllen,2012ApJ...744..105P,2012ApJ...755L..20P,
  2012Sci...338.1314P} are limited by the available computing power.

Before 2009, \EAH{} only searched data from the Laser Interferometer
Gravitational-Wave Observatory
\citep[LIGO;][]{Abramovici17041992,barish:44,0034-4885-72-7-076901}.  So
far these searches have not found any sources, but have set new and
more sensitive upper limits on possible continuous gravitational-wave (CW)
emissions \citep{2009PhRvD..79b2001A, 2009PhRvD..80d2003A, 2013PhRvD..87d2001A}.  These
searches are ongoing, with increasing sensitivity arising from
improved data analysis methods \citep{PletschAllen} and
better-quality data \citep{0264-9381-26-11-114013}.

In 2009, \EAH{} also began searching radio survey data from the
305-meter Arecibo telescope in Puerto Rico. This is the world's
largest and most sensitive radio telescope, and has discovered a
substantial fraction of all known pulsars.  Beginning in 2010 December
a similar search using data from Parkes Observatory in Australia was
also started; the differences from the Arecibo search and some results
are described in \cite{2013arXiv1302.0467K}.

Starting in summer 2011, \EAH{} also began a search for isolated
gamma-ray pulsars in data from the \textit{Fermi} satellite's LAT
\cite{2009ApJ...697.1071A}; this will be described in future
publications.

The Arecibo data are collected by the Pulsar ALFA (PALFA) Consortium
using the Arecibo L-band Feed Array
(ALFA\footnote{\url{http://www.naic.edu/alfa/}}).  For the pulsar
survey, ALFA output is fed into fast, broad-band spectrometers (see
Section~\ref{ss:dataacquisition}); further down the data analysis
pipeline (see Section~\ref{ss:preprocessing}) this enables
compensation for the dispersive propagation of pulses from celestial
sources.

The computing capacity of \EAH{} is used to search the spectrometer
output for signals from neutron stars in short-period orbits around
companion stars. This is a poorly-explored region of parameter space,
where other radio-pulsar search pipelines lose much or most of their
sensitivity. The detection of these pulsars with standard Fourier
methods is hampered by Doppler smearing of the pulsed signal caused by
binary motion during the survey observation
\citep{1991ApJ...368..504J}.

Previous searches \citep{1990Natur.346...42A,2000ApJ...535..975C} have
utilized ``acceleration searches'' \citep{1991ApJ...368..504J}, which
correct for the part of the binary motion which can be modeled as a
constant acceleration along the line-of-sight.  These
computationally-efficient techniques are effective when the
observation time is short compared to the orbital period.  Thus, they
are insensitive to the most compact systems
\citep{2002AJ....124.1788R}. In contrast, the computing power of
\EAH{} enables a full demodulation to be carried out, giving
substantially increased sensitivity to signals from pulsars in compact
circular orbits with periods below $\sim 1$\,hr.

In 2010 August, \EAH{} announced its first discovery of a new neutron
star \citep{2010Sci...329.1305K} which appears to be the
fastest-spinning ``disrupted recycled pulsar'' (DRP) so far found
\citep{MNR:MNR16970}. In the same month, \EAH{} also discovered a
48~Hz pulsar in a binary system \citep{Knispel1952}.  Further
Einstein@Home discoveries in Parkes Multi-Beam Pulsar Survey (PMPS)
are described in \cite{2013arXiv1302.0467K}.  As of 2013 January, Einstein@Home
has discovered almost 50 radio pulsars.

This paper has two purposes.  First, it provides a full description of
the \EAH{} radio pulsar search and post-processing pipeline. Second,
it provides a detailed description and full timing solution for the
first \EAH{} discovery, the 40.8~Hz pulsar PSR~J2007+2722
\citep{2010Sci...329.1305K}.

The paper is structured as follows. Section~\ref{s:EinsteinAtHome}
presents a general description of the \EAH{} computing project,
including its motivation, its history, and its technical design and
structure.  Section~\ref{sec:palfasurvey} is an brief overview of the
PALFA survey, including its history, the data taking rates, and data
acquisition system. Section~\ref{s:EinsteinAtHomeRPS} is a detailed
technical description of the \EAH{} search for radio pulsars, starting
from the centralized data preparation, through the distributed
processing on volunteers' computers, and centralized
post-processing. Section~\ref{s:Discovery} describes the discovery of
the first \EAH{} radio pulsar,
PSR~J2007+2722. Section~\ref{s:followup} is about the subsequent
follow-up investigations and studies, including observations at
multiple frequencies, and accurate determination of the sky position
through gridding and timing.  We also discuss the evolutionary origin
of PSR~J2007+2722.  This is followed in Section~\ref{s:conclusion} by
a short discussion and conclusion.

Unless otherwise stated, all coordinates in this paper are in the
J2000 coordinate system, and $c$ denotes the speed of light.

\section{The Einstein@Home Distributed Computing Project}
\label{s:EinsteinAtHome}

\subsection{Volunteer Distributed Computing}
The basic motivation for volunteer distributed computing is simple:
the aggregate computing power owned by the general public exceeds that
of universities, and public and private research laboratories, by two
to three orders-of-magnitude. Scientific research whose progress is
limited or constrained by computing can benefit from access to even a
small fraction of these resources.  This type of research includes both
numerical simulation and Monte-Carlo-type exploration of parameter
spaces, that make no (direct) use of observational data, and
data-mining and data-analysis efforts which perform deep searches
through (potentially very large) observational data sets.

Worldwide, there are more than one billion personal computers (PCs)
which are connected to the Internet.  These PCs typically contain
x86-architecture central processor units (CPUs) and substantial
disk-based and solid-state storage. Many of these systems also contain
graphics processor units (GPUs) which can perform floating point
calculations one to two orders-of-magnitude faster than a modern CPU
core.

The raw computational capacity of each of these consumer computers is
similar to that of the systems used as building blocks for computer
clusters or research supercomputers.  In fact modern research
computers are made possible {\it only} by the economies of scale of
the consumer marketplace, which ensures that the basic components are
inexpensive and widely available.  But research machines typically
consist of hundreds or thousands of these CPUs; volunteer distributed
computing offers access to hundreds of thousands or millions of these
CPUs.

\subsection{Constraints on Suitable Computing Problems}
Volunteer distributed computing is only a suitable solution for some
computing and data analysis problems: there are both social and
technical constraints. To attract volunteers, the research must
resonate with the ``person in the street''. It must have clear and
understandable goals that appeal to the general public and that excite
and maintain interest. Experience shows that at least four areas have
these qualities: medical research, mathematics, climate/environmental
science, and astronomy and astrophysics.

The technical constraints arise because the computers are only
connected by the public Internet.  This is very different than
research supercomputers, which typically have low-latency high-speed
networks which enable any CPU to access data from any other CPU with
nanosecond latencies and GB/s bandwidth.  In contrast, the latency in
volunteer distributed computing can be fifteen orders-of-magnitude
larger; a volunteer's computer may only connect to the Internet once
per week!  The average available bandwidth is also much smaller,
particularly for data distributed from a central (project) location.
For example if a project is distributing data through a 1Gb/s public
Internet connection to 100k host machines, the average bandwidth
available per host is 10~kb/s, six orders-of-magnitude less than for a
research facility. 

The main technical constraints on the computing problem are therefore
as follows:

\noindent
(1.) It must lie in the class of so called ``embarrassingly parallel''
problems, whose solution requires no communication or dependency
between hosts.

\noindent
(2.) It must have a high ratio of computation to input/output. For
example if the project distributes data through a single 1Gb/s network
connection, and the application requires 1~MB of data per
CPU-core-hour, then at most 360k host CPU-cores can be kept
fully-occupied on a $24 \times 7$ (round-the clock) basis.

\noindent
 (3.) It must use only a small fraction of available RAM (say 100~MB)
so that the operating system (OS) can quickly swap tasks, providing
normal interactive computer response for volunteers.

\noindent
(4.) It must be capable of frequent and lightweight checkpointing
(saving the internal state for later restart) using only a small
amount of total storage (say 10~MB), so that it can snatch idle
compute cycles but stop processing when the volunteer is using the
computer or turns the computer off.

\noindent
(5.) The code that will run on volunteer's hosts must be mature enough
to be ported to several different OSs, and to run reliably on
volunteers computers.

\noindent
In short, volunteer distributed computing is not a panacea: it can
only be used to solve {\it some} computing problems.

\subsection{Trends in Computing Power and GPUs}
The latest trend in computing is the move to systems containing large
numbers of processing cores.  This is largely in response to the
fundamental physical limits that arise in manufacturing integrated
circuits.  For more than 40 yr, the computing power available at
fixed cost has doubled every 18 months. This was a consequence of
``Moore's law'', a heuristic observation that the number of components
on an integrated circuit grew exponentially with time.  This trend was
made possible by the shrinking of the ``process size'' (the size of
the smallest components on an integrated circuit) along with a
corresponding increase in clock speed and a decrease in operating
voltage.  Operating voltages can no longer be decreased because they
have approached the band-gap energy, and process sizes, currently at
22nm, have been shrinking more slowly than in the past.  They are
expected to decrease to about 10nm, but can not get much smaller; the
inter-atomic spacing in a silicon lattice is 0.7nm.  To get more
computing power at reasonable cost, the only approach is to put large
numbers of cores onto a single chip.

Fortunately the consumer marketplace has a demand for such systems:
they are called GPUs and are used for high-quality rendering of graphics
and video.  The evolution of television from radio broadcasting to
transmission over the Internet is now underway, and it is expected
that over the next decade this will be an important driving force
behind further growth in Internet capacity and graphics capability in
consumer computers.  Already more than 25\% of Einstein@Home host
machines contain GPUs, and we expect that this will approach 100\%
within the coming three years.

Current-generation GPUs have 500 or more cores, each capable of
simultaneously doing one floating-point addition and one
floating-point multiplication per clock cycle. The two leading
manufacturers of such systems (NVIDIA and AMD/ATI) both provide
application programming interfaces (APIs) that permit GPUs to be used
for general-purpose computing.  Thus, over the coming decade, if GPU
capacity is accessed and exploited, volunteer distributed computing
should continue to provide ``Moore's law scaling'' and to provide
access to more computing cycles than more traditional approaches.

In the longer-term, tablet devices and smartphones will probably
provide the bulk of the computing power.  Their CPUs and GPUs are very
power-efficient, though typically an order-of-magnitude slower than
laptop or desktop computers. However very large numbers are being
marketed and used.  These devices are often idle while connected to
charging stations; during this time they represent a significant
computing resource.

\subsection{The Einstein@Home Project} \label{subsec:EAHproject}
\EAH{} was formally launched at the American Association for the
Advancement of Science meeting on 2005 February 19, as one of the
cornerstone activities of the World Year of Physics 2005
\citep{stone:18}.  Members of the general public, whom we refer to as
``volunteers'', ``sign up'' by visiting the project Web site
\url{http://einstein.phys.uwm.edu} and downloading a small executable,
which is available for Windows, Mac and Linux platforms.  It takes a
couple of minutes to install on a typical home computer or laptop
(which is then technically refereed to as a ``host''). After that,
when the host is otherwise idle, it downloads observational
astrophysics data from one of the \EAH{} servers, and analyzes it in
the background, searching for signals.  The results of the analysis
are automatically uploaded to a project server, and more work is
requested.  The system is designed to operate without further
attention from the volunteer, although it is also highly configurable
and can be tuned for specific needs if desired.  The collective
computing power is on par with the largest supercomputers in the
world.

The \EAH{} project also incorporates additional features intended to
attract, inform, motivate and retain volunteers.  These include {\it
  message boards} where volunteers can exchange messages with other
volunteers and project personnel and scientists; granting {\it
  computing credits} as a symbolic ``reward'' for successful computing
work; the ability to form {\it teams} to compete for computing
credits; {\it informational Web pages} describing the science and
results; and access to dynamic Web pages that allow volunteers to
track the individual computing jobs done by their computers.

There are a number of such volunteer computing projects world-wide.
They search for signs of extra-terrestrial life (SETI@Home,
\citet{SETI}), study protein-folding (Folding@Home, \citet{FAH}),
search for new drugs (Rosetta@Home, \citet{Rosetta}), search for large
Mersenne prime numbers (GIMPS\footnote{The home page of the Great
  Internet Mersenne Prime Search (GIMPS) is
  \url{http://www.mersenne.org/}.}), simulate the Earth's climate evolution
(ClimatePrediction.net, \citet{Climate}) and so on.

\EAH{} is one of the largest of these projects; to date, hosts
registered by more than 330\,000 people have returned valid results to
\EAH{} and have delivered more than {\it one billion} CPU hours. There
are \EAH{} volunteers from all 193 countries recognized by the United
Nations; currently, more than 100\,000 different computers, owned by
more than 55\,000 volunteers, contact the \EAH{} servers each week,
requesting work and uploading results.

The aggregate computing power of \EAH{} is shown in real-time on a
public server status page\footnote{The Einstein@Home server status
  page gives a real-time display of the number of active hosts, the
  number of active volunteers, and the average CPU power.  It may be
  found at \url{http://einstein.phys.uwm.edu/server\_status.html}.}.
As of 2013 January, it delivers an average of more than one Petaflop
of computing power; according to the current (2012 November) Top-500
list, there are only 23 computers on the planet which can deliver more
computing power on a peak
basis\footnote{\url{http://www.top500.org/lists/2012/11/}} (the
time-average is necessarily lower).  To help understand the scale, it
is useful to provide some cost comparisons.  Simply providing the
electrical power needed to support this amount of computation would
cost three to six million U.S. dollars per year. The costs of hardware and administration would be
substantially greater.\footnote{One can use the Amazon Cloud calculator
  to estimate the monetary costs of replacing Einstein@Home CPU cycles
  with equivalent commercial ``cloud computing'' CPU cycles. For
  example in the last week of 2010 October, Einstein@Home hosts did
  $35\,711$ CPU-weeks of computing.  The hosts are thus the equivalent
  of about 35k CPU cores operating around the clock.  At that time,
  using the Linux/small and Linux/large resources, and leaving out any
  data transfer or storage costs, the estimated cost for the
  Amazon/US-Standard cloud was \$2.2M/month and \$8.7M/month without
  monitoring.}  (Note: at the time of the PSR J2007+2722 discovery in
2010 August, there were about 250\,000 registered volunteers, and
\EAH{} delivered about 200~Tflops of computing power.)

The original and long-term goal of \EAH{} is to search
GW data to find the continuous-wave signals emitted by
rapidly-rotating neutron stars.  The search is an integral part of the
coordinated world-wide effort to make the first direct detections of
GWs.  These were predicted by Einstein in 1916, but
have never been directly detected. A new generation of instruments,
the LIGO in the
USA, VIRGO in Italy, GEO in Germany, and the KAGRA Large-Scale
Cryogenic Gravitational-Wave Telescope Project in Japan, offers the
first realistic hopes of such a detection.  GWs
produced by rapidly spinning neutron stars are one of the four main
targets for these detectors, but because the signals are weak, and the
source parameters (sky position, frequency, spin-down rate, and so on)
are not known, the sensitivity of the search is limited by the
available computational power \citep{bc1,bc2}.

\EAH{} has carried out and published the most sensitive ``blind''
all-sky searches using data from the best GW
detectors. While these searches have not yet detected any signals,
they continue to be a principal target of the project.  Upper limits
obtained from \EAH{} have been published using data from the LIGO instrument's fourth and fifth science runs
\citep[S4 and S5;][]{2009PhRvD..79b2001A,2009PhRvD..80d2003A,2013PhRvD..87d2001A}. \EAH{}
is also re-searching the full S5 and S6 data sets using a new method
that has been proved optimal, for conventional assumptions about the
nature of the instrumental and environmental noise sources
\citep{PletschAllen,2010PhRvD..82d2002P,2011PhRvD..83l2003P}.

Since 2009, \EAH{} has also been searching electromagnetic data from
the Arecibo Observatory, looking for radio pulsars in short-period
orbits around companion stars. As explained in
Section~\ref{s:introduction}, this is an unexplored region of
parameter space, where existing search methods lose much or most of
their sensitivity.

Searches for binary radio pulsars can be characterized by the ratio of
phase-coherently analyzed observation time $T$ to orbital period
$P_\text{orb}$ of the pulsar. There are three cases.  (1) For orbital
periods long compared to the observation time, i.e.\ $T/P_\text{orb}
\lesssim 0.1$, the signal can be well described assuming a constant
acceleration and ``classical'' acceleration searches are a
computationally efficient analysis method \citep{2002AJ....124.1788R}
with only small sensitivity losses. (2) If multiple orbits fit into a
single observation, i.e.\ $T/P_\text{orb} \gtrsim 5$, then sideband
searches, defined in \citet{2003ApJ...589..911R}, provide a
computational short-cut with a factor two to three loss in sensitivity
\citep{2002A&A...384..532J, 2003ApJ...589..911R} compared with the
optimal matched filter coherent search. (3) The intermediate range
$0.1 \lesssim T/P_\text{orb} \lesssim 5$ is accessible with high
sensitivity by time-domain re-sampling with a large number of orbital
parameter combinations (\textit{templates}).

The \EAH{} search is characterized by case (3) above; matched
filtering is used to convolve observational data with large numbers of
templates. These methods and the construction of optimal template
banks have been thoroughly investigated in the context of
GW data analysis \citep{PhysRevD.53.6749,
  PhysRevD.60.022002, PhysRevD.76.082001, 2009PhRvD..79b2001A,
  2009PhRvD..80d2003A} and are adopted here. \EAH{} uses a time-domain
re-sampling scheme to search for radio pulsars in compact binary
orbits \citep{benthesis}. It features a fully-coherent stage, which
removes the frequency modulation of the pulsar signal arising from binary
motion in circular orbits; full details are given in
Section~\ref{subsec:analysis}. The number of trial waveforms is so
large that the required computational resources can only be obtained
with volunteer distributed computing.

\subsection{The Berkeley Open Infrastructure for Network Computing}

Like the majority of volunteer computing projects, \EAH{} is built on
the Berkeley Open Infrastructure for Network Computing (BOINC)
platform.  BOINC was created in 2002 to provide a general-purpose
software infrastructure for this purpose, including all the necessary
server, client-side, and community functions.

Volunteer computing differs from traditional ``grid computing'' or the
use of dedicated clusters, because resources are unreliable, insecure,
and sporadically available, and are donated by participants who are
anonymous and unaccountable.  This creates special requirements for
infrastructure software. BOINC's fundamental design principle is that
{\it every aspect of the volunteer computing system is unreliable
  (perhaps even maliciously so) apart from the central project
  servers}.  To address this intrinsic unreliability, BOINC uses
redundant computing to verify the correctness of results.

For scientists, BOINC is a tool-kit to create and operate volunteer
computing projects.  BOINC provides (1) server software that
distributes work, collects results, and keeps track of hosts, (2) a
client (run on volunteered hosts) that communicates, manages
computation and storage, and displays screen-saver graphics, and (3)
generic Web pages to show account information to volunteers, and to
provide ``community services'' such as message boards, teams, and chat
forums.  Each project runs its own servers, can support multiple
applications with different executables, and is independent of other
projects.

For volunteers, BOINC's design allows participation in multiple
projects, and provides individual control over how the resources are
allocated among them.

\EAH{} was an ``early adopter'' of the BOINC infrastructure, and its
developers have made extensive contributions to BOINC, particularly in
the scheduling system, which determines what work to send to host
computers.  To meet some of the special needs of \EAH{}, BOINC was
also enhanced and extended in a way that made those new features
available to the entire BOINC community.

\subsection{BOINC Internals}

A BOINC project like \EAH{} has two sides: the client side, consisting
of the volunteered host computers (called ``hosts'') and the server
side, which are the computers owned and administered by the project
(called ``the project servers'').  The \EAH{} project servers are
geographically distributed; some are at the University of Wisconsin --
Milwaukee (UWM) and some are at the Albert Einstein Institute (AEI) in
Hannover, Germany.

\subsubsection{BOINC Client Side}
\label{ss:boincclientside}
The ``BOINC Client'' is the most important program running on the
host. This program does not itself do any scientific computation.
Instead, it manages and administers the running of application
executables supplied by one or more projects such as \EAH{}, which the
volunteer has signed up for.  The BOINC client communicates with the
different project servers by sending and receiving small XML files
called ``scheduler requests''and ``scheduler replies''.  When it
detects that the host is idle, it requests tasks from a project,
downloads any needed input data and executables from the project
servers, verifies that they have the correct md5 sums and signatures,
and run the tasks (either from the start, or from a previously-saved
checkpoint).  The BOINC client uses scheduling algorithms to determine
when to run a particular task from a particular project, and when new
tasks and/or data are needed.  It manages the uploading of completed
work, reports the exit status (and any errors) from the executable,
monitors tasks to be sure they are not using too much CPU time,
memory, or disk space, and signals tasks when it is time to
checkpoint.

The executables which the BOINC Client runs on host machines are
called ``applications''; they do the scientific work.  In the case of
\EAH{} they read data files containing instrumental or detector
output, search it for candidate signals, and write the most
significant candidates to a file; a full description is given in
Section~\ref{subsec:analysis}.

When instructed by the BOINC Client, applications checkpoint: they
save enough information to return to the current state in the
computation, so that if interrupted the computation can be completed
without starting from the beginning. The \EAH{} application
checkpointing is described in Section~\ref{ss:checkpointing}.

BOINC application programs are very similar to conventional C-language
programs; however they are linked against a BOINC application library,
which handles the interaction with the BOINC Client.  The library
provides replacements for standard input and output functions: for
example {\tt FILE *fopen() } is replaced by {\tt FILE
  *boinc\_fopen()}. These subroutines interact with the BOINC Client
to ensure that input data are obtained from the project server, and
output data are properly returned to the server.  Another important
library subroutine is {\tt int boinc\_time\_to\_checkpoint()}.  This
must be periodically called by the application, and returns a non-zero
value if the application should checkpoint.  The routine {\tt void
  boinc\_fraction\_done()} must be periodically called by the
application to report the fraction of work completed; the argument is
typically the ratio of the outermost loop-counter to the total number
of iterations.  The last essential library routine is {\tt void
  boinc\_finish()}, which is called by the application to report its
exit status.  The argument is zero if the application completed
correctly, or a non-zero error code if a run-time problem was
encountered.

\subsubsection{BOINC Server Side}
For \EAH{}, the BOINC project servers are located in four 19-inch
equipment racks in a computer server room in the UWM Physics
Department; similar components are located in the Atlas Cluster room
at the AEI. There are also a handful of data download mirrors, located
at participating academic institutions in the USA and Europe.

\begin{figure}[h]
\begin{center}
\ifcase \bwswitch
\includegraphics[width=1.0 \columnwidth]{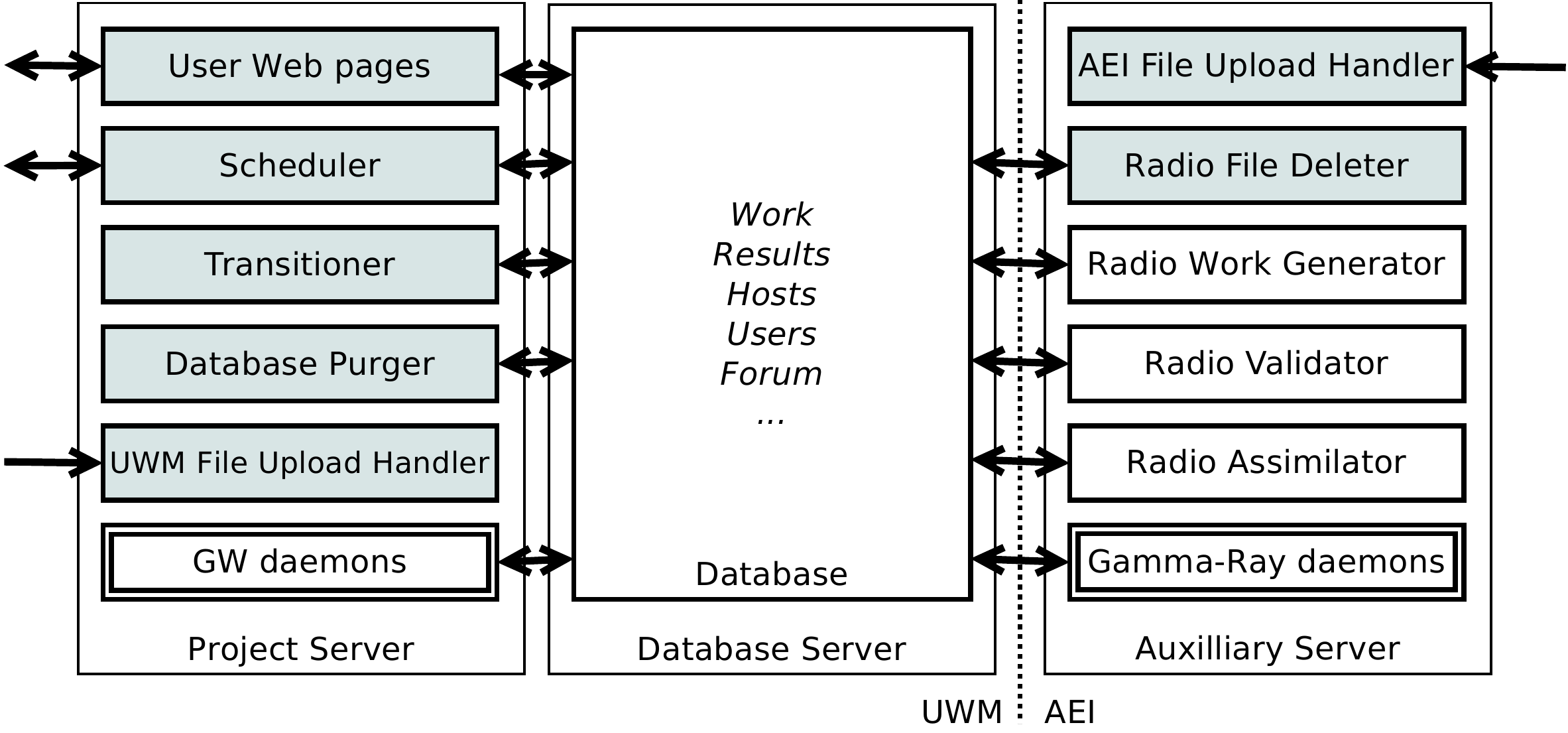}
\or
\includegraphics[width=1.0 \columnwidth]{f1_bw.pdf}
\fi
\end{center}
\caption{A schematic of the most important processes running on the
  \EAH{} project servers, located in the USA (at UWM) and in Germany
  (at AEI).  They communicate with a single central database, that
  ensures coordinated operation and ties the parts together.  The
  components in gray boxes are generic to all BOINC projects and come
  directly from the BOINC software distribution.  The components in
  white boxes are specifically adapted or written for \EAH{}.  The
  gravitational-wave (GW) and $\gamma$-ray workunit generator, file
  deleter, validator, assimilator are not listed individually but
  simply labeled ``daemons''. The arrows pointing externally represent
  network communication with BOINC clients and volunteers. The
  download servers (which provide astrophysical data to the BOINC
  clients) are not illustrated.
\label{f:BOINC_server}}
\end{figure}

The programs/processes running on the \EAH{} project servers are
typical of all BOINC projects, and are illustrated in
Figure~\ref{f:BOINC_server}. Each box denotes an independent computer
program; in the case of \EAH{} these are running on three different
computers at two locations.  As shown in the figure, some of the BOINC
components are generic: the same for all BOINC projects.  Other
components are custom-made for \EAH{}.

The programs are coordinated through a single central MySQL database,
which runs on a high-end server, and is the ``heart'' of the
project. The most important database tables are the User Table, which
has one row for each registered volunteer, the Host Table, which
has one row for each host computer that has contacted the \EAH{}
project, the Work Table, which has one row for each Workunit
(described later), and the Result Table, which has one row for each
separate instance of the Workunit, that is completed, in progress, or
not yet assigned to a specific host.  (For validation purposes, more
than one result is obtained for every workunit, so separate tables are
used for work and results.)  There are other tables which are less
central and not described here, for example the Forum Table contains
community message board items posted by project staff or volunteers.

The majority of the other Project Server components are long-running
background processes.  They typically query the database for a
particular condition, take some action if needed, then sleep for some
seconds or minutes.  For example the \textit{Validator} checks a
database flag to see if there is a workunit with a quorum of completed
results.  If so, it compares the results as described in
Section~\ref{subsec:validation} to see if they agree.  If they agree,
it labels one of these as the ``correct'' (canonical) result, grants
``computing credits'' to the volunteers whose hosts did the work, and
marks the workunit as completed. If the results do not agree, it sets
a flag in the database, which will then be seen by the transitioner,
which will in turn generate a new result for that same
workunit.\footnote{The name ``result'' is misleading.  When first
  created, a ``result'' has not yet been assigned to a host; it is
  simply a line in the database Result Table, and should be thought of
  as the potential result of some future computation.  Only later,
  after the ``result'' has been assigned to a host, and the host has
  carried out the computation and returned its output to the server,
  does the ``result'' actually represent the result of a completed
  computation.} Another example is the \textit{Workunit Generator},
which creates the rows in the work table.  Each row contains the name
and version of the application program to run, the correct command
line arguments and input file name(s), estimates of the required
CPU-time and memory size required, and so on.

An additional set of project server components communicates with
hosts.  The \textit{File Upload Handler} receives completed results
from the BOINC client, through the normal HTTP port 80.  This ensures
that any host which has Web access can be used to run \EAH{}. The
\textit{Scheduler} parses the XML scheduler request files from the
BOINC client.  These typically contain requests for new work, or
report completed work that has been uploaded as just described.  The
Scheduler then queries the database to find new work suitable for the
host, or updates the database to mark that a result has been obtained,
and sends an XML reply to the host.  On \EAH{}, Scheduler requests
typically arrive at the project server at a rate of several Hz.

\subsubsection{BOINC Workflow and Validation}
As explained above, the fundamental design principle of BOINC is that
{\it everything} is unreliable, even maliciously so, with the
exception of the Project Servers.  Thus, when work is sent to hosts, a
correct result might be returned, an incorrect result might be
returned, a maliciously ``falsified'' result might be returned, or the
host machine and its work might simply vanish, never again contacting
the Project Server.  In this hostile environment, BOINC achieves
reliability through replication and validation.

To implement this, the components shown in Figure~\ref{f:BOINC_server}
operate as a state machine. Initially a workunit is created (formally:
a row in the Work Table) by the workunit generator.  The transitioner
then creates a {\it quorum} of ``unsent results''. These are rows in
the Result Table, not yet assigned to hosts.  During its the first
year of operation, \EAH{} used a quorum of three; since then it has
used a quorum of two: to be recognized as valid, ``matching'' results
must be returned from hosts owned by at least two different
volunteers.  The scheduler receives requests from hosts, and
eventually assigns the results to suitable host machines owned by
different volunteers.  The results are then marked with the identity
of the host and with a deadline that is typically two weeks in the
future.

If the computation for the two results is finished and returned to the
server within the deadline, then they are compared by the validator
(described in more detail in Section~{\ref{subsec:validation}}).  If
they agree, then one of the results is chosen as the canonical result,
both hosts and volunteers are credited, and the workunit is over.  If
the results do \textit{not} agree, or if one of the results did not
run to completion and generated a non-zero exit code, or if a result
is not returned to the server by the deadline, then the transitioner
generates {\it another} result (again, a row in the Result database
table) which is subsequently sent by the scheduler to yet another host
owned by yet another user.  This process continues, until a
quorum of valid results is obtained.

To date, in the \EAH{} search of the PALFA dataset, approximately
176~million results have been generated and completed.

\section{The PALFA Survey}
\label{sec:palfasurvey}
The PALFA Survey \citep{2006ApJ...637..446C} was proposed and is
managed by the PALFA Consortium, consisting of about 40 researchers
(including students) at about 10 institutions around the world.  Since
2004, operating at 1.4~GHz, it has been surveying the portion of the
sky that is visible to Arecibo (zenith angle less than $20^\circ$)
within $\pm 5^\circ$ from the Galactic plane.  To carry out a complete
survey will require about $47\,000$ separate pointings of the 7-beam system,
or about $330\,000$ separate beams of data.

Within our Galaxy it is estimated that approximately $20\,000$ normal
radio pulsars and a similar number of millisecond pulsars (MSPs) beam
toward Earth.  The PALFA survey, and the High Timing Resolution
Universe survey (HTRU-North:
\citealt{2011AIPC.1357...52B,2010tsra.confE..66N}; HTRU-South:
\citealt{2010MNRAS.409..619K}) are the final step before a full census
of Galactic radio pulsars is obtained with next-generation telescopes
such as the Square Kilometer Array \citep[SKA;][]{2004NewAR..48.1413C}.
Taking into account achievable sensitivities and radio scattering
limitations, approximately half of these objects are plausibly
detectable with SKA \citep{2004NewAR..48.1413C,2009A&A...493.1161S}.  Approximately 1\% of these
potentially-observable radio pulsars are double neutron-star (DNS) binaries,
and about two-thirds of the MSPs are in binaries with white-dwarf
companions.  About one-quarter of all of these systems are within the portion
of the sky visible to the Arecibo telescope.  The PALFA survey was
initiated to find these pulsars, and to identify the rare systems that
give high scientific returns and act as unique physical laboratories.

Radio pulsars continue to provide unique opportunities for testing
theories of gravity and probing states of matter otherwise
inaccessible to experimental science.  Of particular interest are
pulsars in short-period orbits with relativistic companions, ultrafast
MSPs with periods $P < 1.5$~ms that provide important constraints on
the nuclear equation of state \citep{2006Sci...311.1901H}, MSPs with
stable spin rates that can be used as detectors of long-period
($\gtrsim$~years) GWs \citep{2004NewAR..48..993K}, and
objects with unusual spin properties, such as those showing
discontinuities (``glitches'') and apparent precessional motions, both
``free'' precession in isolated pulsars \citep{1990ApJ...348..226N,
  2000Natur.406..484S, 2011AIPC.1379...23J, 2012MNRAS.420.2325J} and
geodetic precession in binary pulsars \citep{1989ApJ...347.1030W,
  2002ApJ...576..942W, 2003ApJ...589..495K}.  Long period pulsars
(periods $\gtrsim 2$~s) are of interest for understanding their
connection with magnetars \citep{2003ApJ...591L.135M,
  2013MNRAS.429..113H}.  Pulsars with translational speeds (revealed
through subsequent astrometry) in excess of 1000~km~s$^{-1}$ constrain
both the core-collapse physics of supernovae
\cite[e.g.]{2005ApJ...630L..61C,2012MNRAS.423.1805N,2013A&A...552A.126W}
and the gravitational potential of the Milky Way
\citep{2005ApJ...630L..61C, 2009ApJ...698..250C}.

There is also a long-term payoff from the totality of pulsar
detections, which can be used to map the electron density and its
fluctuations, and map the Galactic magnetic field. In the same vein,
multi-wavelength analyses (including infrared, optical and high energy
observations) of selected objects provide further information on how
neutron stars interact with the Interstellar Meduim (ISM), on supernovae-pulsar statistics,
and on the relationship of radio pulsars to unidentified sources found
in surveys at other wavelengths.

\subsection{Importance of, and Expected Numbers of, Pulsars in Short-orbital-period Binaries}
\label{ss:notnumbered}

Strong-field tests of gravity using pulsars have a notable history.
The Hulse-Taylor binary PSR~B1913$+$16, a DNS with a 7.75~hr orbital
period, loses orbital energy via gravitational radiation precisely as
predicted by general relativity \citep{taylor}.  Measurements of
post-Newtonian orbital effects permit the neutron star masses to be
measured to high precision, and provide high-precision tests of the
consistency of general relativity \citep{1989ApJ...345..434T}.  The
shorter 2.4~hr orbital period of the double pulsar J0737$-$3039
provides even better tests of general relativity
\citep{2009CQGra..26g3001K}.  There are strong incentives to search
for binaries with still shorter orbital periods; such compact systems
would provide further stringent tests of general relativity.  But
short orbital-period systems containing active radio pulsars are rare,
so any new discoveries are extremely important.

It is not difficult to estimate the number of short orbital-period DNS
in the Galaxy.  We only need an estimate for the DNS Galactic merger
rate, and a formula for the lifetime of a DNS system as a function of
its orbital period $P_\textrm{orb}$.  Estimating the DNS Galactic
merger rate is not easy \citep{2005ASPC..328...83K,
  2005ApJ...633.1076O, 2008ApJ...672..479O}; current estimates
\citep{2010CQGra..27q3001A} are $R \sim 10^{-4\pm 1}$~yr$^{-1}$. The
GW inspiral time for a circular system of two 1.4
solar-mass neutron stars starting from orbital period $P_\textrm{orb}$
is $\tau = \tau_0 (P_\textrm{orb}/P_0)^{8/3}$, with $\tau_0 \approx
7.1~\rm Myr$ and $P_0 = 1~\rm hr$ \citep{peters,
  1964PhRv..136.1224P}. Thus the expected period for the most compact
DNS in our Galaxy is determined by $R \tau = 1$, implying that the
shortest-period DNS in our Galaxy should have a period $P = P_0(R
\tau_0)^{-3/8} = 5~\rm minutes$ (the above range of $R$ values yields
shortest expected periods from 2~minutes to 12~minutes). The only assumptions
are that the orbital eccentricity is small at the shortest expected
orbital period, and that most DNS systems are born with orbital
periods short enough that their inspiral time is much less than the
Hubble time, 13~Gyr.  Both assumptions are reasonable: some discussion
of the first may be found in Section~\ref{subsec:paramconstraints}.

To estimate of the number of short orbital-period DNS systems one
might expect to find in PALFA data, we also need to know what fraction
of these systems beam towards Earth.  Equation~(15) of
\citet{1998MNRAS.298..625T} predicts beaming fractions of 30\%-40\% for
pulsars having period less than $\approx 200$~ms; 20\% seems a
reasonable compromise between short-period pulsars (which tend to have
broader beams) and long-period pulsars that have narrower beams.

To be detectable in PALFA data, the pulsars must not only beam toward
Earth, they must also lie in the part of the sky visible to PALFA.
Simulations of the DNS population show that these systems are
concentrated toward the Galactic plane and the Galactic center
\citep{2010MNRAS.406..656K}. While Arecibo can see the inner Galaxy,
it can not point closer than $30^\circ$ to the Galactic center; we
estimate that $\approx 25$\% of the DNS population falls within the
sky area covered by PALFA. Thus, multiplying the beaming and coverage
factors, we conclude that $\approx 5$\% of all DNS systems should be
detectable in PALFA data. This number agrees well with a similar
estimate for the detectability of DNS in the PMPS \citep{2011MNRAS.413..461O}.

If 5\% of Galactic DNS systems are detectable in the PALFA survey, the
merger rate of this subset is $0.05~R$; setting $0.05~R\tau = 1$
increases the expected value of the shortest orbital period by a
factor of ${0.05}^{-3/8} \approx 3.1$. Thus we expect there to be a
DNS system visible in the PALFA survey with an orbital period of
$\approx 16$~minutes (the range of $R$ values given above yields
shortest-expected orbital periods ranging from 7 to 37~minutes).  Since
the probability distribution of intervals between events in a Poisson
process is exponential, there is a $1-e^{-1} \approx 63\%$ probability of
finding a system with a period shorter than the expected value we have
calculated. There is a $1-e^{-2} \approx 86\%$ probability of finding a system
with a period shorter than {\it twice} this expected value.

One can derive similar ranges by scaling from the observed numbers of
longer-period systems.  Estimates
\citep{2003Natur.426..531B,2011MNRAS.413..461O} indicate that the
Galaxy may contain as many as $2\, 000$ DNS binaries, with periods $<
10~\rm hr$, of which $\sim 20$\% would beam toward us\footnote{The
  formula in the previous paragraph overestimates the number of
  systems with periods of 10~hr, because such systems are formed with
  eccentric not circular orbits, emit gravitational radiation more
  rapidly, and decay faster.}.  Using the period/lifetime scaling
relationship above (modulo assumptions about birth orbital periods,
whose probability distribution must be convolved with that due to GW
emission) there should then be about 50 DNS systems with periods
smaller than the 2.4-hr period of the double pulsar J0737-3039, or
about 10 that beam toward us.  These numbers then suggest that there
will be $\approx 1$ object beamed toward Earth with a 1-hr period or
less, consistent with our estimate in the previous paragraph.  Given
the uncertainties, there is a reasonable chance that such a DNS binary
can be found in the PALFA survey.

In addition, some neutron-star/white-dwarf binaries will also spiral
in from GW emission while the MSP is still active as a radio pulsar
\citep{1997ApJ...475L..29E}. Given that these systems are far more
numerous than DNS binaries, and that pulsars in
neutron-star/white-dwarf binaries are longer-lived MSPs,
there should be a sizable number visible in PALFA data with
orbital periods less than 1 hr.

Although the prospects are not encouraging, it would be very exciting
to discover a radio pulsar in orbit about a black hole.  This would
likely consist of a normal neutron star with a canonical magnetic
field $\sim 10^{12}$~G; the neutron star would probably be
``canonical'' rather than ``recycled'' because the more massive black
hole progenitor would have formed earlier \citep{2005ApJ...633.1076O,
  2008ApJ...672..479O}. Unfortunately the relatively short
radio-emitting lifetime of canonical pulsars compared to recycled
pulsars, along with the expected smaller absolute number of
neutron-star/black-hole binaries compared to DNS binaries, suggests
that the number of detectable objects in the Galaxy is small.

\subsection{Data Acquisition Spectrometers: WAPPS and Mocks}
\label{ss:dataacquisition}
As briefly described in Section~\ref{s:introduction}, data are taken
with ALFA: a seven feed-horn, dual-polarization, cryogenically-cooled
radio camera operating at 1.4 GHz \citep{2006ApJ...637..446C}.  The
polarizations are summed, to produce an radio frequency signal centered on $\sim
1.4$~GHz.  This is then fed to fast, broad-band autocorrelation
spectrometers.  Until 2009 April, the PALFA survey used correlator
systems, the Wideband Arecibo Pulsar Processors
\citep[WAPPs;][]{2000ASPC..202..275D} to compute and record correlation
functions every $\Delta t = 64~\mu$s.  These mix a 100~MHz bandwidth
to baseband and calculate the autocorrelation for 256 lags.  The
correlation functions are recorded to disk as two-byte integers
combined with appropriate header information in a custom format.  The
\EAH{} analysis used data sets of $2^{22}$ samples, covering
268.435456~s.

The $64~\mu$s sample interval was chosen because many pulsars have
small duty cycles $W/P \ll 1$ (where $W$ is the pulse width and $P$
the spin period) yielding $\sim P/ W$ harmonics that can be combined
into a test statistic (the harmonic sum).  The fast sampling retains
sensitivity to spin periods as short as $P \sim 1$~ms combined with
duty cycles as small as $W/P \sim 1/16$.  If it were not for the
practical constraints of hardware and data volume, even faster
sampling would be desirable.

The complete set of autocorrelation functions for a single 268~s
pointing is recorded in 12 files, each $\approx 2$~GB in size.  Each
set of three files contains the data for two beams. (The last set of three
files contain one ``phantom'' beam of zeros, or a copy of another
beam.)

Since 2009 April, PALFA has used broader-band higher-resolution Mock
spectrometers that incorporate digital polyphase filter
banks.\footnote{ Details of the Mock spectrometers may be found on the
  following NAIC web page:
  \url{http://www.naic.edu/~phil/hardware/pdev/pdev.html}} The Mock
spectrometers cover a frequency bandwidth of 300~MHz, from 1.175 to
1.475~GHz in 1024 channels, with a sample time of $64~\mu$s and a
dynamic range of 16 bits per sample. The operational plan is to cover
the entire survey region ($330\,000$ beams) with this
higher-resolution system.

The Mock spectrometers acquire data with 16-bit resolution, which is
more than we need. To reduce the burden of transfer and storage, data
are rescaled to 4-bit resolution at Arecibo Observatory.  To help
preserve weak pulsar signals in Gaussian-like noise, the
rescaling-algorithm clips outliers (typically arising from
RFI). For each 1~s chunk of data, the median $\mu$ and rms
$\sigma$ are computed for each channel. The data are clipped to the
range ($\mu-2.5\sigma, \, \mu+3.5\sigma$), the floor is subtracted,
then the data are rescaled to 4 bits. The floor subtraction also
flattens the 1~s average bandpass response. The offset and
scaling factors (per channel, per chunk) are saved in the data
structure, and could be used to approximate the original 16-bit data
if desired.

The WAPP data were originally acquired and stored in 16-bit format. In
2011, to reduce the storage volume, it was also reduced to 4-bit
format. The expected total data volume from the complete PALFA survey
is expected to be about 700~TB.

\subsection{Historical Data Acquisition and Processing Rates}

In order to understand how \EAH{} can be used for analysis of PALFA
data, we need to compare the current and historical data acquisition
rates to the \EAH{} data processing rate.  On average, PALFA has been
granted about 265~h of telescope time per year.  About 12\% of the
time is used for follow-up confirmation and initial timing of
newly-discovered pulsars.  Overhead (telescope slewing, calibration)
consumes another 12\%. So about 200~hr of actual survey data are
obtained each year.

\begin{table}[h]
\begin{center}
  \begin{tabular}{c|r|r|r|r}
    \tableline
    Calendar &  Inner  & Total         & Beams  & Beams  \\
    Year     &  Time   &  Time         & acquired &  analyzed \\
    \tableline
    2004 & 69~h    & 108~h  & $15\, 149$~P  \\
    2005 & 278~h   & 365~h  & $25\, 320$~W  \\
    2006 & 250~h   & 360~h  & $28\, 649$~W  \\
    2007 & 72~h    & 143~h  & $11\, 275$~W  \\
    2008 & 182~h   & 184~h  & $ 6\, 640$~W  \\
    2009 & 180~h   & 186~h  & $ 6\, 832$~M & $ 6\, 130$~W \\
    2010 & 249~h   & 275~h  & $10\, 066$~M & $60\, 032$~W \\
    2011 & 175~h   & 434~h  & $24\, 710$~M & $ 7\, 430$~M \\
    2012 & 83~h    & 334~h  & $14\, 126$~M & $27\, 861$~M \\
    \tableline
    & & & $15\, 149$~P &          \\
     Totals    &  $1\,538$~h   & $2\,389$~h  & $71\, 844$~W & $66\, 162$~W \\
         &         &        & $55\, 734$~M & $35\, 291$~M \\
   \tableline
  \end{tabular}
\caption{Annual observation times and data collection volumes for the
  PALFA blind search survey at the Arecibo Telescope, and for \EAH{}
  data processing. ``Inner time'' denotes observations towards the
  inner Galaxy; ``total time'' also includes pointings in the opposite
  direction, towards the outer Galaxy.  ``W'' and ``M'' indicate WAPP
  or Mock spectrometer data; ``P'' indicates pre-survey WAPP data, not
  analyzed by \EAH{}.
  \label{tab:datavolume}}
\end{center}
\end{table}

The annual telescope time (inner Galaxy and total) and data collection
volumes are shown in Table~\ref{tab:datavolume} from the beginning of
the PALFA survey in 2004. The numbers are lower in years when there
were no (commensural) observations antipodal to the inner
Galaxy. Painting work in 2007 and platform repairs in 2010 also
reduced observing time.  The fourth column lists the number of beams
of blind-search survey data acquired in that year, and the
spectrometer used. If everything works correctly, seven beams are
acquired in parallel for each telescope pointing.  The last column
shows the number of beams processed by the \EAH{} data analysis
pipeline\footnote{During much of 2011, \EAH{} was occupied with
  re-processing data from the Parkes Multi-Beam Pulsar (PMPS) survey
  carried out in 1997-2004.  Hence the number of PALFA beams processed
  was small. The results of the PMPS search are reported in
  \cite{2013arXiv1302.0467K}.}. The overall processing speed of the
\EAH{} data analysis pipeline is discussed in
Section~\ref{ss:processingspeed}.  As shown in
Table~\ref{tab:datavolume}, as of the end of 2012, after nine years of
operation, the PALFA survey had acquired $142\, 767$ beams of
blind-search survey data.\footnote{This count does not include data
  collected for confirmation or follow-up observations.}

The accounting of beams of WAPP survey data searched by \EAH{} is as
follows.  The $15\,149$ beams of 2004 WAPP data were taken in a
pre-survey (p1944) mode.  These were not searched by \EAH{} because
they had a shorter time-baseline than the p2030 data that followed,
and the sky pointings were repeated in the p2030 pointings. Of the
original $71\, 844$ beams of WAPP p2030 data, 995 beams were not
transferred to AEI, and $70\, 849$ beams were transferred to AEI. Of
these, $2\,102$ beams were not sent for pre-processing because the
corresponding data file counts were incorrect; $68\, 747$ beams were
sent to pre-processing.  Of these, $1\, 591$ beams could not be
pre-processed because of data corruption or scaling or similar issues;
$67\, 156$ beams were sent to \EAH{} hosts for processing. Of these,
$994$ beams had enough errors during run-time that the corresponding
workunits errored-out or were canceled.  Hence $66\, 162$ beams of
WAPP data were fully-searched by \EAH{}.

As of 2013 January 1, \EAH{} had analyzed a total of $101\,453$
beams ($66\, 162$ WAPP and $35\, 291$ Mock); it is currently
processing about 160 beams of Mock data per day (see
Section~\ref{ss:processingspeed} for details).  Provided that
sufficient telescope time is granted, the survey will continue and
will eventually be extended to higher Galactic latitudes. We expect
the extension to higher latitudes to increase the yield of
MSPs, since they are distributed more widely and their
detection is inhibited by multi-path propagation (interstellar
scattering) that is stronger at low Galactic latitudes.

\subsection{Data Storage and Movement}
\label{ss:datastorage}
Data are recorded to RAID storage systems at the Arecibo
Observatory. Disks containing the data are then shipped to the Cornell
Center for Advanced Computing (CAC), where the raw data are archived
on RAID storage systems for use by the PALFA Collaboration.  For the
\EAH{} search, the data are transmitted over the Internet using
GridFTP\footnote{GridFTP is a high-performance, secure, reliable data
  transfer protocol optimized for high-bandwidth wide-area networks,
  distributed with the Globus toolkit.
  \url{http://www.globus.org/toolkit/docs/latest-stable/gridftp/}}
from CAC to the AEI in Hannover, Germany.
At AEI, they are stored on a Hierarchical Storage Management system.

\section{The Einstein@Home Radio Pulsar Search}
\label{s:EinsteinAtHomeRPS}

The following is a detailed description of how the E@H radio pulsar search works.

\subsection{Preparation of the PALFA Data}
\label{ss:preprocessing}

\subsubsection{WAPP Data}

Before being sent to host machines, data are prepared in a series of
pre-processing steps. The first step is Fourier transformation of the
autocorrelation functions. This produces dynamic power spectra with
256 frequency channels of $390\, 625$~Hz spanning 100~MHz. The
channelization allows compensation for the dispersive propagation of
any pulses from celestial sources.

At AEI, preprocessing is performed separately for each group of three
files containing the autocorrelation functions for two beams.  A
script {\tt preprocess.sh} calls the Cornell/ALFA program {\tt
  alphasplit} to split the files into two sets of three files, each
containing data from a single beam.  For each beam, the script then
calls {\tt filterbank} from the SigProc package.\footnote{SigProc is a
  radio pulsar detection and signal analysis package developed and
  maintained by Duncan Lorimer. The package itself and documnetation
  can be found at \url{http://sigproc.sourceforge.net/}.}  This
reads the three files containing data for that beam.  The output is a
small text header, and a 4~GB file containing $2^{22}$ time samples of
a dynamic power spectra with $256$ channels; power is represented as a
4-byte float.  The header is combined with the data using {\tt
  addheader}; the resulting files (one per beam) are the input to the
\EAH{} Workunit Generator.

\subsubsection{Mock Data}

The first step in the preparation of the Mock data combines two
overlapping sub-band files into a single file with no redundant data,
covering a 300\,MHz bandwidth with 960~channels. The Mock data used
for the \EAH{} pipeline consist of two 4-bit psrfits files for each
beam. Each file covers a bandwidth of 172.0625\,MHz in 512~channels,
one file contains data from a band centered on 1450.168\,MHz, the
other from a band centered on 1300.168\,MHz.  The sub-band files are
$\approx 1.2$\,GB in size, the combined psrfits file is $\approx
1.9$\,GB.

A RFI mask is then computed using \textsc{presto}\footnote{Presto is a
  radio pulsar detection and signal analysis package, obtainable from:
  \url{http://www.cv.nrao.edu/~sransom/presto/}.}
\citep{2002AJ....124.1788R, 2003ApJ...589..911R} software tools. In
addition, strong periodic RFI is identified and added into a
beam-specific ``zap list''.  The RFI mask is used in the generation of
the work units (see next section), while the zap list is sent to the
\EAH{} hosts with all work units of a given beam.

\subsection{Workunit Generation}
\label{ss:workunitgeneration}

The workunit generator has been described in connection with
Figure~\ref{f:BOINC_server}.  It is an ``on demand'' BOINC server
process that prepares data files and ``processing descriptions'' for
the computational work done on \EAH{} hosts. The workunit generator
reads as input one data file per beam\footnote{For the Mock data, the
  RFI mask is also read in through auxiliary files.}, prepared as
described in Section~\ref{ss:preprocessing}. As output it generates data
files (628 per WAPP beam, 3808 per Mock beam) which are later
downloaded by \EAH{} hosts for analysis.  Each of these files contains
one de-dispersed time series, for a different value of the dispersion
measure (DM). The workunit generator also creates one row in the
database Work Table for each beam and for each DM value; these contain
information such as the command-line arguments for the search
application.

To generate workunits from the WAPP input data files, the data for
each beam are de-dispersed with 628 different DM values, and then
down-sampled by a factor of two to 128\,$\mu$s. For the WAPP data, a
single de-dispersed time series has $2^{21}$ time samples with
32\,bits per sample, yielding 8.3\,MB per time series.

\begin{table}
  \setlength{\extrarowheight}{2pt}
  \begin{tabular*}{\columnwidth}{@{\extracolsep{\fill}}ccc}
    \hline
    \multicolumn{1}{c}{DM range} & \multicolumn{1}{c}{$\Delta \text{DM}$} & number of trial values\\
    \multicolumn{1}{c}{(pc\,cm$^{-3}$)} & \multicolumn{1}{c}{(pc\,cm$^{-3}$)} & \\
    \hline
        $0     \; - \;  212.4$ & $0.6$ & $355$\\
	$212.4 \; - \;  348.4$ & $1$ & $136$\\
	$348.4 \; - \;  432.4$ & $2$ & $42$\\
	$432.4 \; - \; 1002.4$ & $6$ & $95$\\
    \hline
    $0     \; - \;  213.6$ & $0.1$ & $2136$\\
    $213.6 \; - \;  441.6$ & $0.3$ & $760$\\
    $441.6 \; - \;  789.6$ & $0.5$ & $696$\\
    $789.6 \; - \; 1005.6$ & $1.0$ & $216$\\
    \hline
  \end{tabular*}
  \caption{\label{tab:dmsteps} Set of DM trial values used in the
    \EAH{} search of the PALFA WAPP (upper half) and Mock (lower half)
    data.}
\end{table}

The discrete DM values are piecewise linear with four distinct
slopes as shown in Table~\ref{tab:dmsteps}; they range from
0 to a maximum of 1002.4\,pc\,cm$^{-3}$.  Since there are (mostly
inner-Galaxy) pulsars with even larger DM values, we may increase
this maximum in future searches: compact H~II regions can create
significant additional dispersion. The spacing at small DM is set
by the requirement that the ``smearing'' over the entire observed
radio bandwidth arising from
the discreteness of DM is less than one sample time. At larger DMs,
the smearing over a single frequency channel becomes the dominant
effect. Also, the increasing electron density along the line of sight
leads to multi-path scattering and pulse broadening \citep{LorimerKramer},
which creates an effective time-smearing larger than the sampling time.
Work by \citet{2004ApJ...605..759B} derived a heuristic relationship
between this pulse broadening and DM; the pulse broadening increases
slightly faster than quadratically with DM. The increasing DM spacing
shown in Table~\ref{tab:dmsteps} is obtained by requiring that the
time-smearing arising from DM discreteness is smaller than the
effective pulse broadening from single-channel smearing and multi-path
scattering. Further details may be found in Section~2.4.2 and 3.7.2
of \citet{benthesis}.

For the generation of workunits from the Mock data, 3808 different
trial DM values up to 1005.6\,pc\,cm$^{-3}$ are used, determined with
the \textsc{DDplan.py} tool from \textsc{presto} and shown in
Table~\ref{tab:dmsteps}. The de-dispersion is done with other tools
from the same software suite, using the previously mentioned RFI masks
to replace broad- and narrowband RFI bursts by constant values. Mock
data are not down-sampled, so there are $2^{22}$ samples per
de-dispersed time series. We initially used a dynamic range of 8\,bits
per sample but halved it to 4\,bits early in 2012 to reduce Internet
bandwidth. The de-dispersed time series generated from Mock data
currently have file sizes of 2.1\,MB.

The workunits cannot all be generated at once.  This would overload
the \EAH{} database server with huge number of rows in the Work and
Result Tables; the resulting time-series data files would also
overflow the \EAH{} download storage servers. So the Workunit
Generator is automatically run ``on demand'' when the amount of unsent
work drops below a low-water mark; it is automatically stopped when
the amount of work reaches a high-water mark.  In this way, the
project typically maintains a pool of tens of thousands of unassigned
results.

To reduce the load on the \EAH{} database server and increase the
run-time per host, up to eight de-dispersed time series are bundled
into a single work unit, as discussed in
Section~\ref{ss:processingspeed}.

\subsection{Signal Model and Detection
  Statistic}\label{subsec:sigmodel}

In searching for possible signals hidden in noise, a model for the
signals is required.  Here, we describe the model used for the signal
from a constant-spin-rate neutron star in a circular orbit with a
companion star.

The phase model $\Phi$ for the fundamental mode of the signal emitted
by a uniformly rotating pulsar in a circular orbit of radius $a$ can
be written in the form \beq \Phi\left(t;\mathbf{\Lambda} \right) =
2\pi f\left(t + \frac{a\sin\left(i\right)}{c}
\sin\left(\Omega_\text{orb} t + \psi\right)\right) +
\Phi_0,\label{eq:fullphase} \eeq where $f$ is the apparent spin
frequency of the pulsar\footnote{This model accurately describes the
  rotation phase of the pulsar for some minutes, which is sufficient
  for the \textit{detection} process. For longer-term observations
  (see Section~\ref{ss:timingmodel}) a more complete and accurate
  phase model is required, for example including additional terms to
  describe a slow secular spin-down. With longer observations,
  parameters such as the frequency $f$ can be determined with great
  precision; by convention it is then defined with respect to time at
  the solar system barycenter at a particular fiducial time.}, $t$ is
time at the detector, and $a\sin\left(i\right)$ is the length of the
pulsar orbit with inclination angle $i$ projected onto the line of
sight. The orbital angular velocity $\Omega_\text{orb}$ is related to
the orbital period $P_\text{orb}$ via
$\Omega_\text{orb}=2\pi/P_\text{orb}$. The angle $\psi$ denotes the
initial orbital phase and $\Phi_0$ is the initial value of the signal
phase.  $\mathbf{\Lambda}$ denotes the ensemble of signal phase
parameters $\mathbf{\Lambda} = \{ f, a\sin\left(i\right),
\Omega_\text{orb}, \psi, \Phi_0 \}$.

The time-domain radio intensity signal is a sum of instrumental and
environmental noise ${\mathcal N}(t)$ and a pulsar signal formed from
harmonics of this fundamental mode \beq
s\left(t;\mathbf{\Lambda}\right) \equiv {\mathcal N}(t) +
\sum_{n=1}^{\infty}
s_n\left(t;\mathbf{\Lambda}\right) \label{eq:sigmono} \eeq where the
intensities of each harmonic are given by \beq
s_n\left(t;\mathbf{\Lambda}\right) \equiv \Re \bigl[ \mathcal A_n
  \exp\left[i n \Phi\left(t;\mathbf{\Lambda}\right)\right]
  \bigr].\label{eq:sigmonon} \eeq The $\mathcal A_n$ are the complex
amplitudes of the different signal harmonics; their values are
determined by (or define) the profile of the observed de-dispersed
radio pulse.

We define a detection statistic $\mathcal P_n$ for the $n$th harmonic
through correlation of the radio intensity with the $n$th normalized
signal template $\exp\left[-i n
  \Phi\left(t;\mathbf{\Lambda}\right)\right]$ for the putative signal.
This detection statistic is optimal in the Neyman-Pearson sense:
thresholding on it minimizes the false-dismissal probability at fixed
false-alarm probability \citep{PhysRevD.66.102003}. It can also be
obtained by maximizing a signal-to-noise ratio (S/N), under the
assumption that the initial phase $\Phi_0$ is unknown and has a
uniform probability distribution; see Appendix B of \citep{2005PhRvD..71f2001A}.

In a search for pulsars, the parameters $\mathbf \Lambda$ are not
known, and so that precise point in parameter space might not be
searched.  However the signal will still appear at a nearby point
${\mathbf \Lambda}'$, for which \beq \mathcal
P_n\left(\mathbf{\Lambda},\mathbf{\Lambda}'\right) = \left|
\frac{1}{T}\int_0^T \!\!\!\!\dm\!t\;s\left(t;\mathbf{\Lambda}\right)
\exp\left[-i n
  \Phi\left(t;\mathbf{\Lambda}'\right)\right]\right|^2.\label{eq:detstatisticintegral}\eeq
Note that $ \mathcal P_n$ is independent of $\Phi_0$ and $\Phi_0'$
because of the maximization described above.  Therefore from here
onward we use $\mathbf \Lambda = \{ f, a\sin\left(i\right),
\Omega_\text{orb}, \psi \}$ to denote a point in the four-dimensional
search parameter-space.

If there is no pulsar signal, or it is very weak, the expected value
of this detection statistic is proportional to the power spectrum of
the instrumental noise in the neighborhood of frequency $nf$.  On the
other hand, if the pulsar signal is strong (in comparison with the noise, so $\mathcal N(t)$ can be neglected), then the expected value is
\begin{align}
& \langle \mathcal P_n\left(\mathbf{\Lambda},\mathbf{\Lambda}'\right) \rangle \approx  \label{eq:expectedvalue} \\
& \left| \frac{\mathcal A_n}{2}\right|^2
\left| \frac{1}{T}\int_0^T \!\!\!\!\dm\!t
    \;\exp\left[i n\left(\Phi\left(t;\mathbf{\Lambda}\right) -
        \Phi\left(t;\mathbf{\Lambda}'\right)\right)\right]\right|^2. \nonumber
\end{align}
This assumes that the observation time $T$ is much
longer than the pulsar period: $fT \gg 1$.

If the instrumental/environmental noise $\mathcal N$ is Gaussian
\footnote{For some beams, the noise $\mathcal N$ contains strong RFI
  and is non-Gaussian. However there are many clean beams where this
  is not the case.  For contaminated beams, the event selection
  procedures described in Section~\ref{subsec:thresholding} also has a
  mitigating effect. In any case, using \textit{lower} thresholds
  based on the assumption of Gaussian noise is justified: RFI does not
  weaken real pulsar signals but instead creates stronger false
  alarms.}, then the detection statistic $\mathcal P_n$ is described
by a non-central $\chi^2$ distribution with 2 degrees of freedom, one
coming from each of the real and imaginary parts of the integrand in
Equation~\eqref{eq:detstatisticintegral}.  The strength of the pulsar
signal determines the non-centrality parameter: in the absence of a
pulsar signal the non-centrality parameter is zero.

The detection statistics $\mathcal P_n$ for different values of $n$ may be
combined to form other detection statistics.  If the pulse profile
were known in advance, a particular weighted sum would be
optimal. Since in practice for blind searches this is not the case, we
need to make some arbitrary choices about what statistics to
construct, and how many such statistics to construct.

To design statistics, we simply assume that radio pulsars have
profiles that resemble a Dirac delta-function, truncated to some
finite number of harmonics. A delta-function has equal weights in all
the amplitudes ($|\mathcal A_n|$ independent of $n$) so we have chosen
to use statistics that equally weight the $\mathcal P_n$ up to some
maximum harmonic.  This choice also makes it simple to characterize
the false alarm probability associated with the resulting statistic.

Thus we define five detection statistics $S_0, \dots , S_4$ by
incoherently summing the values of $\mathcal P_n$ \beq S_L \equiv
\sum_{n=1}^{2^L} \mathcal P_n . \label{eq:statdef} \eeq The statistic
$S_0$ is proportional to the power in the fundamental harmonic of the
pulsar rotation period; the statistic $S_4$ equally weights the power
in the first 16 harmonics.
In the noise-only case  the probability distribution of $S_L$ is 
\beq
  p\left(S_L\right) \dm\! S_L =
  \chi^2_{2N}\left(2 S_L\right) \dm\!
  \left(2 S_L\right),
\label{eq:chisquaredist}
\eeq
which is a chi-square distribution with $2N = 2^{L+1}$ degrees of
freedom.

The false-alarm probability $p_\text{FA}$ is the probability that
$S_L$ exceeds some threshold value $S_L^*$ in the absence of a signal.
This is given by the area under the tail of the probability
distribution $p_\text{FA} = Q_{2N}\left(2 S_L^* \right)$, where \beq
Q_{2N}\left(x\right) = \Gamma\left(x;2N \right) =
\frac{1}{\Gamma\left(2N\right)}\int_x^\infty\!\! \dm\!y\, y^{2N - 1}
e^{-y}\label{eq:imcompgamma}\eeq is the complement of the cumulative
$\chi^2$ distribution function: the incomplete upper Gamma function.
This may be easily computed by means of analytical or numerical approximations.

The detection statistic is unlikely to assume large values in random
Gaussian noise; large values are indications that a pulsar signal may
be present (or that RFI is providing a significant background of
non-Gaussian noise).  We define the significance of such a candidate
as \beq \mathcal S\left(S_L\right) \equiv
-\log_{10}\left(p_\text{FA}\right)\label{eq:sig}.  \eeq A candidate
with significance of (say) 30 has a probability of $10^{-30}$ of
appearing in Gaussian random noise.

\subsection{Template Banks}\label{subsec:templatebanks}

In a search for unknown new pulsars, as explained before
Equation~\eqref{eq:detstatisticintegral}, one evaluates the detection
statistics $S_L({\mathbf \Lambda}')$ at many points in the parameter
space $\mathbf{\Lambda} = \{ f, a\sin\left(i\right),
\Omega_\text{orb}, \psi \}$.  In order to enhance the statistical
likelihood of detection (to maximize the S/N) one
would like to evaluate this quantity at precisely the correct point in
parameter space ${\mathbf \Lambda}' = {\mathbf \Lambda}$ where the
pulsar is located.  But this is impossible, since the pulsar
parameters ${\mathbf \Lambda} $ are not known before discovery!
 
In a practical search, $S_L({\mathbf \Lambda}')$ is calculated for
many different values of ${\mathbf \Lambda}'$.  These ``trial values''
of the unknown pulsar parameters must be spaced ``closely enough''
that not too much S/N is lost from the mismatch
between ${\mathbf \Lambda}$ and ${\mathbf \Lambda}'$.  However if they
are spaced too closely, precious computer cycles are wasted, because
$S_L({\mathbf \Lambda})$ and $S_L({\mathbf \Lambda}')$ are correlated
if $\Delta {\mathbf \Lambda} = {\mathbf \Lambda} - {\mathbf \Lambda}'$
is small.

The set of points in the parameter space ${\mathbf \Lambda}$ where the
detection statistic is evaluated is called a {\it template grid} or
{\it template bank}. An optimal grid will maximize the probability of
detection at fixed computing cost; in general it will {\it not} be a
simple regular Cartesian lattice with uniform spacings along each
axis.  Within the GW detection community, substantial
research work has shown how to construct optimal or near-optimal
template grids \citep[][; H. Fehrmann \& H. Pletsch 2013, in preparation]{PhysRevD.53.6749, PhysRevD.60.022002, 2009PhRvD..80j4014H, 2009PhRvD..79j4017M}; we make use of
those ideas and methods here.

The most important tool for setting up a template bank is the {\it
  metric} \citep{PhysRevD.53.6749} on the search parameter space.  To
simplify matters, consider only the detection statistic $S_0 =
\mathcal P_0$ for the fundamental harmonic of the pulsar.  The metric
measures the loss of the expected strong-signal detection statistic
which arises if the parameters of the search point ${\mathbf
  \Lambda}'$ are mismatched from those of the putative signal
${\mathbf \Lambda}$.  It follows immediately from
Equation~\eqref{eq:expectedvalue} that this loss is described by a
quadratic form in $\Delta {\mathbf \Lambda}$, since the second
modulus-squared term on the right-hand-side (rhs) is maximized (at unity) if the
signal and search parameters match exactly ($\Delta {\mathbf \Lambda}
= 0$).  Thus the fractional loss of detection statistic (called the
{\it mismatch} $m$) must be quadratic in $\Delta {\mathbf \Lambda}$ as
one moves away from this maximum:  \beq
m\left(\mathbf{\Lambda},\mathbf{\Lambda}'\right) = 1 - \frac
  {\langle \mathcal P_0\left(\mathbf{\Lambda},\mathbf{\Lambda}'\right)\rangle} 
  {\langle \mathcal P_0\left(\mathbf{\Lambda},\mathbf{\Lambda}\right) \rangle} = g_{ab} \Delta
\Lambda^a \Delta \Lambda^b + O( \Delta {\mathbf \Lambda}^3).
\label{eq:mismatch}
\eeq Here the indices $a$ and $b$ label the four parameter-space
coordinates $f$, $a\sin\left(i\right)$, $\Omega_\text{orb}$, and
$\psi$, and we adopt the Einstein summation convention where repeated
indices (in this case $a$ and $b$) are summed.  We assume the strong
signal limit, so $\langle \mathcal P_o \rangle $ is defined as in
Equation~\eqref{eq:expectedvalue}.

It is straightforward to show that $g_{ab}$ is a positive-definite
symmetric quadratic form: a metric of signature $(+,+,+,+)$.  The
components of the metric can be computed directly from the phase model
Equation~\eqref{eq:fullphase}.  A short calculation yields \beq g_{ab} =
\langle\partial_a \Phi\partial_b \Phi\rangle_T - \langle\partial_a
\Phi\rangle_T \langle\partial_b \Phi\rangle_T ,\eeq where the angle
brackets denote a time-average \mbox{$\langle \mathcal G \rangle_T \equiv
\frac{1}{T}\int_0^T \mathcal G \left(t\right)\, \dm\!t$} and
$\partial_a$ denotes the partial derivative with respect to the $a$'th
component of $\mathbf \Lambda$.

If the mismatch is small (positive, but much less than unity) then the
surface of constant mismatch is a ellipsoid in parameter space. The
problem of efficient template bank construction is to cover the
desired part of parameter space with the smallest possible number of
these ellipsoids for a given \textit{nominal mismatch} $m_0$.  For a
general (non-constant, as here) metric this template bank is not
regular or uniform.

The quadratic approximation in Equation~\eqref{eq:mismatch} is inaccurate
for typical \EAH{} mismatches ($m_0 = 0.2$ or $0.3$).  For these
values, the region of parameter-space covered by a template is
banana-shaped rather than ellipsoidal; see
Figure~\ref{fig:paramspacewedge}. Thus, in creating template banks,
mismatches are computed using the exact definition
Equation~\eqref{eq:mismatch} rather than the metric approximation.
Nevertheless, the metric is still useful, as described below.

\subsection{Parameter space searched by Einstein@Home}
\label{subsec:paramconstraints}

In order to carry out a search the parameter space must be covered
with a suitable template bank.  Thus, one must decide what region of
parameter space to cover: what range of pulsar spin frequencies,
orbital periods, etc.\ should be searched?  With unlimited computing
resources, one could search the entire physical parameter space. In
practice, \EAH{} has finite computing power, so we can only search
some part of parameter space. Just as an intelligent gambler needs to
decide whether to play blackjack or poker, we need to decide where (in
parameter space) to invest our precious compute cycles.  What
parameter-space regions are most likely to yield a scientific pay-off?

The region to search is astrophysically motivated and targets the
\EAH{} search to the most likely range of putative pulsar orbital
parameters and spin frequencies. We constrain the search parameter
space by setting a probabilistic limit on projected orbital radii, and
by an upper limit on spin frequencies.

As described in Section~\ref{subsec:EAHproject}, standard acceleration
searches lose sensitivity where $P_\text{orb} \lesssim 10T$. For the
PALFA data, this is $P_\text{orb}\lesssim 45$\,minutes. Since other search
pipelines within the PALFA collaboration use standard accelerations
searches, the \EAH{} search was set up to complement these
efforts. Thus, the longest orbital period in the \EAH{} search is
chosen to be $45$\,minutes (plus one template for an isolated system).

The lower limit on $P_\text{orb}$ is determined by the available
computing power: as we show below, the computing cost grows
rapidly as the minimum orbital period decreases.  We choose
$P_\text{orb}\gtrsim 11$\,minutes, significantly increasing sensitivity to
pulsars in compact binary systems.

Even for these short orbital periods, for the purposes of detection,
we can neglect relativistic corrections $\mathcal O((v/c)^2)$ to the
phase model \eqref{eq:fullphase}, because they correspond to less than
a single cycle of phase error. In the worst case, the value of
$(v/c)^2\approx 4 \times 10^{-6}$ for $P_\text{orb} = 660$\,s and
$a\sin(i)=0.2$\,lt-s. Thus, the additional phase accumulated over
$T=268$\,s for a signal at $f=400$\,Hz is $\Delta \Phi \approx
fT(v/c)^2 \approx 0.4 < 1$\,cycles.  This corresponds to an acceptable
worst-case 19\% loss in detection statistic.

Our search, described by the phase model
Equation~\eqref{eq:fullphase}, assumes circular orbits. However as
described in Section~\ref{subsec:templatebanktests} the search is still
sensitive to pulsars in orbits with eccentricities $e \le 0.1$.  Both
theoretical arguments and extrapolation from known pulsars in binaries
suggests that by the time they evolve to the short periods that are
the new feature of the \EAH{} search, their orbits will be
circularized by the emission of gravitational radiation.

We now review the arguments and expectations regarding orbital
eccentricity $e$.  The majority of known pulsar/white-dwarf binaries
have very small orbital eccentricities ($e\lesssim
\text{few}\times10^{-4}$) \citep{2008LRR....11....8L}.  Known
DNS systems typically have larger orbital
eccentricities, but their orbital periods are much longer than the
target values for \EAH{}. These systems will evolve by the emission of
GWs, which over time circularizes the orbits
\citep{peters, 1964PhRv..136.1224P}.  If the known DNS systems from
\citet{2008LRR....11....8L} are evolved until their orbital periods
drop to $11$\,minutes, they are well described by a circular phase model:
the evolved eccentricities $e_{11}$ at an $11\,$minute orbital period are
very small compared to the present-day values.  This is not
surprising: binaries formed with short periods and large $e_{11}$
would decay rapidly through emission of gravitational radiation.  With
the exception of PSR~B1913$+$16 ($e_{11} = 0.0302$) and
PSR~B2127$+$11C ($e_{11} = 0.0416$), we find that
$e_{11}\lesssim0.005$ for all known DNS systems.  Highly evolved
pulsars in such systems are therefore detectable by the \EAH{} search
as show in Section~\ref{subsec:templatebanktests}.

Mass transfer in X-ray binaries also circularizes the orbits of radio
pulsars in compact binaries. As \citet{2009Sci...324.1411A} have
shown, X-ray binaries can become visible as binary radio pulsars after
the accretion stops and radio waves from the pulsar can escape the
system and reach Earth. The orbits of these systems are quickly
circularized during the phase of mass transfer
\citep{2004Sci...304..547S}.  For example, the X-ray binary with the
shortest known orbital period (about 11 minutes) is X$1820-303$
\citep{1987MNRAS.225P...7S}. If the mass transfer stopped and a radio
pulsar emerged, it would have an almost perfectly circular $11\,$minute
orbit. Such objects would probably not be found by an acceleration
search, but might be detected by \EAH{}.

The constraints on the projected orbital radius $a\sin(i)$ are
determined by the expected ranges of pulsar and companion masses. We
allow the maximum allowed value of $a\sin(i)$ to depend on pulsar and
companion masses and on the orbital period. From Kepler's laws we find
\beq \nonumber a\sin(i) \leq \alpha F\left(m_\text{c,max},
m_\text{p,min}\right) \Omega_\text{orb}^{-\frac{2}{3}}
,\label{eq:asinimax} \eeq where $m_\text{c,max}$ is the maximum
companion mass and $m_\text{p,min}$ is the minimum pulsar mass. The
function \beq F\left(m_\text{c,max}, m_\text{p,min}\right) =
\frac{G^\frac{1}{3} m_\text{c,max}}{c(m_\text{p,min} +m_\text{c,max})
  ^\frac{2}{3}} \eeq is a mass-dependent scaling factor, where $G$ is
the gravitational constant. The parameter $0\leq\alpha\leq 1$ bounds
the orbital inclination angles: for given masses $m_\text{p,min}$ and
$m_\text{c,max}$, and given $\alpha$, this condition defines an upper
limit on the projected orbital radii as a function of the orbital
angular velocity.  For the \EAH{} search we selected $\alpha = 0.5$,
$m_\text{p,min} = 1.2$\,M$_\odot$ and $m_\text{c,max} =
1.6$\,M$_\odot$.

\begin{figure}
\begin{center}
\ifcase \bwswitch
\includegraphics[width=\columnwidth]{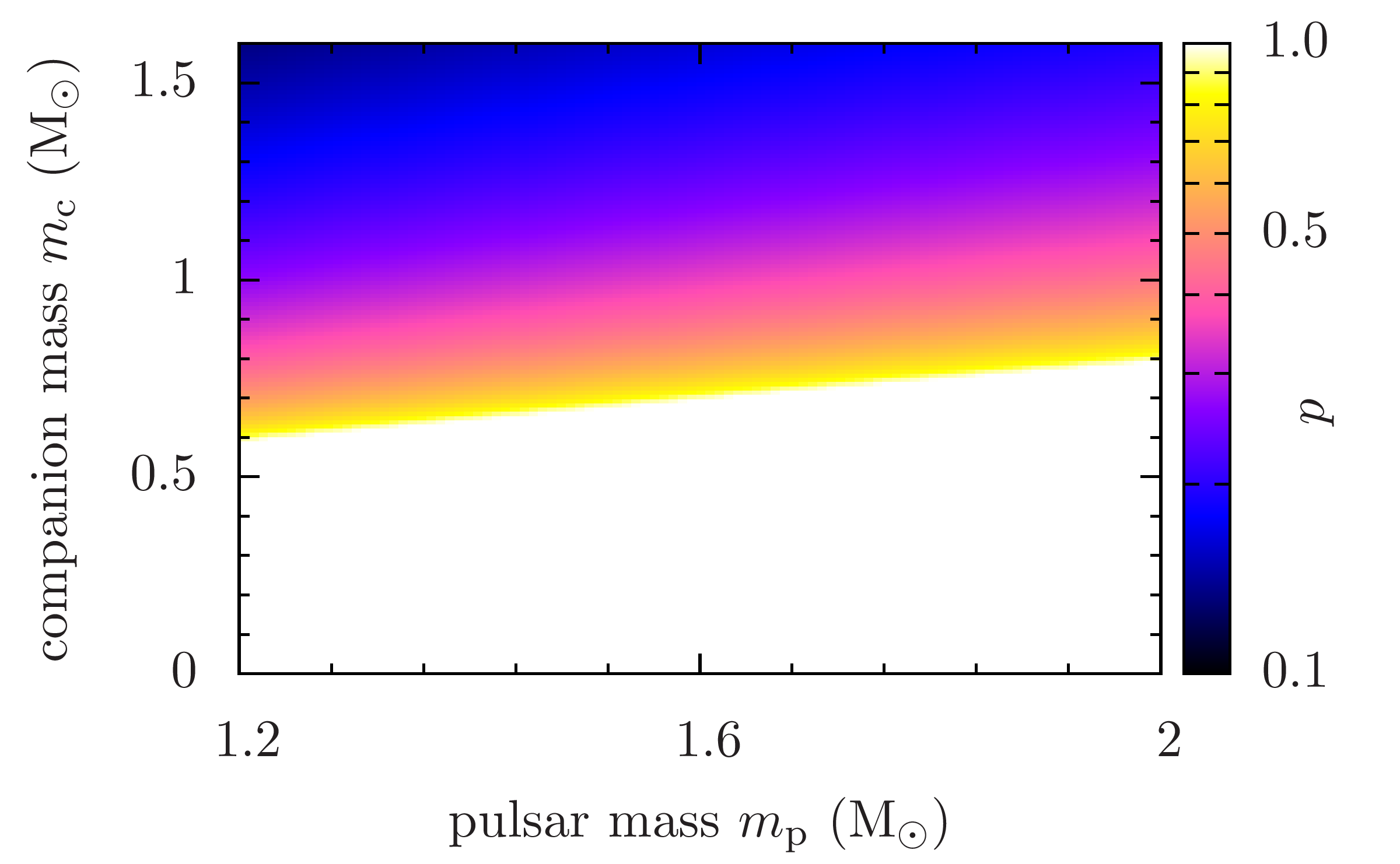}
\or
\includegraphics[width=\columnwidth]{f2_bw.png}
\fi
\caption{The fraction $p$ of total solid angle covered by the \EAH{} search parameter
space, Equation~\eqref{eq:fracsolidangle}. The horizontal axis shows the pulsar mass and the
vertical axis the companion mass. All systems in the
white region are detectable for any inclination angle, elsewhere only a fraction $p$ has
favorable orbital inclinations.}
\label{fig:detectfraction}
\end{center}
\end{figure}

We can use Equation~\eqref{eq:asinimax} to calculate the fraction $p$ of
the total possible solid angle $4\pi$ steradians in which the normal
vector to the orbital plane may lie.  The distribution of possible
orbital inclination angles is uniform in $\cos\left(i\right)$ and thus
the fraction $p$ of systems with inclination angles between $0$ and
$i$ is $p = 1 - \cos\left(i\right) = 1-\sqrt{1-\sin(i)^2}$. For
arbitrary pulsar ($m_\text{p}$) and companion ($m_\text{c}$) masses,
we may write the orbital radius as $a = F\left(m_\text{c},
m_\text{p}\right) \Omega_\text{orb}^{-\frac{2}{3}}$. Inserting this in
the left-hand-side of Equation~\eqref{eq:asinimax} yields $\sin(i)\leq \alpha
F\left(m_\text{c,max}, m_\text{p,min}\right)/F\left(m_\text{c},
m_\text{p}\right)$. From this, the fraction $p$ follows \beq p =
1-\sqrt{1-\alpha^2 \left(\frac{F\left(m_\text{c,max},
    m_\text{p,min}\right)} {F\left(m_\text{c},
    m_\text{p}\right)}\right)^2}.\label{eq:fracsolidangle} \eeq This
quantity, the fraction of orbital inclination vectors covered by the
\EAH{} search parameter space,  is shown in
Figure~\ref{fig:detectfraction}.

The \EAH{} search parameter space is also constrained in maximum spin
frequency $f<f_\text{max}$.  As explained in
Section~\ref{subsec:tempbankconstruction}, the number of orbital
templates grows with $f_\text{max}^3$.  So one must strike a
compromise, choosing a frequency for which \EAH{} can detect a large
fraction of millisecond (and slower) pulsars, while not exceeding the
available computing power.  The search grid is designed to recover
frequency components up to $f_\text{max}= 400$\,Hz\footnote{For a
  pulsar spinning at 100~Hz, this would only recover the power up to
  the fourth harmonic.}.

\begin{figure}
  \begin{center}
  \ifcase \bwswitch
  \includegraphics[width=\columnwidth]{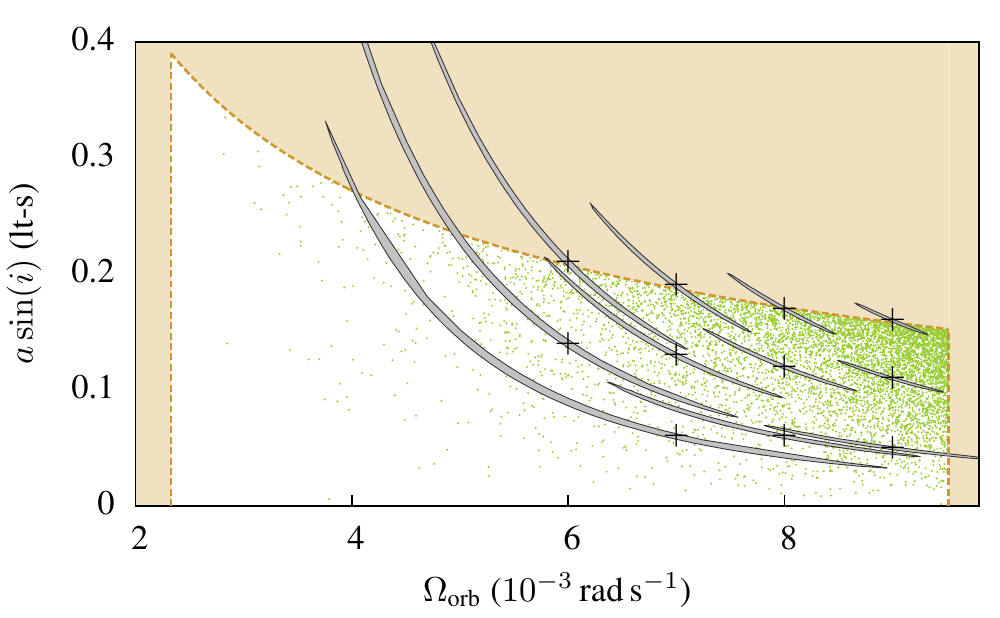}
  \or
  \includegraphics[width=\columnwidth]{f3_bw.pdf}
  \fi
    \caption{The white region is the ``design wedge'' of orbital
      parameter space searched by \EAH{}, as described in
      Section~\ref{subsec:paramconstraints}. The vertical axis is the
      projected orbital radius, the horizontal axis is the orbital
      angular velocity, and the initial orbital phase
      $\psi\in[0,2\pi)$ dimension is suppressed.  The dots are
        the orbital template locations, constructed as described in
        Section~\ref{subsec:tempbankconstruction}.  For a few orbital
        templates (located at the black crosses) the region of
        mismatch $m<0.3$ at fixed $\psi$ for $f=f_\text{max}$ is shown
        in gray. As discussed in Section~\ref{subsec:templatebanks} the
        template coverage regions are banana-shaped, not ellipsoidal.
        }
    \label{fig:paramspacewedge}
  \end{center}
\end{figure}

The constraints above define a wedge of orbital parameter space, shown
in Figure~\ref{fig:paramspacewedge}.

The shorter PALFA data sets spanning $T=134$\,s have different
parameter space constraints. The orbital period range was halved, to
$\SI{5.5}{minutes} \leq P_\text{orb} \leq \SI{22.5}{minutes}$, which also sped
up the overall data analysis. We re-invested this gain into searching
for higher spin frequencies $f\leq 660$\,Hz. The constraint on the
projected radius was left as in Equation~\eqref{eq:asinimax}.

In the part of the PALFA survey using the Mock spectrometers,
there also are some observations covering $T=536$\,s. For the \EAH{}
pipeline we only used the first half of these observations.

\subsection{Template Bank Construction for \EAH{}}\label{subsec:tempbankconstruction}

For the \EAH{} search, we have chosen to construct a template bank
which is completely regular and uniform in the frequency dimension.
Thus, our template bank is the direct Cartesian product of a
uniformly-spaced grid in frequency with a three-dimensional {\it
  orbital template bank} in the remaining parameters ${\mathbf
  \Lambda}_\text{orb} = \{ a\sin\left(i\right), \Omega_\text{orb},
\psi \}$.  Having uniform frequency spacing simplifies matters and
allows the use of Fast Fourier Transforms (FFTs) in the
frequency-domain; FFTs are computationally very efficient if the
frequency points are uniformly-spaced.

In this paper \textit{template bank} refers to the four-dimensional
grid, and \textit{orbital template bank} to the three-dimensional
grid.  To construct the orbital template bank, a three-dimensional
``orbital'' metric is obtained by projecting the metric $g_{ab}$ onto
the sub-space $f=$constant.  A detailed calculation of the
four-dimensional metric $g_{ab}$ and the three-dimensional projected
metric may be found in \citet{benthesis}.  

If a metric is constant or approximately constant, then lattice-based
methods \citep{PhysRevD.60.022002} can be employed to generate
templates covering the parameter space.  However the metric here is
not even approximately constant, and alternative methods are needed.
Two simple and efficient methods are random template banks
\citep{2009PhRvD..79j4017M}, and stochastic template banks
\citep{2009PhRvD..80j4014H}.

For a random template bank, template locations are chosen at random
with a coordinate density proportional to the volume element: the
square-root of the determinant of the metric.  The expected number of
templates can be calculated from the proper volume of the search
parameter space and the chosen coverage $\eta$ and nominal mismatch
$m_0$ \citep{2009PhRvD..79j4017M}.

Stochastic template banks are formed in the same way, but then in a
second step, superfluous templates (those closer than the nominal
mismatch) are removed.

For both random and stochastic template banks, the goal is to cover
most, but not all, of the parameter space; the coverage $\eta \le 1.0$
describes the fraction of parameter space which lies within the
nominal mismatch of one of the template grid points.

As described, \EAH{} template banks are a Cartesian product of a
one-dimensional uniform frequency grid with a three-dimensional
orbital template bank.  This affects the construction of the orbital
template bank in three important ways.

First, the orbital template bank must be created for the highest
frequency used in the search. This is because the same orbital
template bank is used at all frequencies.  Thus its spacing (mismatch)
must be the finest needed at any frequency.  The spacing is finest at
the highest frequency $f_\text{max}$, because the expected detection
statistic Equation~\eqref{eq:expectedvalue} depends upon the
difference in phase, which varies most rapidly at the highest
frequency.  The total number of orbital templates required at a given
mismatch and coverage grows like $f_\text{max}^3$ because the grid
coordinate spacings are proportional to $1/f_\text{max}$ in each of
the three dimensions.

\begin{figure}
  \begin{center}
  \ifcase \bwswitch
  \includegraphics[width=\columnwidth]{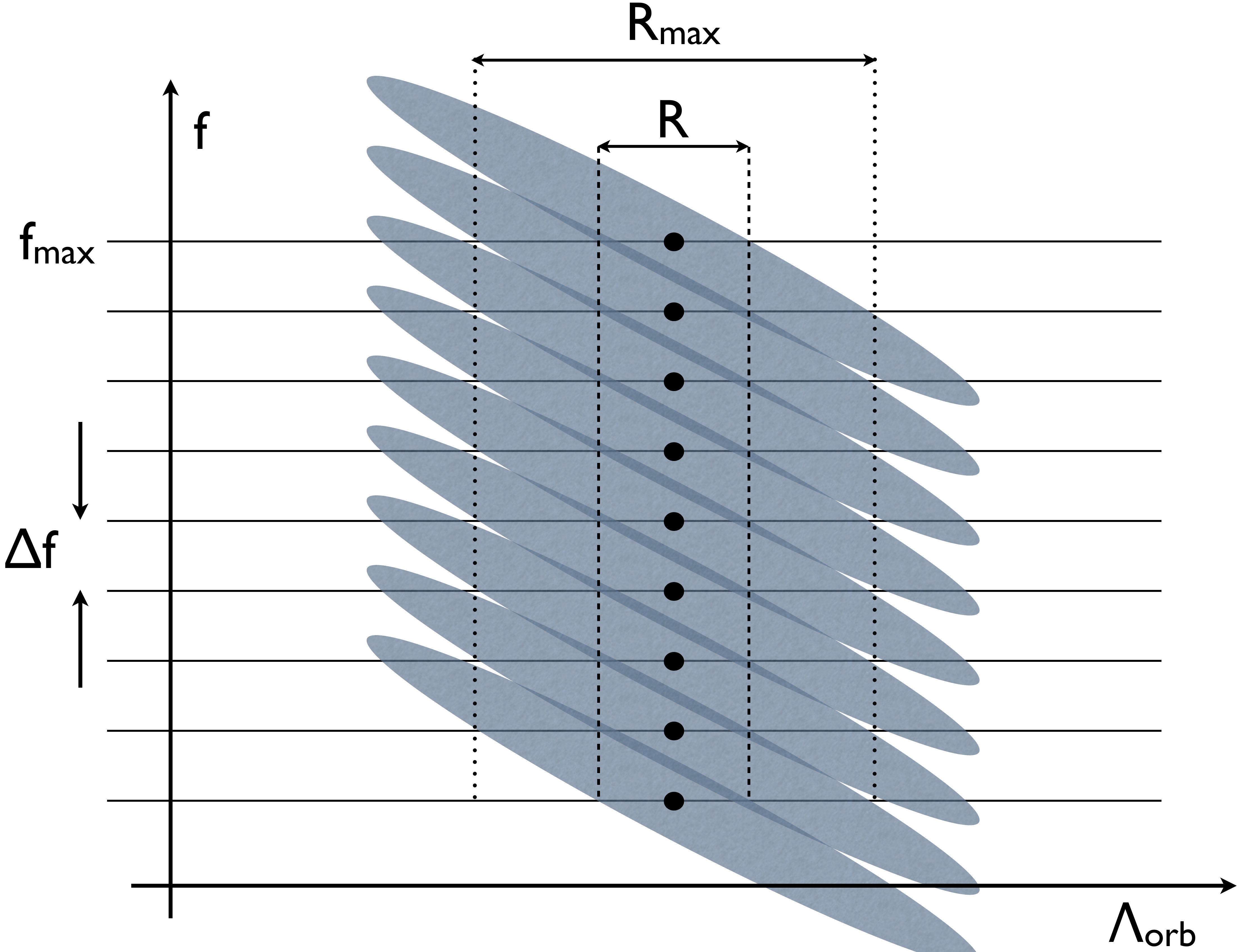}
  \or
  \includegraphics[width=\columnwidth]{f4_bw.pdf}
  \fi
    \caption{\label{fig:redundant} A schematic of template coverage in
      the parameter space with coordinates $\mathbf \Lambda$.  The
      vertical axis is frequency $f$, the horizontal axis denotes the
      orbital parameters $\mathbf \Lambda_\text{obs} $, and the
      horizontal lines denote frequency bins, separated by $\Delta f
      \approx 1\,$mHz and extending up to $f_\text{max} =
      400\,$Hz. The dark dots show template locations; the ellipse
      denotes the coverage region (mismatch $m=0.2$) of one orbital
      template.  Because the four-dimensional grid is a Cartesian
      product, the orbital template is reproduced at each frequency
      bin, separated by $\Delta f$. At fixed frequency a single
      template covers a small region $R$ of orbital parameters.
      However the ``cookie cutter'' copies of the templates cover a
      much larger region $R_\text{max}$ of orbital parameter space,
      obtained by minimizing the mismatch over frequency and orbital
      parameters.  The four-dimensional mismatch $m=0.3$ is allowed to
      be somewhat larger, hence $R_\text{max}$ includes a small amount
      of parameter-space outside the orbital templates. This
      illustration is only schematic because the coverage region of a
      template is not elliptical in shape (see
      Figure~\ref{fig:paramspacewedge}) and can extend over more than
      a hundred frequency bins.}
  \end{center}
\end{figure}

Second, this affects how mismatches are computed between two orbital
templates, in creating a stochastic bank, as illustrated in
Figure~\ref{fig:redundant}. Because the orbital templates are
reproduced at every frequency bin, a given orbital template covers a
larger region of the orbital parameter space than that defined by its
overlap with the surface $f=f_\text{max}$.  The orbital and frequency
parameters are degenerate: one can recover most of the detection
statistic at the incorrect orbital parameter value, provided that the
frequency value is also mismatched.  If the frequency and orbital
parameters are denoted $\mathbf \Lambda = \{f , \mathbf
\Lambda_\text{orb} \}$, then the mismatch between two orbital
templates is
\beq m({\mathbf \Lambda}_\text{orb}, {\mathbf
  \Lambda}_\text{orb}') \equiv \min_{f'} m(\{ f_\text{max}, {\mathbf
  \Lambda}_\text{orb} \}, \{ f', {\mathbf \Lambda}_\text{orb}' \}
). \label{eq:minimization}
\eeq
In practice, the minimum does not occur for $f'$ widely separated from
$f_\text{max}$, so one does not need to search a very large
range. Typically for $f_\text{max} = 400\,$Hz the range needed is less
than $\pm 150\,$mHz.

Third, the mismatch in the four-dimensional parameter space may be
larger than that in the three-dimensional space; in this work the
corresponding values are $0.3$ and $0.2$.  

As previously described, \EAH{} uses five distinct detection
statistics $S_0, \dots, S_4$, which weight contributions up to the
sixteenth harmonic of the pulsar spin frequency. However we use
\textit{the same template bank} for all of these.  The template banks
are designed using only the detection statistic $\mathcal P_0 = S_0$.
Since that statistic only measures the power in the fundamental mode
of the pulse profile, it corresponds to building a search optimized
for sinusoidal pulse profiles.  Thus in constructing and testing
template banks, we only use noise-free simulated pulsar signals whose
intensity profile varies sinusoidally at the spin frequency.

Because it was quick and easy, \EAH{} initially used a random orbital
template bank with $22\,161$ templates.  However after approximately ten
months of operation, this was replaced by a stochastic orbital
template bank containing $6\,661$ templates.  This required an
investment of computer time and human effort, but was justified
because the orbital template bank is used in the analysis of every
de-dispersed time series.

\subsection{Parallel Construction of Stochastic Template Banks}\label{subsec:effparstoch}

It required about $200\,$khr of dedicated computer cluster time to
produce a stochastic bank which was about one-quarter the size of the
initial random template bank.  This reduced the total \EAH{} computing
time by a factor of two, saving hundreds of millions of CPU hours.
The parallelized construction algorithm for
metric-assisted\footnote{The (square root of the determinant of the)
  metric is used to determine the coordinate-density of grid points in
  a random bank.  However in computing mismatches the full detection
  statistic (rather than the quadratic approximation in
  Equation~\eqref{eq:mismatch}) is used.}  stochastic template
placement is described in more detail in Section~3.5 of
\citet{benthesis}; we summarize it here.

Begin by fixing the desired mismatch $m_0$ (here $m_0 = 0.2$).  To
describe the algorithm, it is useful to define operations on
\textit{template banks}.  As before, a template bank (denoted $A$ or
$B$) is a set of distinct points in parameter space (denoted $a$ or
$b$). A template bank $A$ is called \textit{non-overlapping} if for
all distinct points $a,a' \in A$ one has $m(a,a') > m_0$.

The algorithm works by combining pairs of template banks to produce
new ones.  For the description, it is helpful to define a
\textit{merge and prune operation} which we denote $P$.  This
operation takes as arguments (or inputs) two non-overlapping template
banks, and returns (or produces) a single non-overlapping template
bank: \beq P(A,B) = A \cup \left\{ b \in B \; | \; m(a,b) > m_0
\text{\ for\ all\ } a \in A \right\}.  \eeq It is easy to see that if
$A$ and $B$ are both non-overlapping, then $P(A,B)$ is also
non-overlapping.  It is also easy to parallelize into independent
parts.

The algorithm begins with $2^p$ non-overlapping template banks, and
proceeds through $p$ reduction steps, each of which halves the number
of template banks. Each step takes as its input $2^j$ non-overlapping
template banks, and produces as its output a set of $2^{j-1}$
non-overlapping template banks. To carry out a reduction step, the
template banks are grouped into $2^{j-1}$ pairs $A,B$, and each pair
is replaced by $P(A,B)$.  These have increasingly higher coverage at
fixed nominal mismatch.  This procedure continues until a single bank
remains, which is the final output of the procedure.

The algorithm can be trivially parallelized, because a single merge
and prune operation can be trivially parallelized. If the
non-overlapping template bank $B$ is partitioned into $n$ disjoint
pieces $B = \bigcup_{i=1}^n B_i$, then \beq P(A, B) = \bigcup_{i=1}^n
P(A, B_i).  \eeq This also holds if the partition is not disjoint, but
is computationally less efficient.

In practice, the template bank $B$ is partitioned into roughly
equal-sized pieces so that the merge and prune operations take similar
time. The number $n$ of partition elements is selected so that the
compute time required by the merge and prune operations (proportional
to the product of the number of templates in each argument: $|A||B_i|$)
is independent of the reduction level.

For the \EAH{} search, we construct a template bank using $\mathcal
O(1000)$ CPU-cores of the Atlas computer cluster \citep{AtlasRef}.
The number of partitions $n$ is chosen so that the merge and prune
operations $P(A, B_i)$ take about one hour.

The initial input is $2^p = 1024$ non-overlapping template
banks. These are produced as random template banks, each containing
$M=100$ templates, corresponding to $\eta \approx 0.01$ at mismatch
$m_0 = 0.2$.  Then all $M(M-1)/2$ inter-template mismatches
\eqref{eq:minimization} are computed and templates closer than
mismatch $m_0 $ are removed.

We compute the coverage of the final template bank with Monte-Carlo
simulation (or integration). We begin with a large number of simulated
noise-free signals at random points $\mathbf \Lambda$ in the parameter
space. As discussed earlier, these have pulse profiles containing only
the fundamental mode $\mathcal A_i = 0 \text{\ for\ } i>1$: the
detection statistic is $\mathcal P_0 = S_0$.  For each signal, the
mismatch $m$ is computed for all templates, and the minimum is
recorded.  The coverage $\eta$ is the fraction of simulated signals
with mismatches $m<m_0$.

The coverage can also be monitored in the prune and merge operations:
when 99\% coverage has been achieved, $|P(A,B)|$ contains 1\% of the
points from $|B|$. If sufficient coverage has been achieved the reduction
procedure can be terminated ``early'' (before the reduction index
$j=0$).  In this case, one of the $2^j$ remaining template banks is
arbitrarily chosen as the output.

\subsection{Template Bank Verification}\label{subsec:templatebanktests}

We constructed a template bank with $\eta = 90$\% coverage and nominal
mismatch $m_0=0.3$ as described above. For the PALFA data spanning
$T=268\,$s it covers the region of parameter space described in
Section~\ref{subsec:paramconstraints} with $6\,661$ orbital templates.
For data spanning $T=134$\,s, the bank (which now goes to shorter
orbital periods and higher frequencies) contains $7\,113$ orbital
templates. In both cases a single template with $a\sin(i)=0$ was added
by hand to facilitate the detection of isolated pulsars by the \EAH{}
pipeline. The obtained stochastic orbital template bank is shown in
Figure~\ref{fig:paramspacewedge}.

\begin{figure}
  \begin{center}
  \ifcase \bwswitch
  \includegraphics[width=\columnwidth]{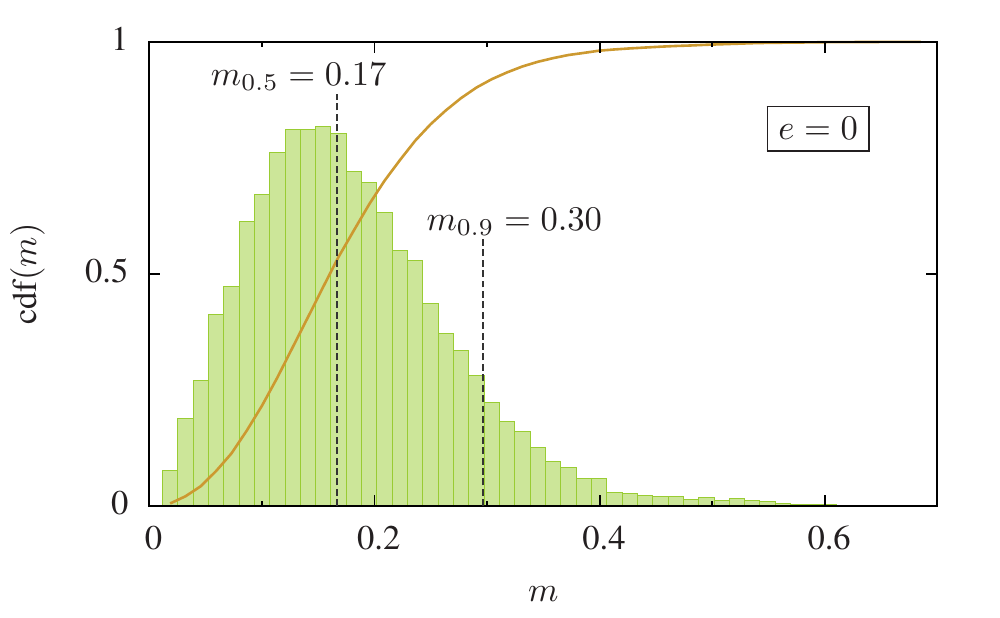}
  \or
  \includegraphics[width=\columnwidth]{f5_bw.pdf}
  \fi
  \caption{Test of the \EAH{} template bank for pulsars in circular
    orbits with a $T=268\,$s data span. The bars show a
    histogram of the mismatch distribution for
    $20\,000$ noise-free signals from simulated pulsars in random
    circular orbits. The curve shows the cumulative
    distribution function (CDF) of the mismatch. The median $m_{0.5}$
    and the 90\%-quantile of the mismatch distribution $m_{0.9}$ have
    been highlighted. The template bank covers 90\% of the parameter
    space with mismatch $m<0.3$.}
  \label{fig:mismatchdist}
  \end{center}
\end{figure}

Monte-Carlo integration (as described in the previous section) was
used to verify that the template banks have the specified coverage and
nominal mismatch. This was done using $20\, 000$ noise-free signals at
$f=f_\text{max}$ with random orbital parameters and a sinusoidal pulse
profile as previously discussed.  The resulting mismatch distribution
(minimum over all templates) is shown in
Figure~\ref{fig:mismatchdist}. It demonstrates that the template bank
has the desired coverage $\eta=0.9$ at nominal mismatch $m_0 = 0.3$.
We note that the median mismatch $m_{0.5} = 0.17$ is significantly
smaller than the nominal mismatch $m_0$.

We used the same method to test if pulsars in elliptical orbits could
be detected by the \EAH{} pipeline.  These signals lie outside our
parameter space, which includes only circular orbits. Thus it was
unclear how well pulsars in eccentric orbits could be detected by
\EAH{}. We again created $20\, 000$ simulated signals at
$f=f_\text{max}$ with random orbital parameters, but with non-zero
orbital eccentricity $e$.  Separate tests were done, with
eccentricities $e=10^{-4}, 10^{-3},10^{-2},0.025, 0.05$, and
$0.1$. The longitude of the periastron was fixed at $\omega = 0$ in
all runs\footnote{Allowing $\omega$ to vary would not change the
  results much: even at the largest eccentricity $e=0.1$ the
  elliptical-orbit phase models have properties similar to the $\omega
  = 0 $ ones.}.  As before, the mismatch was minimized over all
templates in the bank.

For $e \le 0.025$, there was no significant change in the mismatch
distribution: the median and the 90\%-quantile were similar to those
obtained for circular orbits. Thus the \EAH{} search can detect
pulsars in orbits with $e\le0.025$ without significant sensitivity
losses.

\begin{figure*}
  \begin{center}
  \ifcase \bwswitch
  \includegraphics[width=0.49\textwidth]{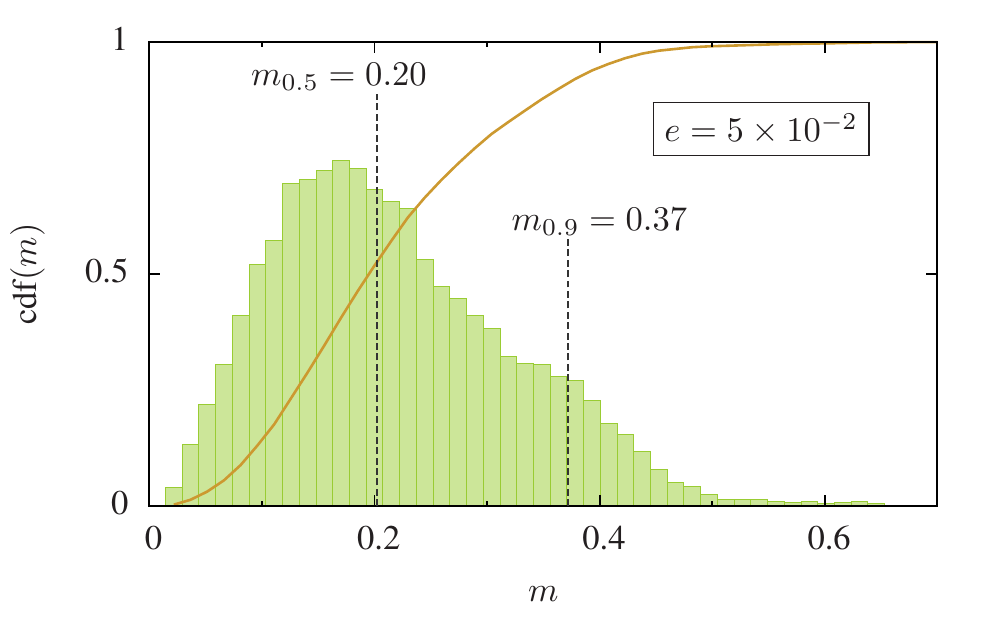}
  \hfill
  \includegraphics[width=0.49\textwidth]{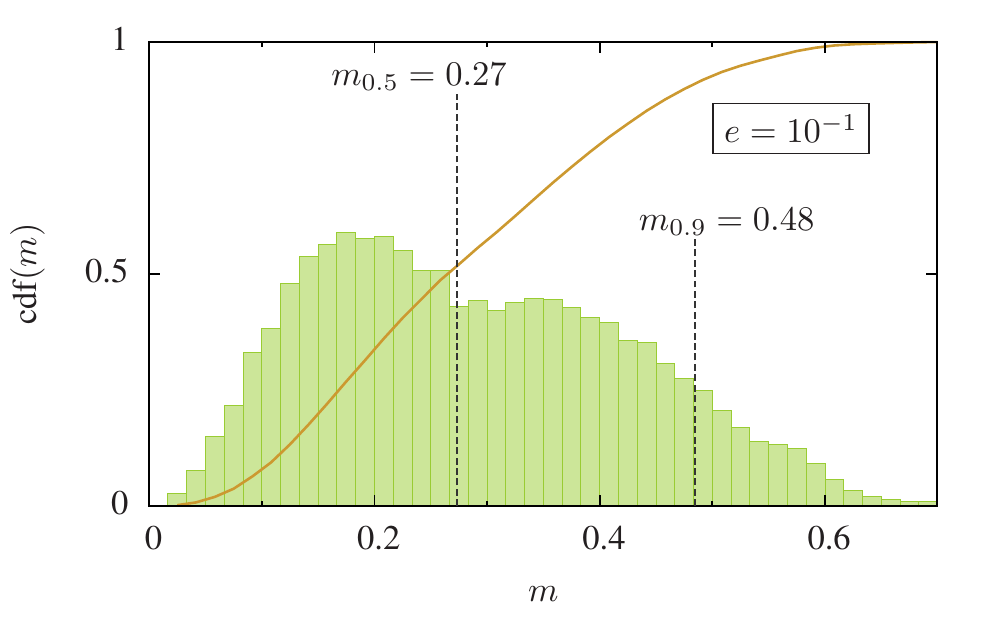}
  \or
  \includegraphics[width=0.49\textwidth]{f6a_bw.pdf}
  \hfill
  \includegraphics[width=0.49\textwidth]{f6b_bw.pdf}
  \fi  
  \caption{Test of the \EAH{} template bank for simulated pulsars in
    eccentric orbits. The left panel shows the results for $e=0.05$,
    and the right panel those for $e=0.1$, respectively. The
    bars show histograms of the mismatch distribution obtained from
    $20\,000$ simulated noise-free signals. The curve shows
    the CDF of the mismatch. The median $m_{0.5}$ and the 90\%-quantile
    of the mismatch distribution $m_{0.9}$ are highlighted. For
    eccentricities of 0.001 and 0.025, there is no significant loss of
    sensitivity compared with the circular orbit tests.
 }
  \label{fig:ecctests}
  \end{center}
\end{figure*}

For $e=0.05$ and $e=0.1$, as shown in Figure~\ref{fig:ecctests}, the
simulations show clear deviations from the mismatch distribution for
circular orbits. The distribution shifts to higher mismatches,
reaching e.g.\ $m_{0.9} = 0.48$ for $e=0.1$. In this case, for $10\%$
of the target signals, about half of the detection statistic
(squared S/N) is lost: detection is still possible, but
the search is less sensitive.

\subsection{Client Search Code}
\label{subsec:analysis}

The client search code is the part of the Einstein@Home radio pulsar
search pipeline which runs on the volunteers' hosts and does the bulk
of the computing work.  Its input is de-dispersed time-series radio
intensity data as described in Section~\ref{ss:workunitgeneration}.
The client search code computes the detection statistics $S_0, \dots,
S_4$ at each template grid point in parameter space, and then returns
back to the \EAH{} server a list of ``top candidates'': the points in
parameter space where the detection statistic was largest.

The client search code is distributed under the GPL 2.0 license and is
publicly available from
Einstein@Home\footnote{\url{http://einstein.phys.uwm.edu/license.php}},
as are binary executables optimized for the complete range of
supported OSs and CPU and GPU types.  Further details of
these optimizations are given in Sections~\ref{subsubsec:cpucode} and
~\ref{subsubsec:gpucode}.

Below, we give a detailed description of how the client search code
operates.  It carries out five main steps:

\noindent
I. The time-series data are uncompressed and type-converted.

\noindent
II. The data are shifted into the frequency domain and whitened,
frequency bins affected by RFI are
``zapped'', and the data are shifted back into the time domain.

\noindent
III. For each orbital template, this new time series is re-sampled in
the time-domain to remove the effects of the orbital motion.

\noindent
IV. The detection statistics $S_0, \dots, S_4$ are computed in the
frequency-domain using an FFT, searched over frequency for the largest
values, and five lists of top candidates are maintained.

\noindent
V. When the iteration over orbital templates is complete, the lists of
top candidates are merged and the most significant candidates are
returned to the \EAH{} server.

\begin{figure}
  \begin{center}
  \ifcase \bwswitch
  \includegraphics[width=0.75\columnwidth]{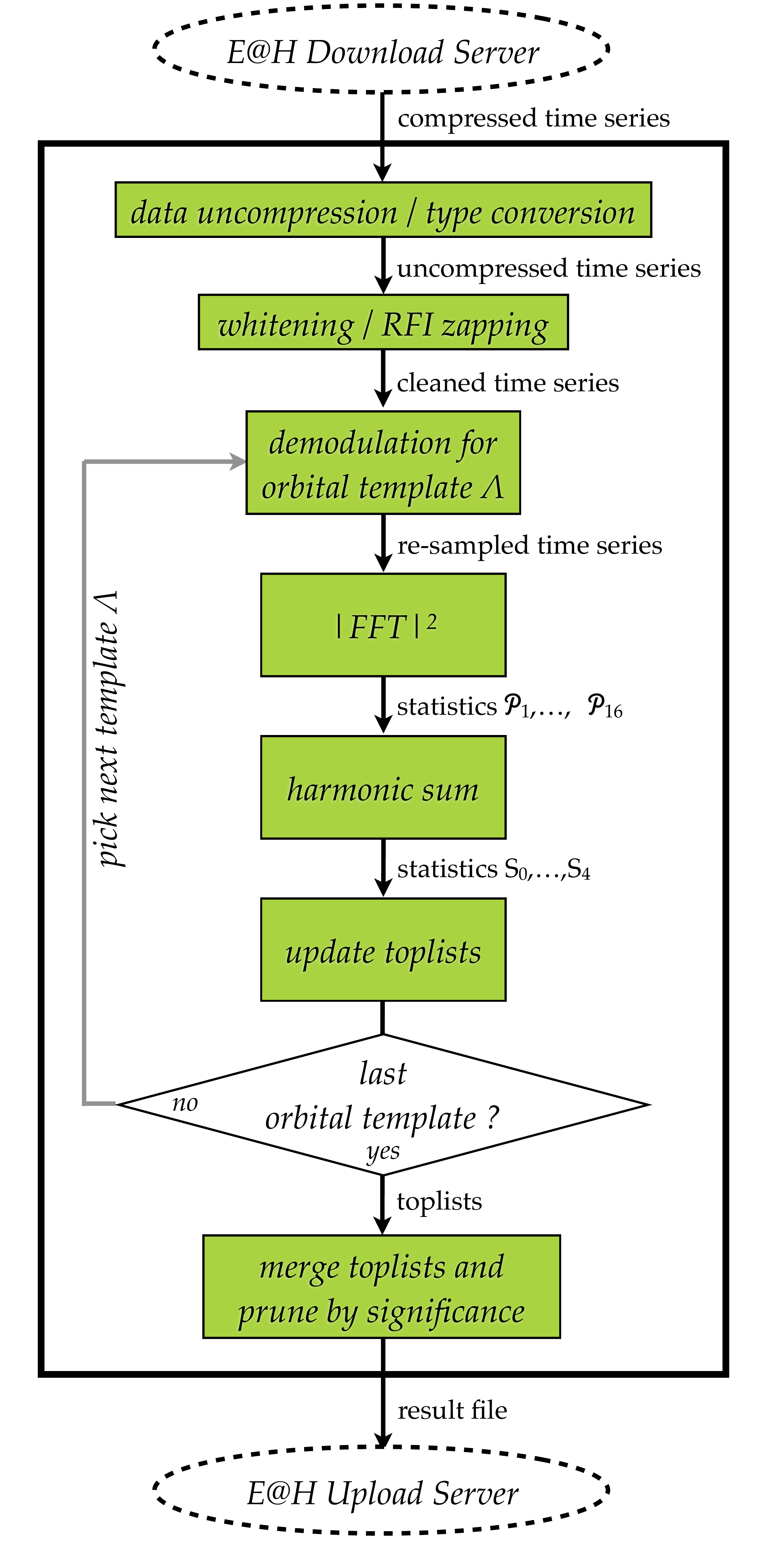}
  \or
  \includegraphics[width=0.75\columnwidth]{f7_bw.pdf}
  \fi
    \caption{Data analysis on the \EAH{} hosts, as described in
      Section~\ref{subsec:analysis}. The client search code (rectangular
      box) receives a de-dispersed time series as input from the
      \EAH{} download server. The data are searched with a large number
      of orbital models, then a list of the most
      statistically-significant candidates is returned to the \EAH{}
      upload server.}
      \label{fig:pipeline}
  \end{center}
\end{figure}

\noindent
These steps are schematically illustrated in Figure~\ref{fig:pipeline}
and described in detail below.

\subsubsection{(I) Data Uncompression / Type Conversion}

The uncompression and type conversion is done immediately after the
client search code receives its input: one of the 628 WAPP (3808 Mock)
different de-dispersed time-series data sets described in
Section~\ref{ss:workunitgeneration}.  In producing these, the original
16-bit or 4-bit instrumental data are converted to floating-point format
for de-dispersion on the \EAH{} server. To reduce the network bandwidth
required to transmit it to the host, the time-series is down-sampled
to 4-bits (and if a significant compression factor can be achieved,
compressed with \textit{gzip}).  The first action of the client search
code is to uncompress the data if required, and then convert it back
into IEEE-754 single-precision floating point representation.  A
factor from the data file header is used to set the overall
scale. This is only needed to avoid dynamic-range problems, and is
irrelevant in what follows.

\subsubsection{(II) Whitening / RFI Zapping}

The next stage of client processing is to whiten the time domain data.
Whitening is necessary because instrumental noise and RFI can result
in a very colored data spectrum.  If the detection statistic were
computed from this colored data, it would be impossible to compare the
statistical significance $\mathcal S$ for templates at different
frequencies.  In addition, the detection statistics $S_1,\dots,S_4$
would be dominated by the ``noisiest'' frequency band which appeared
in the harmonic sum, and their statistical distribution would no
longer be described by the $\chi^2$ distribution of
Equation~\eqref{eq:chisquaredist}, which would make it impossible to
compare the statistical significance of different candidates.

To whiten the time-domain data (time-span $T$) they are first padded
with $2T$ of zeros to produce a time-series of length $3T$.
The data (which have had the mean removed) are then converted into the
frequency domain using an FFT.  The individual frequency bins have a
frequency width $\Delta f \approx 1\,$mHz; their contents are complex
Fourier amplitudes.  The modulus-squared of each amplitude (a
periodigram) is computed bin by bin, then replaced with a running
median value $\mathcal M$ using a sliding boxcar window of width $\pm
500$ bins (covering $\pm0.62\,$Hz). Finally, the data are whitened by
multiplying the amplitude in each bin by
$\sqrt{\ln\left(2\right)/\mathcal M}$. (For Gaussian data this
normalization yields real and imaginary parts that are zero-mean
unit-variance Gaussians.) The first and last 500 bins are not whitened
and are excluded from further analysis.

We use the term ``zapping'' to describe the process of replacing data
in frequency bins that are contaminated with RFI with random Gaussian
noise. Zapping is needed because RFI introduces regular (periodic)
variations into the radio intensity that can mimic pulsar signals and
would dominate the candidate lists if not removed. In
Section~\ref{subsec:thresholding} we describe the ``toplist
clustering'' technique that is used to create a list of top
candidates.  For most beams, these candidates are not dominated by
RFI, although for certain beams the most significant candidates are
from RFI.

The frequency bands to be zapped were selected from a database of
candidates generated by the Cornell pulsar search pipeline
(J. Deneva \& J. Cordes 2013, private communication).
The basic idea is that if an
apparently periodic signal appears in many different sky positions
(beams) at different observation times, it can not be a radio pulsar,
but must be due to RFI.

The Cornell candidate database contained $2\,030\,604$ candidates up
to frequency \SI{7.8125}{kHz} and over the complete range of trial DMs
up to \SI{1000}{pc\,cm^{-3}}.  In the database, $654\,468$ of these
candidates had been flagged as arising from RFI. These candidates were
binned in frequency bins of width $\approx 3.7\,$mHz. Frequency bins
containing more than 200 candidates were then broadened by a
fractional amount of $1.05 \times 10^{-4}$ to account for Doppler
shift in frequency arising from Earth's orbital motion. Overlapping
frequency bands were then merged to obtain a set of non-overlapping
bands, and frequency intervals of $\pm 0.25\,$Hz around the
first three harmonics of the power-line frequency (60, 120, and 180 Hz)
were added. For the \EAH{} search, the relevant part of the zap list is
transmitted to the host along with the search executable.

\begin{figure}
  \begin{center}
  \ifcase \bwswitch
  \includegraphics[width=\columnwidth]{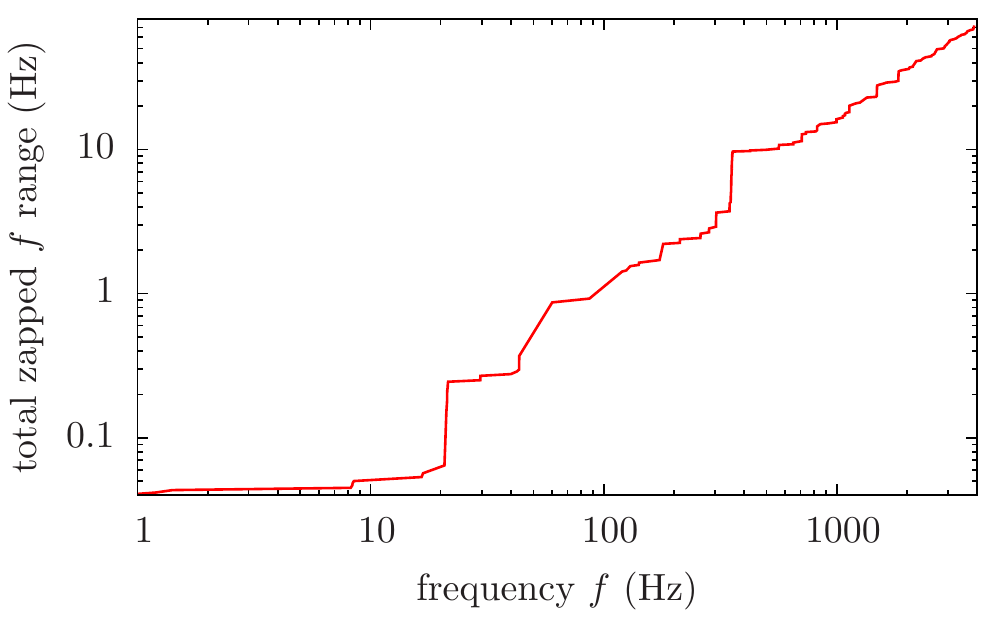}
  \or
  \includegraphics[width=\columnwidth]{f8_bw.pdf}
  \fi
    \caption{The total frequency bandwidth zapped by the fixed zap
    list used in the \EAH{} search as a function of the frequency $f$.
    Below $f=100$\,Hz about $1$\,Hz (1\% of the data) is
    zapped.}
      \label{fig:zaplist}
  \end{center}
\end{figure}

The zap list is a two-column table of lower and upper frequency
values, and extends up to the Nyquist frequency \SI{3.90625}{kHz} of
the down-sampled data. The same zap list is used for all beams: it
contains a total of 233 bands covering \SI{72.383}{Hz}, which
represents 1.85\% of the total bandwidth of \SI{3.9}{kHz}.
Figure~\ref{fig:zaplist} shows the total frequency bandwidth zapped
as a function of the frequency. Note that some recent work has
demonstrated that RFI at Arecibo is highly time-dependent, so
using a fixed zap list is not optimal.  In future \EAH{} searches
it may be beneficial to instead use dynamic beam-dependent zap lists.

The \EAH{} search client receives this zap list and replaces the
amplitudes of the corresponding frequency bins in the whitened Fourier
spectrum with computer-random-number-generated zero-mean Gaussian
noise whose real and imaginary parts have unit variance.  Then the
whitened and zapped Fourier amplitudes are inverse-FFT'd to shift the
data back to the time domain. After the inverse FFT the time series is
cut back to its initial length $T$ by removing the previously padded
bins at the end.  This data conditioning is done only once per
de-dispersed time series when the science code is started.  However if
the code is restarted from a checkpoint, the data conditioning is
repeated, since it takes just a fraction of a second; the whitened and
zapped time series is not stored on the Einstein@Home hosts.

Whitening is done before and not after zapping, because typical RFI
corrupts at most a handful of bins and so does not significantly bias
the median estimator used for the whitening normalization.

\subsubsection{(III) Orbital Demodulation}

The client search code now begins to step through the orbital
templates one-by-one.  For each orbital template with orbital
parameters $\mathbf \Lambda$, the detection statistics $S_L$ of
Equation~\eqref{eq:statdef} are computed on the full frequency grid with
spacing $\Delta f = 1/3T$.

The detection statistics can be efficiently computed in the
frequency domain. To do this, the time-series is first re-sampled
so that instead of being indexed by uniform steps of time $t$ at the
telescope, it is indexed by uniform steps of time $t'$ at the binary
system's barycenter.  This demodulation is done by replacing the
$k$'th sample of the time series at time $t = k \Delta t$ by the
sample closest (nearest neighbor) to time $t'(t)$.  The time
coordinate $t'$ at the binary system's barycenter is defined by the
condition $\Phi(t, \mathbf \Lambda) \equiv 2 \pi f t'$. The definition
of the pulsar spin phase Equation~\eqref{eq:fullphase} then implies \beq t'
= t + \frac{a\sin\left(i\right)}{c} \sin\left(\Omega_\text{orb} t +
\psi\right).\eeq Offsets in time $t'(t=0) \ne 0$ are dropped. They
correspond to constant phase offsets $\Phi_0$, on which the detection
statistic in Equation~\eqref{eq:detstatisticintegral} do not depend.

This transformation means that the phase which appears in the
exponential of the detection statistic
Equation~\eqref{eq:detstatisticintegral} becomes $\exp(- 2 \pi i n f t')$.
Then Equation~\eqref{eq:detstatisticintegral} simply becomes a Fourier
transform\footnote{The systems we search for have non-relativistic
  orbital velocities $v_\text{orb}/c \ll 1$, so the factor $dt/dt' = 1
  + O(v_\text{orb}/c)$ that appears when changing integration
  variables is close to unity and may be neglected.}: the detection
statistic $\mathcal P_n$ is the squared-modulus of the Fourier
amplitude of the re-sampled time series in the $n$th frequency bin.

Before the re-sampled time series is FFT'd to compute the detection
statistics, it is padded with its mean value in the same way as
described earlier: to a total time interval of $3T$. This lessens the
reduction of the detection statistic for putative pulsar signals with
frequencies that do not fall exactly at the center of a Fourier
frequency bin; the maximum loss is 8.8\% \citep{benthesis}.

\subsubsection{(IV) Detection Statistic Computation}

The client search code internally maintains five different candidate
lists (called ``toplists'') corresponding to the detection statistics
$S_0$ through $S_4$.  Here ``candidate'' denotes the point in
parameter space as well as the value of $S_i$.  The $i$th toplist
includes the 100 candidates with the largest values of $S_i$ having
\textit{distinct} values of fundamental frequency $f$.  The toplists
are initialized with null entries ($S_i = 0$) and then updated as
follows.

After the time domain data have been demodulated for an orbital
template and FFTd, five arrays are created, indexed by frequency $f$,
which contain $S_0$ through $S_4$. Note that these statistics are
obtained by combining values of $\mathcal P$ for harmonically-related
frequency bins. This ``harmonic summing'' can be quite compute
intensive, in part because it requires striding over widely separated
parts of the frequency-domain arrays, summing elements. For computational
efficiency, the number of required summations are minimized by re-using
the $S_L$ with smaller $L$ to compute those with larger $L$ \citep{LorimerKramer}.

The array containing the detection statistic $S_i$ is then stepped
through, element by element.  If the statistic $S_i(f)$ is less than
the smallest statistic currently on the $i$th toplist, then the next
element is considered.  Otherwise, the toplist is searched to see if
it contains a candidate \textit{at the same fundamental frequency}
$f$.  If not, the toplist candidate with the smallest detection
statistic is replaced with the new, higher-statistic candidate. If so,
then the candidate at the same frequency is replaced with the new
candidate if and only if the new candidate has a larger value of the
detection statistic than the existing candidate. This procedure
ensures that the 100 entries on each toplist are for 100 distinct
frequencies.

The comparison process required to insert new candidates in the
toplist can be quite compute-intensive.  To speed it up, the
comparison is only carried out for values of the detection statistic
that lie above a predefined threshold. The threshold is
data-independent: it is the largest statistic value expected in
Gaussian noise for the relevant number of ``independent trials''
(roughly speaking, this is the number of orbital templates $\times$
the number of frequency bins). Further details may be found in
Section~\ref{subsec:thresholding}

\subsubsection{(V) Result Files}
When the loop over orbital templates is finished, the search code
computes the statistical significance Equation~\eqref{eq:sig} of the 500
candidates stored in the five toplists. These are then winnowed
further: the 100 candidates with the largest statistical significance
are selected, sorted into canonical order, and returned to the \EAH{}
server in a single \textit{result file}.  The remaining 400 candidates
are dropped.

Each de-dispersed time series generates one fixed-format ASCII text
result file.  Five lines contain identifiers for the volunteer and the
computer that did the computation, the date that the computation was
completed, and similar information. The remaining 100 lines are for
the most significant 100 candidates winnowed from the toplists.

Each candidate line contains seven white-space-separated values: the
spin frequency $f$ in Hertz, the orbital period $P_\text{orb}$ in
seconds, the projected orbital radius $a \sin(i)$ in light-seconds,
the initial orbital phase $\psi$ in radians, the detection statistic
$S_L$, the statistical significance $\mathcal S$ defined by
Equation~\eqref{eq:sig}, and the number of harmonics $2^L$.

\subsection{Thresholding and Candidate Selection}
\label{subsec:thresholding}

As discussed above, candidates are only checked against toplist
entries if their statistics $\mathcal P_n$ exceed certain
thresholds. These thresholds are computed from a false-alarm
probability, provided as command-line parameter to the search code.

The false-alarm probability for each orbital template in any
de-dispersed time series is set to $p_0=0.08$. For 6661 orbitals
templates and in pure Gaussian noise data, we expect $6661\times 0.08
\approx 530$ candidates to exceed this threshold after all orbital
templates have been searched.  Thus, the search code should always
return $\gg 100$~candidates for each $\mathcal P_n$, after searching
the complete template bank, and fully populate all five toplists.

For easy thresholding during runtime, the global false-alarm threshold
$p$ is converted into a single-FFT-bin false-alarm threshold
$p_\text{single}$ and thresholds on the detection statistics $\mathcal
P_n^*$. The probability of not having a false-alarm in $N_f$
frequency bins in random Gaussian noise is $1-p =
(1-p_\text{single})^{N_f}$. From this, we find $p_\text{single} =
1-(1-p)^{1/N_f}$. The detection statistic threshold $\mathcal P_n^*$,
is determined indirectly by $p_\text{single}=Q_{2N} (2\mathcal
P_n^*)$, where $Q$ is the incomplete upper gamma function as in
Equation~\eqref{eq:imcompgamma}.

We compared these expectations, based on Gaussian noise, with results
from real data, and were able to verify that the \EAH{} search is not
dominated by non-Gaussian noise. The returned candidates in a single
de-dispersed time series typically have $\mathcal S \gtrsim 8.5$,
unless strong pulsar or RFI signals are present. For most beams this
is not the case.  The number of total trials per de-dispersed time
series (neglecting parameter correlations in the detection statistic)
is the product of the number of frequency bins and the number of
orbital templates $N_\text{tot} = N_f \times N_\text{templ} =
3\times2^{21} \times 6661 \approx 4\times10^{10}$. Assuming that the
number of candidates exceeding a particular significance threshold
follows binomial statistics, one expects of order $N_\text{tot} \times
10^{-8.5} \approx 133$ candidates with $\mathcal S \gtrsim 8.2$ from
noise alone. Indeed, the search code always reports $\gtrsim$100
candidates, validating the assumption above.

As described, each beam is analyzed with 628 different DM values for
WAPP data and 3808 different DM values for Mock data, respectively.
For each DM value, the 100 top candidates are returned.  So the search
procedure always returns $ 62 \, 800$ ``candidates'' or  $380\,800$
``candidates'' per beam, respectively, regardless of whether RFI is
present or absent in the beam. Moreover, the 100 candidates for each
DM value are at distinct frequencies. This makes it harder for RFI to
dominate the candidates for a given beam.

There is a consensus among radio astronomers that RFI has become more
severe in the past decade, probably due to the proliferation of
wireless devices such as cellphones and WiFi.  Nevertheless, the
procedures we have described are reasonably effective in mitigating
the effects of this RFI.  The \EAH{} search is not dominated by
non-Gaussian noise, in the sense that a typical beam returns statistic
values in the expected ranges for Gaussian noise. Of course there are
beams which contain strong RFI or strong pulsars for which this is not
the case.

If one looks across the entire search (not beam-by-beam) the top 1\%
of candidates are \textit{not} consistent with Gaussian noise: these
arise from pulsars or RFI.  However if one looks further down the
list, the distribution of statistic values are reasonably consistent
with Gaussian noise. In fact the situation is similar for the Pulsar
Exploration and Search Toolkit (PRESTO) processing pipeline, which is
also used to process PALFA data.  In that pipeline, for each beam, the
200 strongest candidates are followed-up (folded and refined).  For
beams that are strongly affected by RFI, most or all of these
candidates are not consistent with Gaussian noise.  However for the
majority of beams, the bulk of candidates are consistent with Gaussian
noise.

\subsection{Client Search Code Checkpointing}
\label{ss:checkpointing}

The search execution on the host may stop for many reasons.  For
example the volunteer might turn off the computer, or the BOINC client
might stop execution because it appears that the volunteer is busy
using the computer for other purposes.

As described in Section~\ref{ss:boincclientside} the client search
code checkpoints on a regular basis, by default once per minute.  This
checkpointing saves the internal state of the search, and permits it
to be efficiently restarted with very little computing time lost.  The
checkpointing is done by sorting and saving the toplist files, and
then saving a counter which records the last orbital template that was
completed.

When the search is started (or restarted) it carries out the whitening
and zapping steps on the input data, and then checks if a valid
checkpoint file exists. If not, the search begins execution at the
first orbital template as previously described.  However if a valid
checkpoint file is found, then the toplists are initialized from the
stored values, and the loop over orbital templates begins following
the orbital template index recorded in the checkpoint file.

\subsection{CPU Implementation of the Search Algorithm}
\label{subsubsec:cpucode}

The search algorithm is implemented in the C programming
language. Mathematical functions are provided by the standard C math
library with special functions from the GNU Scientific Library
\citep{gsl} and FFT routines from the Fastest Fourier Transform in the
West\footnote{The Fastest Fourier Transform in the West (FFTW) package:
\url{http://www.fftw.org/}}
\citep[FFTW;][]{FFTW05,JohnsonFr08:burrus} library.  The search code is then wrapped
into the BOINC framework \citep{Anderson:2006:DRS:1188455.1188586} as
described earlier. The implementation is single-threaded, i.e., hosts
simultaneously execute one instance on each CPU core that BOINC
allocates to the search.

To produce executable binaries, the Linux applications are compiled
using standard GNU tools. The applications for Mac~OS~X are built
using the Mac~OS~X 10.4 SDK build environment. For Windows, the
applications are cross-compiled on Linux machines using the MinGW
tools\footnote{\url{http://www.mingw.org/}}.  The underlying compiler
in all three cases is the GNU C Compiler; this permits identical
optimizations and execution ordering on all platforms.

\subsection{\EAH{} Processing Speed / Throughput}
\label{ss:processingspeed}

The speed with which \EAH{} can process one beam of PALFA data is
determined by the amount of computing time required for a single
workunit and the number of workunits per observed beam. These have
varied over the years as the processing code was made more efficient;
the number of participating volunteers has also varied.

Individual workunits should not take too long to run on a host:
volunteers become discouraged if the results of their processing do
not quickly lead to successful results and visible computing credits.
The workunits also should not be too short, or the \EAH{} database
gets too large to operate efficiently, and the overhead of uploads,
downloads, and sending new workunits to hosts becomes excessive.  In
general our goal has been to have workunit run-times of between one
hour and one day.  As the application code became faster, we achieved
this by bundling multiple single workunits into larger ones: the
runtime has remained between one hour and one day for the majority of
the hosts, the lower end populated by the GPUs.

The first implementation of the \EAH{} search ran from 2009~March to
2010~February and processed on average $\approx 25$ WAPP beams each
day.  After that, two major code improvements increased the processing
speed by a factor of $\sim 6$, and between 2010~February and
2010~August, \EAH{} processed $\approx 160$ WAPP beams per day. The
first GPU version of the search code increased the processing rate to
more than $300$ beams per day between 2010~September and
2010~December.

The \EAH{} search of the Mock spectrometer data started in 2011~July
and processed on average $\approx 50$ beams per day until
2012~September.  After that date, the processing rate gradually
increased over a period of three months and has been running at around
$160$ beams per day since the end of 2012. As of 2013 February, the
majority of the Mock data (see Table~\ref{tab:datavolume}) has been
analyzed, and the data processing backlog is less than two months.

\subsection{GPU Implementation of the Search
  Algorithm}\label{subsubsec:gpucode}

As previously described, \EAH{} also takes advantage of the GPUs
available on a substantial fraction of host
machines, providing applications for NVIDIA GPUs which support CUDA
version 3.2 or higher, and for AMD/ATI GPUs which support OpenCL
version 1.1 or higher. CUDA and OpenCL are programming models, API
interfaces, and support libraries which enable GPUs to be used for
scientific computation.

The supported GPUs typically execute double-precision floating point
operations very slowly compared to single-precision operations, or do
not support them at all. So the CPU codes were designed so
that all floating-point operations can be performed in single
precision. Tests with simulated pulsar signals were performed to
ensure that this does not degrade the sensitivity of the search.

The code was also designed to have a reasonably small memory
``footprint'', particularly because of limits imposed by
consumer-grade graphics cards. The GPU version requires less than 250
MB of GPU memory, which substantially enlarges the set of GPU cards on
which the code can run.

The overall structure of the GPU code is similar to that of the CPU
version (see Figure~\ref{fig:pipeline}), with the most compute-intensive
analysis offloaded to the GPU.  These are the time-series re-sampling
to remove the effects of orbital motion via demodulation, the FFT and
power spectrum computations, and the harmonic-summing to obtain the
$S_L$ from the $\mathcal P_i$.  For NVIDIA GPUs, the CUDA 3.2
programming framework
\footnote{\url{https://developer.nvidia.com/cuda-toolkit-32-downloads}}
was used to embed calls to CUDA-C code (kernels) executing on the
GPU. On AMD/ATI GPUs, the OpenCL programming framework
\footnote{\url{http://www.khronos.org}} was used for the same purpose.

To maximize GPU utilization, the GPU implementation of the time-series
re-sampling is split into five CUDA kernels to maximize thread
parallelization. To avoid the overhead of memory transfers to the host
CPU, intermediate output is kept in GPU memory as much as
possible. The time-offsets $t-t'$ needed for re-sampling are computed
in parallel, using a lookup table and interpolation to avoid costly
sine/cosine operations. An identical lookup table is pre-computed for
both CPU and GPU hosts to help ensure that their results
cross-validate later in the processing pipeline (see
Section~\ref{subsec:analysis}). We use intrinsic functions to avoid
generating fused multiply-add instructions that could introduce
rounding errors which would also hamper cross-validation. The length
of the modulated time series is computed in a separate kernel, and the
re-sampling itself is done by yet another kernel, using the
time-offsets and the time-series length computed in the
previously. Each time-series sample is computed in parallel by a
separate GPU thread.  A parallel sum-reduction algorithm is then used
to compute the mean of the re-sampled time-series, and a final CUDA
kernel implements the mean-padding of the re-sampled time-series.

To perform FFTs efficiently on NVIDIA GPUs, the NVIDIA \textsc{cufft}
3.2 library\footnote{\url{https://developer.nvidia.com/cufft}} is
used.  The \textsc{cufft} library has an FFTW compatibility mode,
which simplified development and integration with the CPU code.  A
custom CUDA kernel is used to compute the power spectrum in parallel
from the FFT output.  Intermediate as well as the final output (for
the next step) is again kept in GPU memory.

The GPU implementation of harmonic summing differs from the CPU
version: the GPU version re-orders the computations so that hundreds
of processing cores on the GPU can independently perform calculations
in parallel.  Memory caching is needed, because of the low locality
and irregular access strides associated with summing the different
harmonics of $f$.  Caching is done in texture memory; without it the
memory access pattern of the individual threads would be very
inefficient. Write accesses have been eliminated, except for those
associated with the (comparatively rare) signals that might make it
onto the candidate toplist, i.e. detection statistics exceeding the
false-alarm threshold and the weakest toplist signal.

GPU versions of the host applications are provided for Linux, Windows,
and Mac~OS~X operating systems. The Windows version is cross-compiled
under Linux for the same reason as described in
Section~\ref{subsubsec:cpucode}: to improve cross-platform result
validation. This also allows for a tighter integration in the
automated build system of Einstein@Home, but adds complexity because
cross-compilation requires the use of the lower-level CUDA driver API
instead of the higher-level CUDA runtime API.

The OpenCL implementation differs somewhat from the CUDA one. The FFT
library is derived from software developed by
Apple\footnote{\url{http://developer.apple.com/library/mac/\#samplecode/OpenCL_FFT}}.
As provided, the Apple library can only do complex-to-complex FFTs of
arrays whose length is a power-of-two ($2^n$); we extended it to
efficiently do real-to-complex transforms of length $3 \times 2^n$, as
required by the search code. It was also modified to eliminate calls
that approximate trigonometric functions with different accuracy on
different GPUs. This reduces the numerical difference between
different GPU models, making the results more hardware-independent
and simplifying result cross-validation.

OpenCL is a vendor-independent framework and the OpenCL application
also runs on NVIDIA graphics cards that support OpenCL 1.1. Somewhat
surprisingly, we found better numerical agreement between the OpenCL
application running on ATI/AMD GPUs and the CUDA application running
on NVIDIA cards, than between the (same!) OpenCL application running
on both ATI/AMD and NVIDIA GPUs.

\begin{table}
\begin{center}
  \begin{tabular}{r|r|r|r|r|r|r}
    Compute          & CPU    & \% of  & CUDA  & \% of  & OpenCL     & \% of  \\
    operation        & time   &  time  & time  &  time  & time       & time\\ 
    \tableline
Uncompress         &  $ < 1 $~s & $ < 1\% $  &   $ < 1$~s &  $ < 1 $~\%    & $ < 1$~s       &  $ < 1$~\% \\
Whiten                  &  1~s       &  $<1$~\%   & 1 ~s       &   $ < 1 $~\%   & 1 ~s           &  $ < 1$~\% \\
Demodulate               &   898~s    &  13~\%     & 20 ~s      &   14 ~\%       & 123 ~s         &  41 ~\% \\
$\lvert FFT\rvert^2$       &   4022~s   &   59~\%    &  48 ~s     &   32 ~\%       & 59 ~s          &  20 ~\% \\
Harmonic sum           &   1888~s   &   28~\%    &  68  ~s    &   45 ~\%       & 107 ~s         &  35 ~\% \\
Update toplists          &   12~s     &  $< 1$~\%  &  12 ~s     &  8 ~\%         & 12 ~s          &  4 ~\% \\
Merge toplists           & $ < 1 $~s  &  $<1$~\%   &  $<1$~s    &  $<1$ ~\%      & $ < 1$~s       &  $ < 1$~\% \\
\tableline
   Totals                  & 6822~s   & 100~\%     & 150~s    & 100~\%         & 299~s        & 100~\% \\
      \end{tabular}
\end{center}
\caption{\label{t:CPUGPU_comp} Comparison of run-times for the CPU
  (using only one core) and GPU versions of the client search
  application, processing a single de-dispersed time-series through a
  template bank containing 6662 orbital templates. The different rows
  show the execution time spent in the different functional blocks of
  Figure~\ref{fig:pipeline}.  The absolute run-times vary considerably
  for different combinations of CPU and GPU models; we measured it for
  typical consumer-grade hardware. The CPU is an Intel Core 2 Q8200
  (2.33GHz), the CUDA GPU is a NVIDIA GTX 560 Ti and the OpenCL GPU
  is an AMD Radeon HD 7970 (all running on unloaded Linux systems).}
\end{table}

The GPU version of the search application evolved considerably over
time, by incrementally porting more steps of the main loop to code
executing on the GPU.  The first GPU version of the search application
only implemented the FFT step on the GPU, and was limited to a
speed-up of between 2 and 3 compared to the CPU version, because on
the CPU version the FFT step consumes almost two-thirds of the total
CPU run time. The next important step was to port the re-sampling code
to the GPU. This gave an overall speed-up of about 4 compared to the
CPU version, and left the harmonic-summing step dominating the run
time.  When the harmonic summing step was also ported to the GPU, the
overall speed-up factor reached 50 (and even higher on some CPU and
GPU combinations).  Table~\ref{t:CPUGPU_comp} shows typical run time
examples for the current GPU and CPU version of the client search
application and the relative fraction of time spent in each processing
step.

Running one instance of the application, a typical high-end NVIDIA GPU
(for example the GTX 560) achieves up to 85\% utilization\footnote{The
  utilization is reported by NVIDIA's System Management Interface
  \textit{nvidia-smi}; information may be found at
  \url{https://developer.nvidia.com/nvidia-system-management-interface}}.
Provided that the GPU has sufficient memory, BOINC can run two or
three instances in parallel.  This saturates the GPU, achieving more
than 98\% utilization!

\subsection{Validation}
\label{subsec:validation}

As described earlier, any result file uploaded to the \EAH{} servers
must be validated because it could be partially or completely
incorrect, and/or corrupted.  Validation is done on the \EAH{} server,
by comparing the result file to another result file for the same
workunit, generated on another host.  An automatic validator compares
results and rejects those that appear to be corrupted and/or
inconsistent with other results.

The validation process is not trivial; it cannot be based on a simple
binary comparison of the two result files, because the use of
different floating-point libraries, compiler instructions, and
hardware can lead to numerical differences in the results.  Thus,
results from two different hosts might \textit{both} be correct, but
\textit{not} binary identical.  So the comparison process must allow
for numerical differences at a level which is typically of the order
of 1 part in $10^{5}$.

For \EAH{}, the validation process operates in two steps.  The first
step checks a single result file for syntax and internal
consistency. The second step compares two (or if necessary, more than
two) results which have passed the first step against one another.
Most incorrect or invalid results are detected in the first step.

In the first ``syntax and consistency'' step, a result file is checked
to see if it has the fixed seven-column output format with 100 lines
described near the end of Section~\ref{subsec:analysis}. For each
line, the seven fields are individually checked to confirm that they
are valid numbers and lie in pre-defined ranges. The overall ordering
of lines within the file is also checked to confirm that they are
ordered by decreasing significance. If any of these checks fails, then
the result is marked invalid, and another copy of the corresponding
workunit is generated on the \EAH{} server sent to a different
volunteer's computer.  Slightly less than 1\% of results fail to
validate at this stage\footnote{See ``validate error rate'' at
  \url{http://einstein6.aei.uni-hannover.de/EinsteinAtHome/download/BRP-progress/}.}.

In the second step, two or more result files that have passed the
first step are pairwise-checked for mutual consistency. The
validator tries to match each line from one result file to a line in
the other. Two lines ``match'' when the individual values for DM, $f$,
the orbital parameters, the $S_L$, and $\mathcal S$ agree within less
than a fractional error of $10^{-5}$. The number of harmonics $2^L$
must match exactly.

The last lines in the result files are typically near the noise
threshold and because of differences in floating-point accuracy and
rounding on different hosts, they may not correspond to the same
candidates.  Thus the validator permits unmatched lines in the result
files if (within fractional error $10^{-5}$) the corresponding
candidate might not have appeared in the most significant 100 results
in the \textit{other} result file.

If two results both pass the ``syntax and consistency'' step, but are
found to be inconsistent, another instance of the work is generated
and sent to a different client machine.  The process of generating
further instances of the results is repeated until a consistent set
are found, containing two or more results. Those results that are
inconsistent with that set are marked as invalid; slightly less than
0.5\% of results fail validation in this way\footnote{See ``invalid
  result rate'' at
  \url{http://einstein6.aei.uni-hannover.de/EinsteinAtHome/download/BRP-progress/}.}.
If more than twenty results are generated without getting a match,
then warning messages are sent to project personnel, and the workunit
``errors out''.

\subsection{Post-processing}
\label{subsec:postprocessing}

The client search code identifies the 100 most statistically
significant signal candidates in 628 de-dispersed WAPP (3808 Mock)
time series for each telescope beam. Ideally, all significant
candidates should be followed up using the ``raw'' observational
data with the full time resolution. In practice, this is neither
computationally feasible nor necessary, because a real pulsar can
be detected at different DMs, frequencies, and in multiple orbital
templates.

Several different sifting methods are used to reduce the number of
candidates to follow up.  These include overview plots for the
inspection all candidates in a given beam (described below) and an
automated filtering routine, summarized here and described in detail
in \citet{2013arXiv1302.0467K}.

When valid result files (see previous section) for all de-dispersed
time series of a given beam are available on the Einstein@Home upload
servers, a set of overview plots is automatically produced for visual
inspection. Theses show all candidates in a given beam in the
multi-dimensional parameter space of DM, spin frequency, and orbital
parameters, projected into two and three dimensions. Pulsars are
identified by the characteristic patterns they produce.

\begin{figure*}
  \begin{center}
  \ifcase \bwswitch
  \includegraphics[width=\columnwidth]{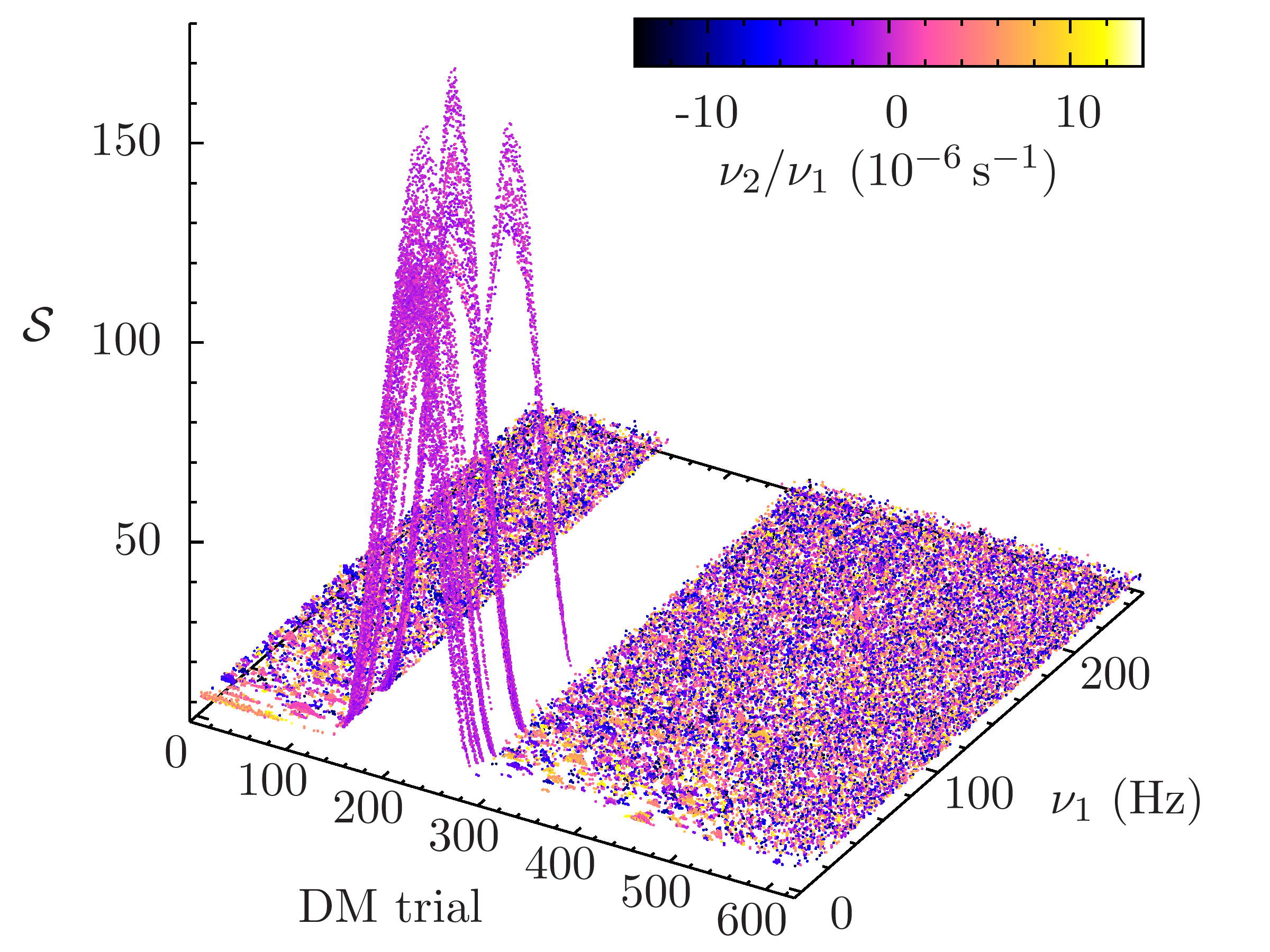}
  \hfill
  \includegraphics[width=\columnwidth]{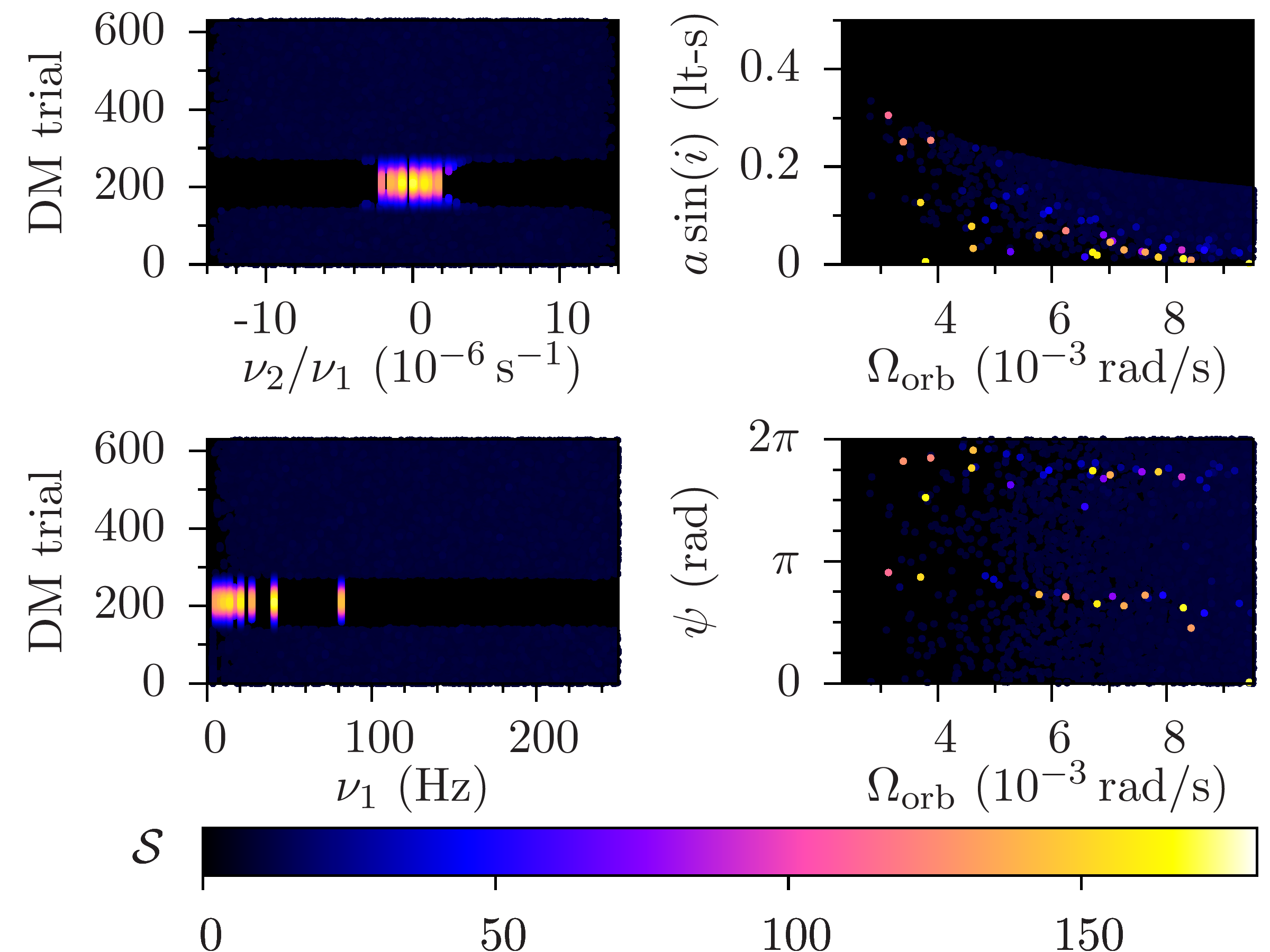}
  \or
  \includegraphics[width=\columnwidth]{f9a_bw.png}
  \hfill
  \includegraphics[width=\columnwidth]{f9b_bw.png}
  \fi
    \caption{Example post-processing overview plots, showing the
      highly-significant detection of PSR~J2007$+$2722. \textbf{Left:}
      this plot shows the significance $\mathcal S$ as a function of the
      DM trial number and the fiducial spin frequency $\nu_1$ (see
      Section~\ref{subsec:postprocessing}) of each candidate. The color-code
      displays the relative change in fiducial spin
      frequency $\nu_2/\nu_1$ from orbital motion. Since the top 100
      candidates are reported for each DM trial and the pulsar is
      detected with very high significance, there are no detections of
      the noise floor in a DM range around the pulsar. \textbf{Right:}
      the four sub-panels show the significance $\mathcal S$
      as a function of different combinations of spin frequency and the orbital
      parameters.}
      \label{fig:ovplot}
  \end{center}
\end{figure*}

A combination of five different plots are used in post-processing. As
an example, Figure~\ref{fig:ovplot} shows the highly-significant
detection of PSR~J2007$+$2722 in the Einstein@Home results. For each
candidate the left-hand panel shows the significance $\mathcal S$ as a
function of the trial DM number and spin frequency. The right-hand
panel shows four projections into subspaces of the parameters.  These
help identify pulsar candidates and provide initial estimates of spin
and orbital parameters.

These plots use coordinates defined in the Appendix of
\citet{2013arXiv1302.0467K}. They are obtained from writing the phase model
\eqref{eq:fullphase} as a power Taylor series in $t$. Then, the
coefficient of the linear term $\nu_1=f\left(1 +
a\sin(i)\Omega_\text{orb}\cos\left(\psi\right)/c \right)$ identifies a
spin frequency. The coefficient of the quadratic term
$\nu_2=-a\sin(i)\Omega_\text{orb}^2f \sin\left(\psi\right)/(2c)$ is
proportional to the Doppler spin-down or spin-up.

Promising candidates are identified from the visual inspection of
these plots. The number of promising candidates is relatively small.
The majority of PALFA beams have none; the most promising beams have
at most a handful.

In the next step, \textsc{PRESTO} software tools are used to fold the
full-resolution filterbank data for all candidates, starting with the
spin-period and DM values identified from the Einstein@Home
results. The \textsc{prepfold} plots are inspected by eye and used to
judge the broadband nature and temporal continuity of the signal.

We also developed an automated routine which filters through the list
of all candidates for a given beam and returns the most promising
candidates. These candidates are then followed up automatically with
different software tools described in detail in
\citet{2013arXiv1302.0467K}.  The automated routine consolidates
candidates at harmonically related frequencies, neighboring DMs, and
similar orbital parameters. The remaining candidates are folded with
\textsc{prepfold} to produce folded pulse profile and other diagnostic
plots, as well as associated ASCII files. These are then filtered by a
second piece of software, which uses these plots and ASCII files to
select the most ``pulsar-like'' candidates
\citep{2013arXiv1302.0467K}.

Using these two post-processing methods, the \EAH{} search of the
PALFA WAPP data made 322~detections of 158~unique radio pulsars. Of
these pulsars, 156 were already known; they were listed in the ATNF
catalog \citep{2005yCat.7245....0M}, or on Web sites maintained by
different ongoing pulsar
surveys\footnote{\url{http://astro.phys.wvu.edu/GBTdrift350/},
  \url{http://www.physics.mcgill.ca/~hessels/GBT350/gbt350.html},
  \url{http://astro.phys.wvu.edu/dmb/},
  \url{http://www.naic.edu/~palfa/newpulsars/}}.  Two of the pulsars,
PSR~J2007$+$2722 and PSR~J1952$+$2630, were new; they appeared during
the non-automated (visual inspection of the overview plots)
post-processing.

\section{Discovery of PSR~J2007+2722}
\label{s:Discovery}

PSR~J2007+2722 was found by project scientists on 2010 July 11 as part
of the routine post-processing described in the previous section; the
corresponding data had been acquired at Arecibo on 2007 February 11.
In the post-processing plots (Figure~\ref{fig:ovplot}) the pulsar
appeared with maximum significance $\mathcal S = 169.7$ at a
dispersion measure DM=$127\, \text{pc\ cm}^{-3}$ and spin frequency of
40.821\,Hz. The orbital parameters at highest significance were
consistent with no orbital modulation, or with an orbital period
longer than the longest orbital period in the template bank. In other
words, it appeared that the pulsar was either isolated, or was in a
long-period binary system. Further PRESTO-based analysis
refined these values and supported the isolated or
long-period interpretation.

The discovery was confirmed with a short Green Bank Telescope (GBT)
observation soon thereafter, following which the pulsar was
(re)observed at Arecibo, Jodrell Bank and Effelsberg.  Details of
later GBT studies are given in Section~\ref{ss:greenbank}.  A full
timing analysis based on dozens of additional observations extending
to late-2012 is given in Section~\ref{ss:timingmodel}.

Because the project database, and the result files themselves, contain
information about the computers that carry out analysis, it is
straightforward to identify the volunteers whose computers provide any
particular result.  As described in the
Section~\ref{subsec:validation} on validation, all \EAH{} work is sent
to computers owned by at least two different volunteers. In this case,
the valid result files containing the statistics of highest
significance for PSR~J2007+2722 were returned by computers owned by
volunteers from Ames, IA, USA and from Mainz, Germany.

The U.S. volunteers were Chris and Helen Colvin.  For security reasons,
the Colvins are not allowed to use their ``work'' computers for
personal email and Web browsing, so they maintain a small mail and Web
server at home.  Since 2006, this home computer has been running
\EAH{} as a background job.  The machine was equipped with an NVIDIA
graphics card, whose GPU did the ``discovery'' processing.

The German volunteer was Daniel Gebhardt, who is the system
administrator for a Musikinformatik group at Universit{\"a}t
Mainz. Gebhardt runs a mail server for the group, which is continuously
powered up, and runs \EAH{} as a background task.

It is notable that the first discovery from the \EAH{} pipeline, which
was designed to find pulsars in binary systems, was an isolated
pulsar, which was not found in either of the other PALFA processing
pipelines. This is not unexpected: as described previously, the \EAH{}
search pipeline contains long-orbital period templates and one
template with infinite period, so it can detect isolated systems. But
why was it not found by the other pipelines?

In fact this is not surprising: the three pipelines in question
(\EAH{}, PRESTO and Cornell) produce statistical outlier candidate
signals that are different owing to re-sampling differences, to
differences in the way orbital motion is treated, and to the way
signals that exceed statistical thresholds are reported.  In total,
each of these pipelines has involved about $10^{15}$ statistical tests
so far, and the initial reduced set of candidate signals is in the
millions.  The three pipelines have different procedures and criteria
for further winnowing these candidate signals into much shorter lists
of viable pulsar candidates worthy of detailed visual inspection and
follow-up observations at the telescope. The three pipelines also
process the data in different order, and at the time of the
PSR~J2007+2722 discovery, all three had data backlogs: the fact of the
matter is that the E@H pipeline found PSR~J2007+27
first. Retrospectively, the pulsar could be seen in the output of one
other pipeline (i.e. when we knew what to look for), while the other
pipeline had not yet processed the relevant beam.  In just the same
way, the other pipelines have also found new pulsars that the E@H
pipeline subsequently also detected.

\subsection{Distance to PSR J2007+2722}

Based on the NE2001 model \citep{2002astro.ph..7156C} for the Galactic
distribution of free electrons,
the DM=$127 \pm 0.4$ value implies a distance of 5.4~kpc. The
uncertainty in distance arising from the 0.4~pc~cm$^{-3}$ error in DM
is negligible in comparison with the NE2001 model uncertainty. We know
of two ways to bound this model uncertainty.

A direct measurement of errors in the NE2001 model can be obtained
from comparisons of NE2001 distance estimates to actual parallax-based
distance measurements (see \citet[][and references
  therein]{2009ApJ...698..250C}). While direct comparisons are only
possible for objects significantly closer than J2007+2722, for objects
within 10$^{\circ}$ of the pulsar, the parallax and DM distances agree
to within 20\%. Thus this direct measurement would suggests errors of
less than 20\% in the 5.4~kpc distance estimate.

To indirectly estimate the NE2001 model uncertainty, we first need to
identify if an H~II region or void perturbs the electron density along
the line of sight, which would increase this uncertainty.  In the case
of PSR J2007+2722, we could not identify any specific H~II region or
source of radio recombination along the line of sight.  The closest
young star cluster on the sky is IRAS 20050+2720, about $20^\prime$ away
from the line of sight and $\sim 0.7$~kpc distant from Earth. IRAS
20050+2720 has no massive stars that could produce a detectable H~II
region \citep{2012AJ....144..101G}.  IRAS and 5-GHz Very Large Array (VLA) images also do
not show any extended emission near the line of sight.  Thus, we
estimate the NE2001 model uncertainties following the approach given
in Section~4.2 and Figure~12 of \citet{2002astro.ph..7156C}.  We
assume that the DM is perturbed by subtle departures from the model at
the level of $\Delta {\rm DM} = 10$ and 20~pc~cm$^{-3}$.  These alter the
inferred distance by $\pm 0.3$ and 0.6~kpc, respectively,
corresponding to $\sim 6$\% and 11\% errors, or a maximum total error
of 17\%.

Choosing the worst case, we conservatively estimate the distance error
to be less than 20\%, and conclude that the distance of PSR J2007+2722
is $5.4 \pm 1.1$~kpc.

\section{Followup Observations and Characterization of PSR~J2007+2722}
\label{s:followup}

\subsection{Accurate Determination of the Sky Position}\label{subsec:detskypos}
\subsubsection{Gridding Observations with the Arecibo Telescope}
\label{ss:arecibogridding}
The initial discovery of PSR~J2007+2722 determined the sky position within
about $2^\prime$: the Arecibo beam radius at 1.4~GHz.  In normal
circumstances, one determines pulsar positions more precisely using
timing measurements over a period of a year or longer.  Carefully
fitting pulse arrival times to a timing model makes it possible to
determine the sky position with an angular error $\delta \gamma \sim
\epsilon P/D = 3 \times 10^{-8}$~radians, where $\epsilon \approx
10^{-2}$ is the typical time-of-arrival (TOA) error, measured as a
fraction of the rotation phase, $P=1/f=25$~ms is the pulsar period,
and $D \approx 10^3$~s is the light travel time across the diameter of
the Earth's orbit.  This corresponds to a position error $\delta
\gamma \sim 6$~milliarcsec; a timing-model position determination to
such accuracy can be found in Section~\ref{ss:timingmodel}

However, the discovery of PSR~J2007+2722 was an important milestone for
Volunteer Distributed Computing, and waiting a year to precisely
determine the sky position using timing was not an
option. Nevertheless, we \textit{were} able to narrow down the sky
position using a combination of methods, in order to search for
associated X-ray or gamma-ray sources and to set a limit on the
magnitude of the spin-down $\dot P$.  (If the TOA measurements cover
much less than one year, then uncertainties/errors in sky position are
degenerate with uncertainties/errors in $\dot P$.)

The first step in determining the sky position of PSR~J2007+2722 more
precisely was with a set of ``gridding'' measurements using the
Arecibo telescope on 2010 July 19. The observations were done in S-band
using the Mock spectrometers to construct five 172~MHz bands (center
frequencies 2136, 2308, 2687, 2859, and 3013~MHz) with 1024~channels
per band and a $65.5~\mu$s~sampling time.  A filter at the upper end
of the S-band receiver bandwidth (band-passes at 2040-2400MHz and
2600-3100MHz) was used to minimize RFI and reduce the half-power beam
width to $2^\prime$.

\begin{table}
\begin{center}
\begin{tabular}{c|c|c|c|}
 Decl. $\backslash$ R.A. & $20^h07^m18^s$ &  $20^h07^m14^s$ & $20^h07^m10^s$\\
\hline  & & & \\
$27^\circ 25^{\prime} 26^{\prime \prime}$   &           &            &         \\
\hline & & & \\
$27^\circ 24^{\prime} 26^{\prime \prime}$   &    2.9    &    15.6    &    22.5  \\
\hline & & & \\
$27^\circ 23^{\prime} 26^{\prime \prime}$   &    19.8   &    97.9    &      \\
\hline
\end{tabular}
\end{center}
\caption{\label{t:arecibogrid} Arecibo gridding measurements used to
  refine the sky position of PSR~J2007+2722.  The pulsar was visible
  in five of the nine pointings; the table entries show the ratio of
  the folded profile peak to the rms noise floor.}
\end{table}

The results of these first gridding measurements are shown in
Table~\ref{t:arecibogrid}.  A square grid of nine pointings was used,
with the center at R.A.~$20^h07^m14^s$ decl.~$27^\circ 24^{\prime} 26^{\prime
  \prime}$, and the adjacent pointings offset by about $\pm 1^\prime$
($\pm 4^s$ in RA and $\pm 1^\prime$ in decl.); the half-power beam contours
overlapped by about $7^{\prime \prime}$ in R.A. As shown in the table, the pulsar
was detected in five of the nine pointings. A weighted average of the
two pointings with the largest S/Ns gave a position estimate
R.A.~$20^h07^m12^s.7$, decl. $27^\circ 23^{\prime} 26^{\prime \prime}$.  We were
confident that the pulsar was inside a $1^\prime$ radius circle about
this point. A weighted average of all five pointings gives a position
estimate differing by about $25^{\prime \prime}$, but might be biased since there
are no observations to the south of the brightest grid point.

\subsubsection{Observations with Westerbork Synthesis Radio Telescope}
\label{ss:westerborkgridding}
\begin{figure}[h]
\begin{center}
\ifcase \bwswitch
\includegraphics[width=0.95\columnwidth]{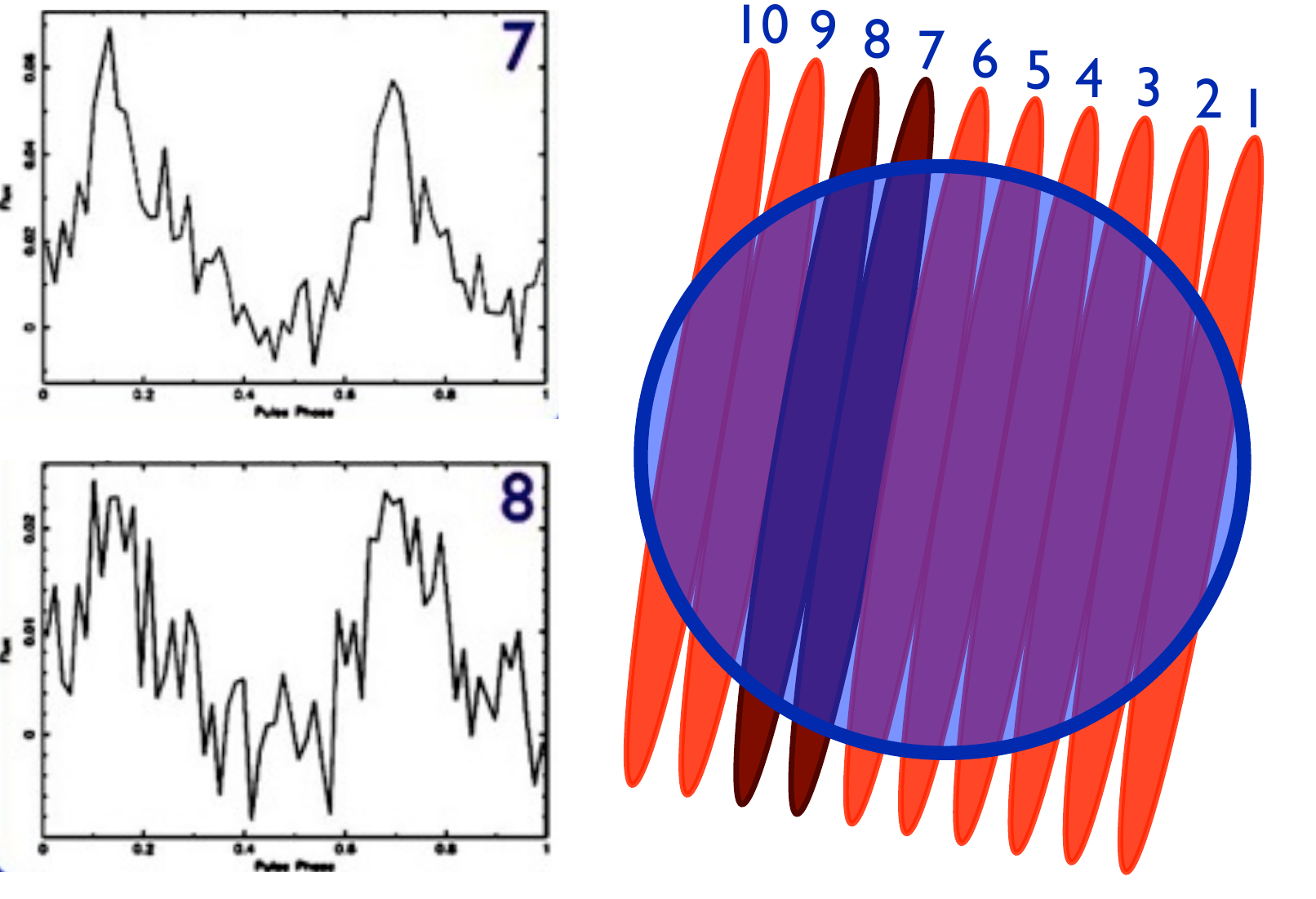}
\or
\includegraphics[width=0.95\columnwidth]{f10_bw.pdf}
\fi
\end{center}
\caption{{\bf Right:} schematic illustration of 10 WSRT fan beams
  overlapped with the $1^{\prime}$-radius error circle obtained from
  gridding observations at the Arecibo observatory.  The fan beam
  ellipses are not to scale: the minor axis is correct but the major
  axis is much longer than shown here. {\bf Left:} folded
  pulse-profiles for fan beams 7 and 8 (horizontal axis is pulse
  phase, vertical axes is normalized flux). PSR~J2007+2722 was not
  detected in any other fan beam.}
\label{f:Gridding}
\end{figure}

To further refine the sky position, observations were made with
Westerbork Synthesis Radio Telescope (WSRT, Netherlands) at central
frequency 1380~MHz with a 160~MHz bandwidth.  WSRT is a linear array
of 14 circular radio antennas, each 25m in diameter, arranged on a 2.7
km east-west line.  Aperture synthesis creates a fan-beam
approximately $12'' \times 30'$ in size, with the long axes along the
north-south direction at transit.  On the evening of 2010 July 19, ten
1180~s observations were made, with the center of each observation
displaced by $12''$, as schematically shown in Figure~\ref{f:Gridding}.
These covered the uncertainty region obtained from the Arecibo
gridding observations.  For each WSRT observation, the data were
de-dispersed and folded with the PSR~J2007+2722 period and DM using
PuMa-II \citep{PuMa-II}, a high time-resolution coherent de-dispersion
pulsar-processing back-end.  We believe this is the first time that
WSRT has been used for pulsar position refinement in this way.

The pulse profile was only convincingly detected in contiguous beams 7
and 8, with respective S/Ns 25 and 20, as shown in
Figure~\ref{f:Gridding}.  Weighted overlapping of fan beams 7 and 8
yields a position-constraint ellipse centered at R.A.~$20^h07^m14^s.5$,
decl.~$27^\circ 23^{\prime} 36^{\prime \prime} $ as shown in Figure~\ref{f:WSRT}. The major and
minor radii are $51^{\prime \prime}$ and $7^{\prime \prime}$; the major axis is rotated
$20^\circ$ clockwise from North.

\subsubsection{Westerbork Imaging and NVSS Catalog Sources}
\label{ss:westerborkimaging}
\begin{figure}[h]
\begin{center}
\ifcase \bwswitch
\includegraphics[width=0.8\columnwidth]{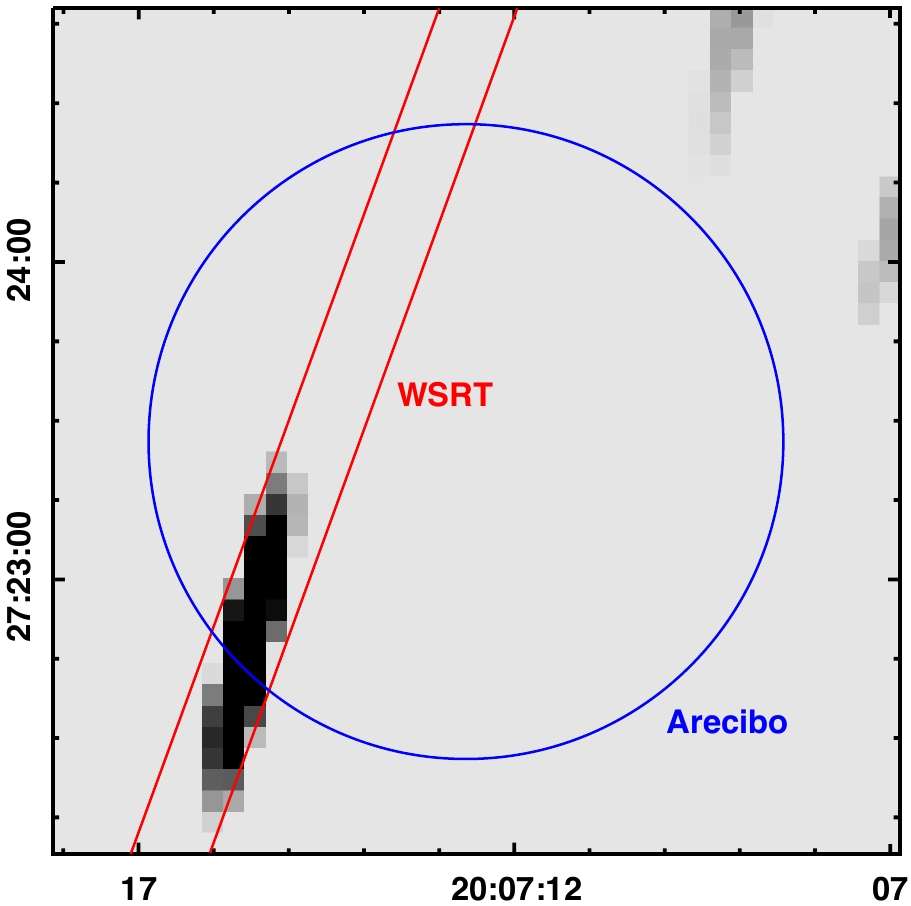}
\or
\includegraphics[width=0.8\columnwidth]{f11_bw.pdf}
\fi
\end{center}
\caption{An image of the WSRT data, along with the error circle
  obtained from Arecibo gridding, and the WSRT error region obtained
  by overlapping fan beams 7 and 8 as described in the text.  The
  imaged source corresponds to a cataloged NVSS source and is
  PSR~J2007+2722.}
\label{f:WSRT}
\end{figure}

Simultaneously with pulsar data, WSRT imaging data were also acquired.
Shown in Figure~\ref{f:WSRT} is the radio image, along with the error
ellipse just described.  A single radio source is visible on the
southern side of the error ellipse, just within the position circle
obtained from the Arecibo gridding.  This source is also listed in the
1.4-GHz National Radio Astronomical Observatory (NRAO) VLA Sky Survey (NVSS) catalog \citep{NVSS}.  The cataloged
source NVSS 200715+272243 has coordinates R.A.~$20^h07^m15^s.86$,
decl.~$27^\circ 22^{\prime} 43^{\prime \prime}.5$, a cataloged flux density of 2.3mJy at 1400~MHz,
and an estimated size less than $3.3^{\prime \prime}$, consistent with the WSRT
image.

\subsubsection{VLA Data Archive}

It was possible to determine the position even more precisely from
archival data.  This part of the sky contains the young star cluster
IRAS 20050+2720 \citep{2011ASPC..448..625G}. The VLA data archive
contains a 1610~s on-source observation of IRAS2005 taken on
1997-08-14 (VLA project code AE0112A, data-set
VLA\_XH97065\_file6.dat); the field of view (FOV) is approximately $16^\prime
    \times 16^\prime$. The data were acquired with the VLA C array
    operating in a 50~MHz bandwidth centered at 4.8601~GHz, in full
    Stokes mode, with a central beam position R.A.~$20^h07^m05^s.859$,
decl.~$27^\circ 28^{\prime}59^{\prime \prime}.77$.

\begin{figure}[h]
\begin{center}
\ifcase \bwswitch
\includegraphics[width=0.99 \columnwidth]{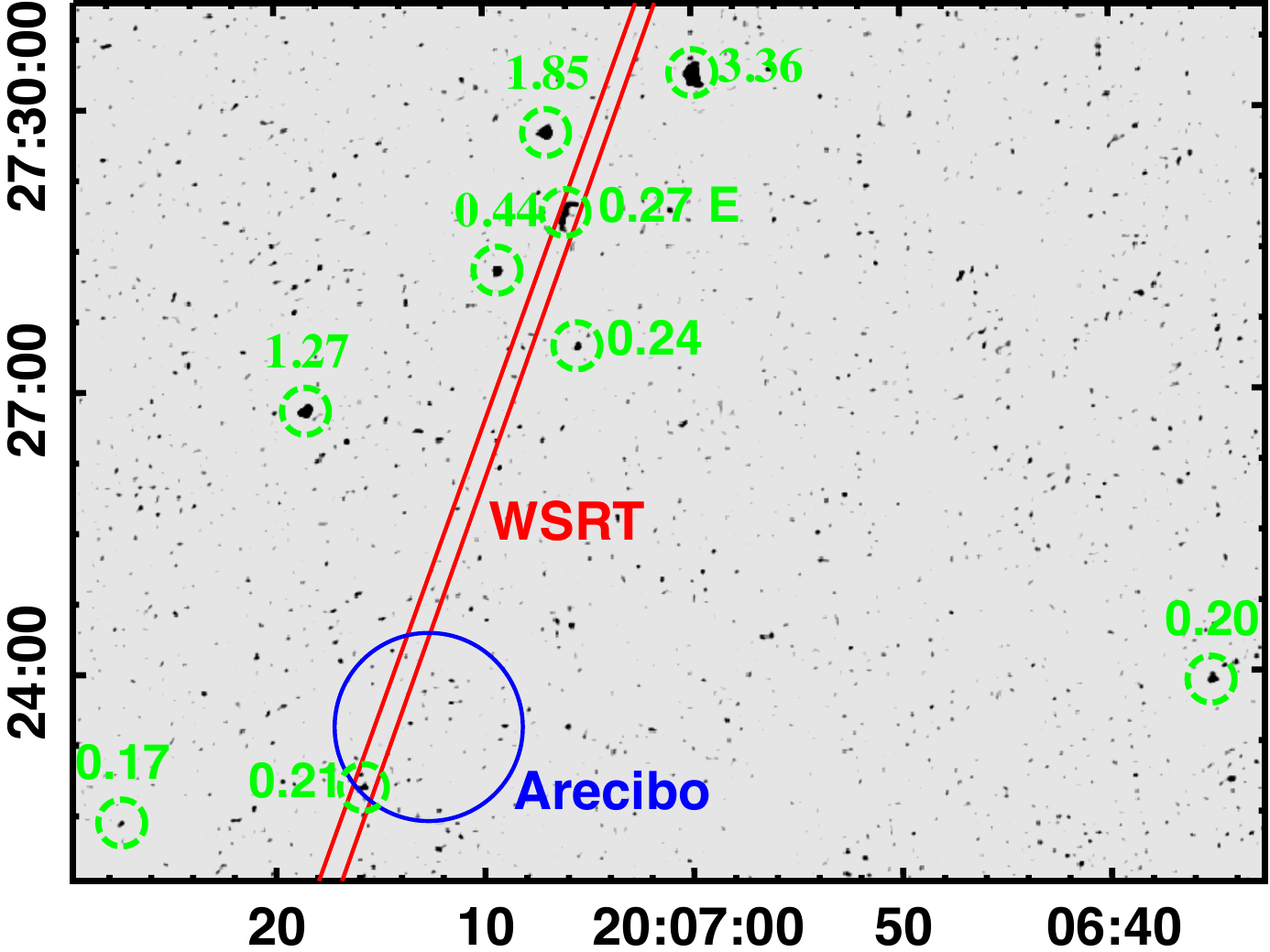}
\or
\includegraphics[width=0.99 \columnwidth]{f12_bw.pdf}
\fi
\end{center}
\caption{An image of (a part of the) archival VLA data at 4.8~GHz.  To
  compensate for the drop-off in sensitivity near the edge of the
  primary beam, the intensity has been divided by a model for primary
  beam response. The intensity has an rms of 42~$\mu$Jy; there are nine
  sources brighter than 170~$\mu$Jy, which are shown in the dashed
  circles.  The source fluxes determined by \textit{MAXFIT} are given
  in mJy, \textit{before} dividing by the beam response; the extended
  source is indicated by ``E''. The larger circle is the
  $1^\prime$-radius source uncertainty region found by the Arecibo
  gridding and the region between the two ``parallel lines'' is the
  relevant portion of the uncertainty ellipse found by the WSRT
  gridding. There is only one source (PSR~J2007+2722) lying in both
  uncertainty regions; it has a unnormalized flux of $210 \,\mu$Jy and
  a normalized flux of $1.2$~mJy. \label{f:VLA}}
\end{figure}

We analyzed the full FOV using \textit{MAXFIT} to characterize the eight
point sources and one extended source which are visible above the
background noise.  Shown in Figure~\ref{f:VLA} is the part of this
data (about $10^\prime \times 14^\prime$) containing the sources, which are
circled.

\begin{figure}[h]
\begin{center}
\ifcase \bwswitch
\includegraphics[width=0.85 \columnwidth]{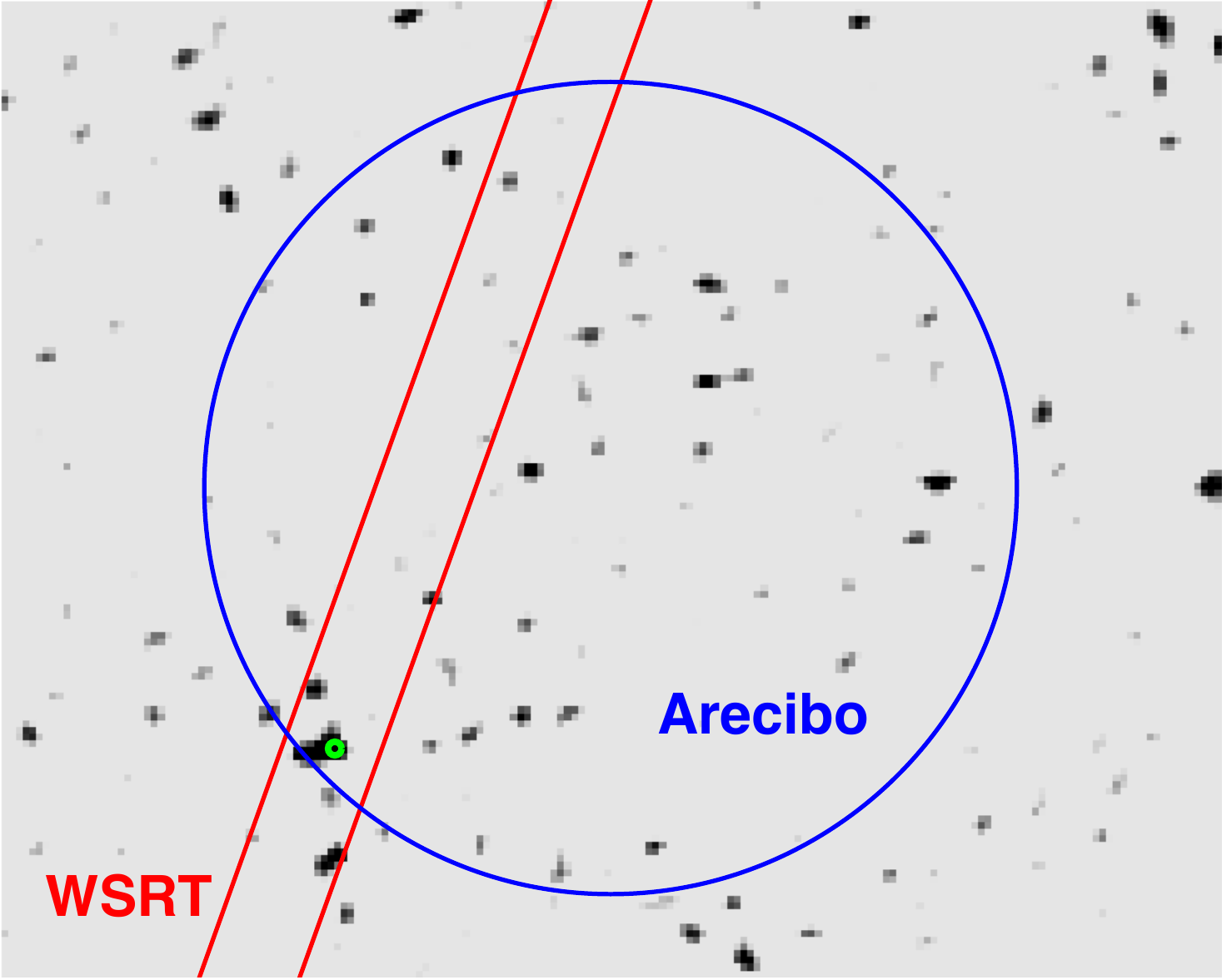}
\includegraphics[width=0.85 \columnwidth]{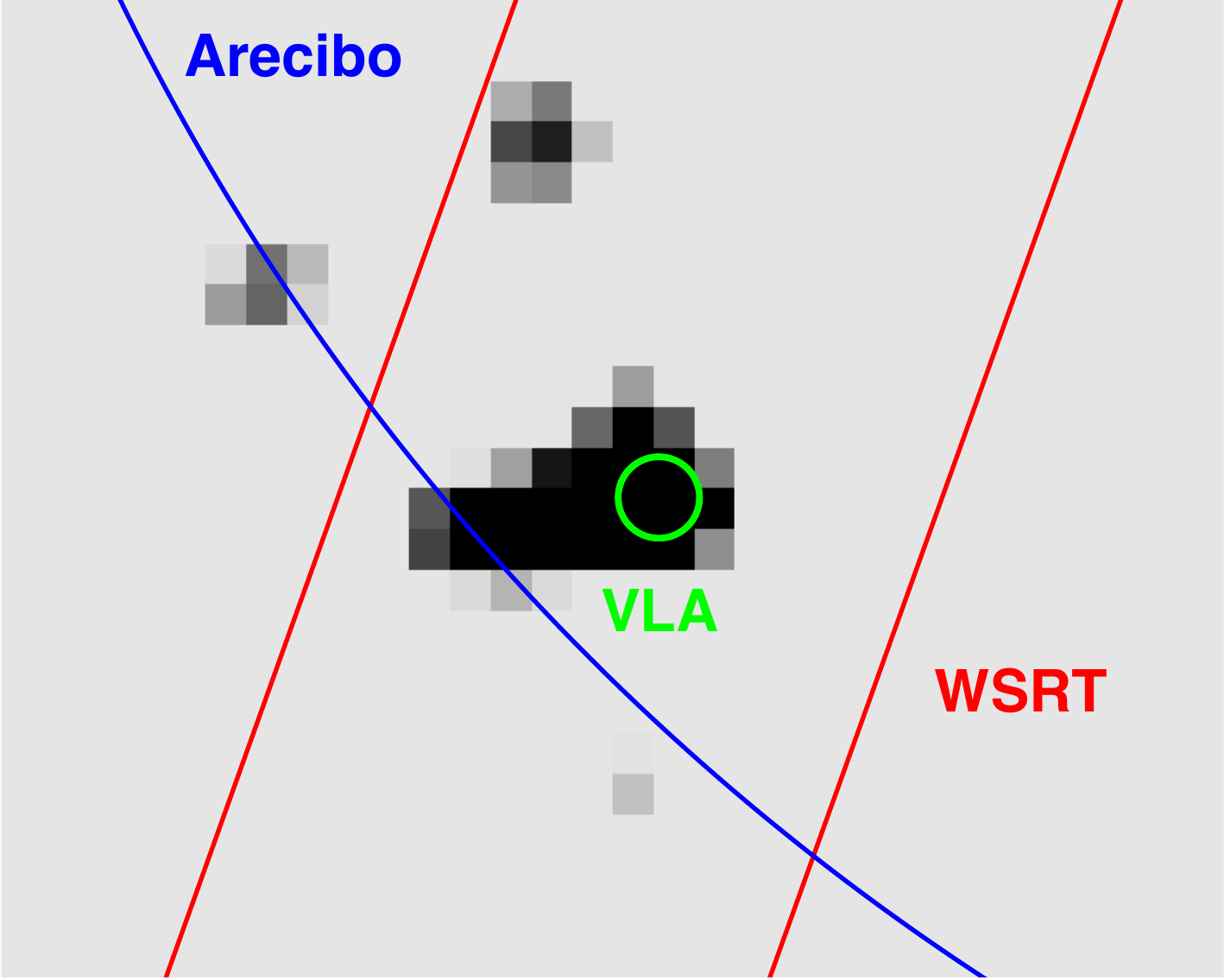}
\includegraphics[width=0.85 \columnwidth]{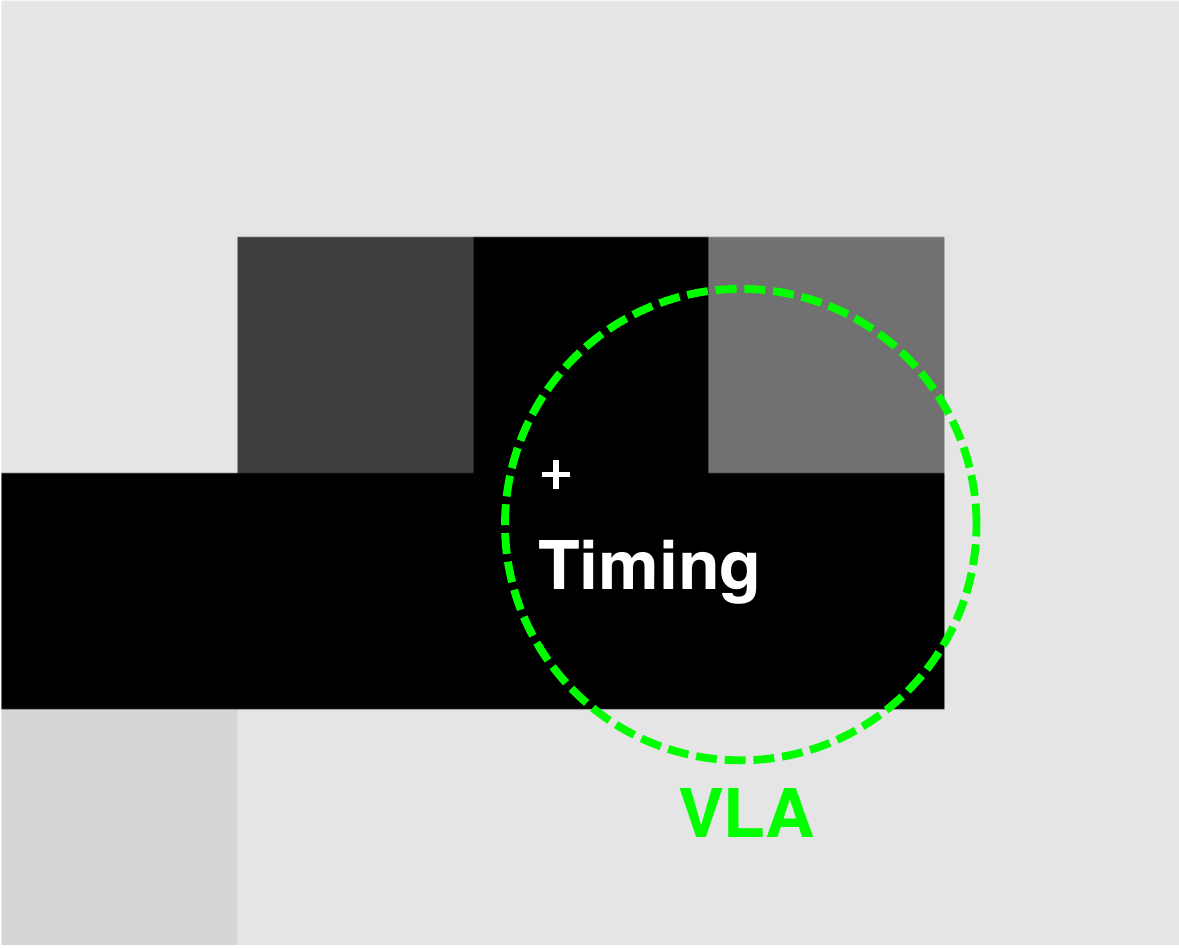}
\or
\includegraphics[width=0.85 \columnwidth]{f13a_bw.pdf}
\includegraphics[width=0.85 \columnwidth]{f13b_bw.pdf}
\includegraphics[width=0.85 \columnwidth]{f13c_bw.pdf}
\fi
\end{center}
\caption{Archival VLA data at 4.8~GHz. {\bf Top:} a $180 \times
  144^{\prime \prime}$ region with the 1'-radius uncertainty region
  from Arecibo gridding and the 13''-wide uncertainty region from WSRT
  gridding; the overlap contains a single VLA source (small circle).
  {\bf Middle:} a $30 \times 24^{\prime \prime}$ zoom showing the 1.2~mJy
  VLA source near the south side of the previous region.  The
  1''-radius circle shows the uncertainty region obtained by fitting a
  Gaussian to the VLA intensity; the discovery publication used this
  as the PSR~J2007+2722 position. {\bf Bottom:} a $5 \times 4^{\prime
    \prime}$ zoom; the cross indicates the location of PSR~J2007+2722
  obtained in this paper by timing analysis.  It lies inside the
  1''-radius VLA uncertainty region.  The intensity scale has been
  changed in the bottom plot to show the brightest VLA pixels.
}
\label{f:zoom}
\end{figure}

As can be clearly seen in Figure~\ref{f:VLA}, only one of these
(point) sources lies inside the uncertainty regions obtained from the
WSRT and Arecibo observations. This is shown in more detail in
Figure~\ref{f:zoom}. The point source has coordinates R.A.~$20^h07^m15^s.77$,
decl.~$27^\circ 22^{\prime} 47^{\prime \prime}.68$ and an uncorrected
flux density of $0.21$~mJy; the primary beam-corrected flux density is
1.2~mJy ($\pm 10$\%) at 4.86~GHz.  (The absolute flux density
measurement is referred to 3C48; the errors arise primarily from
uncertainties in the primary beam model, because the source is close
to the edge of the beam.)  The flux density is consistent with the
normal spectral behavior of similar radio pulsars; we conclude that
this is the correct location of PSR~J2007+2722.

\subsection{Multi-frequency Observations and Emission Geometry}

\subsubsection{Arecibo Observations at 327 and 430 MHz}
\label{ss:AO327}

Early observations of PSR J2007+2722 at 327 and 430\,MHz did not see
convincing evidence of pulsations.  It turned out that the issue was
instrumental: more extended observations at Arecibo Observatory in
Spring 2013 did succeed.

On 2013 April 16, the pulsar was observed for 960\,s from
420-447\,MHz, using the Mock spectrometer in search mode with 1024
channels covering 34\,MHz bandwidth, and a 119\,$\mu$s sample time.
The data were rescaled and converted from 16-bit PSRFITS to 4-bit
PSRFITS, then folded using \textsc{presto} with the ephemeris of
Section\,\ref{ss:timingmodel}.  Channels falling outside the 27\,MHz
receiver bandwidth were discarded. After folding, the profile was
calibrated with respect to the measured system gain (11\,K/Jy) and
noise equivalent temperature of 115K.  Pulsations were observed with a
false alarm probability corresponding to $4.0\,\sigma$.

The noise equivalent temperature was determined from measurements on
April 17, using a diode noise source calibrated against hot- and
cold-loads of known temperature.  These yielded 75K at high Galactic
latitudes and 115K close to the Galactic plane. The 40K difference is
consistent with the 408\,MHz measurements and models of
\citet{1982A&AS...47....1H}: extrapolation to 430\,MHz using spectral
slope $T_{\rm sky} \propto f^{-2.3}$ predicts a 44K Galactic
contribution.

On April 30 the pulsar was observed for 1761~s from
290-359\,MHz.  We used the Puertorican Ultimate Pulsar Processing
Instrument (PUPPI)
backend \footnote{\url{http://www.naic.edu/~astro/guide/node11.html}}
operating in search mode with a 81.92\,$\mu$s sample time. This PUPPI
mode nominally covers 100\,MHz of bandwidth in 4096 channels, but only
2816 channels covering the 69\,MHz receiver bandwidth were recorded.
The data were folded and calibrated as above, using a measured
equivalent noise temperature of 186\,K and gain of 11 K/Jy.
Pulsations were observed with a false alarm probability corresponding
to $5.5\,\sigma$.

The equivalent noise temperature was determined using calibrated
measurements away from the Galactic plane (which yielded 100-105K, and
adding the estimated Galactic background contribution of 83K, obtained
as above from \citet{1982A&AS...47....1H}.

Pulse profiles from these observations are shown in the bottom two
plots of Figure~\ref{f:GBT}; rotation-averaged pulsed fluxes derived
from these are given in Table~\ref{t:flux}.

\subsubsection{Green Bank Telescope Observations}
\label{ss:greenbank}
The GBT carried out follow-up
observations on 2010 July 21, in bands centered at 820, 1500, 2000,
and 8900\,MHz.  Full Stokes data were obtained for the observations
at 1500, 2000, and 8900\,MHz, but the 8900~MHz data was too noisy to
be useful for polarimetry.

All GBT observations of PSR~J2007$+$2722 were carried out using the
Green Bank Ultimate Pulsar Processing Instrument
(GUPPI) \footnote{\url{https://safe.nrao.edu/wiki/bin/view/CICADA/NGNPP}}
in incoherent de-dispersion mode. The observations at 820~MHz used
200~MHz total bandwidth, 2048 spectral channels and 40.96~$\mu$s time
resolution.  For the 1500~MHz and 2000~MHz observations, 800~MHz total
bandwidth, 2048 channels and 25.6~$\mu$s time resolution were used.
At 8900~MHz, the parameters were 512 channels, 800~MHz bandwidth and
6.4~$\mu$s time resolution.

The total observation time at each frequency was approximately 30
minutes.  Along with each pulsar observation, a short amount of data
were recorded with the local calibration-noise source pulsed at 25~Hz.
The equivalent noise source flux in each polarization channel was
determined by observing standard astronomical flux calibration sources
(3C190 was used at 820, 1500, and 2000~MHz; 3C48 at 8900~MHz). The
noise source measurements were then used for polarimetric calibration
(differential gain and phase) and absolute flux calibration of the
pulsar data.  All data processing described in this section was
performed using the
PSRCHIVE\footnote{\url{http://psrchive.sourceforge.net}} software
package \citep{psrchive}.

\begin{figure}
\begin{center}
\ifcase \bwswitch
\includegraphics[width=\columnwidth]{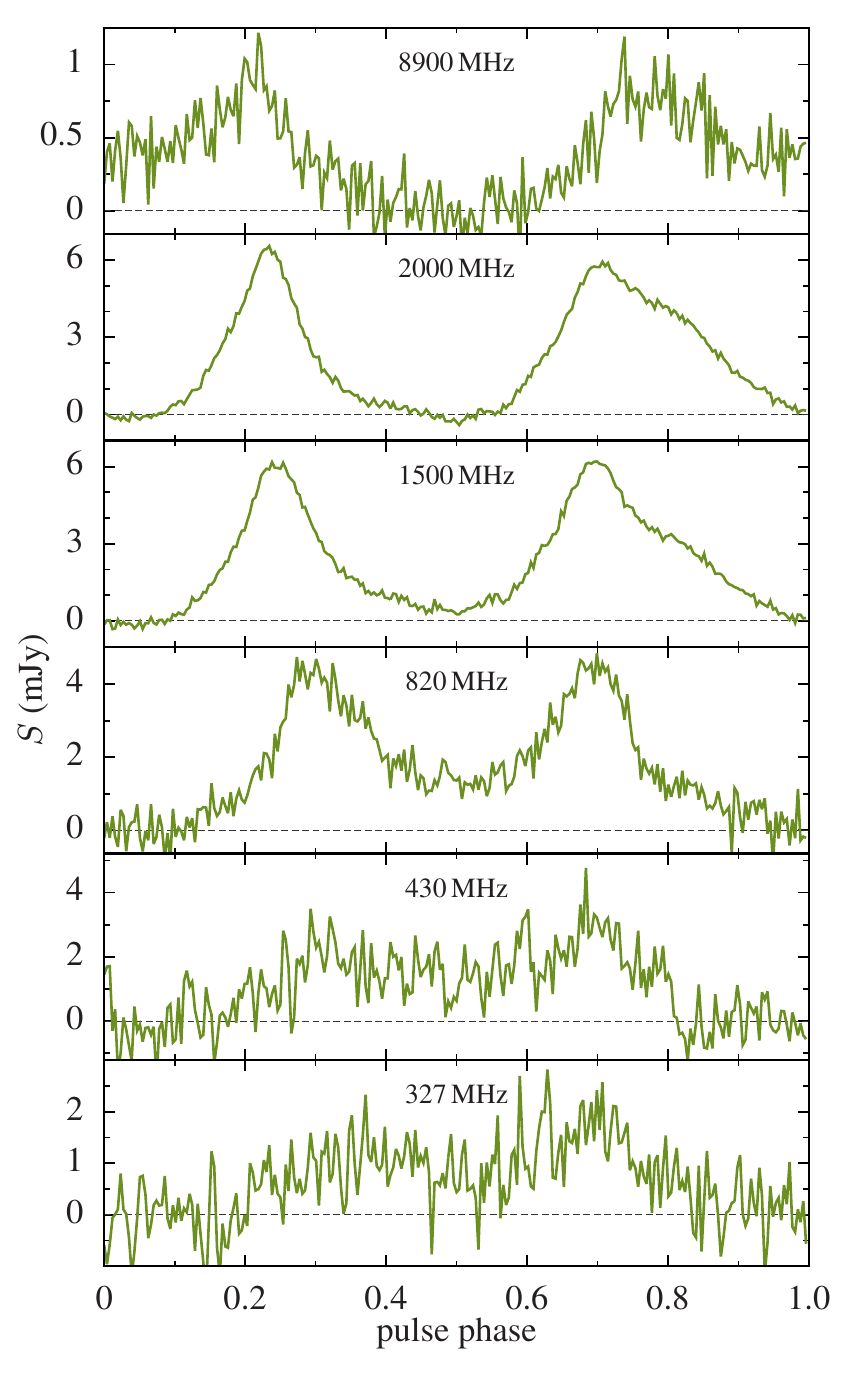}
\or
\includegraphics[width=\columnwidth]{f14_bw.pdf}
\fi
\end{center}
\caption{The pulse profile in mJy at 327, 430, 820, 1500, 2000, and
  8900\,MHz.  The rotation-averaged pulsed flux is given in
  Table~\ref{t:flux}. All the plots show an emission ``bridge''
  between the two pulses, which shifts to outside the peaks at the
  highest frequency.  This is evidence that the pulsar is ``always
  on'', suggesting that the pulsed flux shown in Table~\ref{t:flux} is
  only a fraction of the total flux.
\label{f:GBT}}
\end{figure}

The pulse profile of PSR~J2007+2722 is unusually broad: at 1500~MHz
the full pulse width between the outer half-maxima is $\approx
224^\circ$.  The folded pulse profiles at the four GBT observed
frequencies are shown in the top four plots of Figure~\ref{f:GBT}.
All observations exhibit a double-peaked pulse profile with an
emission bridge between and connecting the two peaks.  The emission
bridge flattens with increasing observation frequency and shifts
location from between the peaks at lower frequency to outside the
peaks at 8900\,MHz. This indicates that some radio emission is present
at all rotational phases in addition to the pulsed emission.

For all frequencies at which the pulsar was detected, pulse-averaged
flux densities were obtained.  In combination with the flux density
from the NVSS catalog and the VLA archival data, the pulsar's flux
density has been measured at eight different
frequencies. Table~\ref{t:flux} summarizes these measurements.

\begin{table}
\renewcommand{\arraystretch}{1.5}
\begin{tabular*}{\columnwidth}{@{\extracolsep{\fill}}cccl}
\hline Frequency (MHz) & Flux Density (mJy) & Pulsed/Total? & Instrument \\
\hline
327 & 0.6 & \textit{P} & Arecibo \\
430 & 1.0 & \textit{P} & Arecibo \\
820 & 1.6 & \textit{P} & GBT \\
1400 & 2.3 & \textit{T} & NVSS Catalog \\
1500 & 2.1 & \textit{P} & GBT \\
2000 & 1.7 & \textit{P} & GBT \\
4860 & 1.2 & \textit{T} & VLA Archive \\
8900 & 0.3 & \textit{P} & GBT \\
\hline
\end{tabular*}
\caption{\label{t:flux} Flux density of PSR~J2007+2722 at different
  frequencies.  The pulsed measurements (\textit{P}) only show the
  (rotation-averaged) component of the flux density that varies with
  pulse phase, referred to the dashed baseline in
  Figure~\ref{f:GBT}. The total measurements (\textit{T}) also include
  the phase-independent part.  The VLA/NVSS flux measurements are
  described in Section~\ref{ss:westerborkimaging}, and the GBT and
  Arecibo measurements in Section~\ref{ss:greenbank}.}
\end{table}

The flux density measurements from the Arecibo Telescope and GBT
observations are only sensitive to the pulsed emission.  Fitting a
single-component power law \beq S\left(\nu\right) = S_{1400}
\left(\frac{\nu}{1400\,\textrm{MHz}}\right)^{\xi} \eeq to measurements
of the pulsed flux density $S$ at frequencies $\nu>$1\,GHz, we obtain
a spectral index $\xi=-1.12(6)$, i.e.\ a relatively flat spectrum
\citep{LorimerKramer}.

The low pulsed-flux density below 1400~MHz is unusual, as pulsar
spectra generally turn over at frequencies around $\sim$100
MHz. Unless the non-pulsed flux dominates the pulsed flux, PSR
J2007+2722 belongs to the small subset of pulsars with GHz-peaked
spectra.  \citet{2011MNRAS.418L.114K} suggest that this behavior could
be due to unusual environments, since PSR B1259-63 exhibits such a
spectrum at periastron. While only 5 such sources have been reported
thus far, \citet{2013arXiv1302.2053B} estimate that they may comprise
up to 10\% of the pulsar population.  However, more such objects are
necessary to draw any reliable conclusions.

\subsubsection{Polarimetry}

\begin{figure}
\begin{center}
\ifcase \bwswitch
\includegraphics[width=\columnwidth]{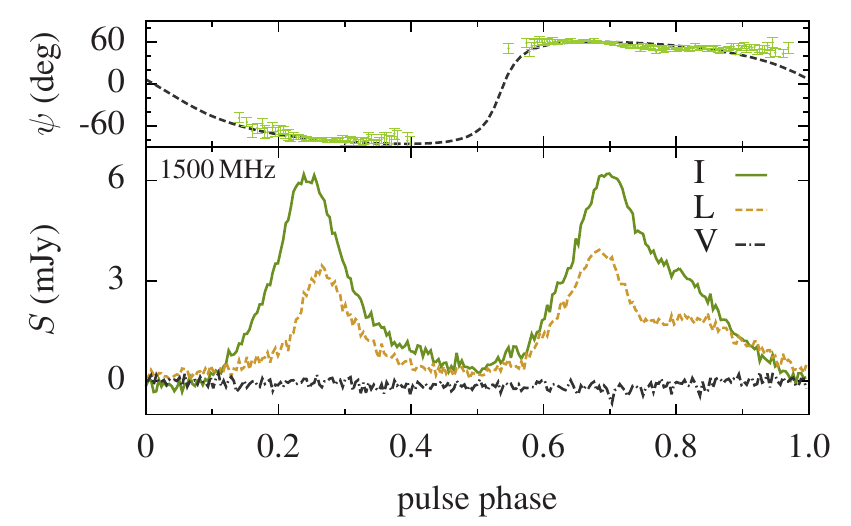}
\includegraphics[width=\columnwidth]{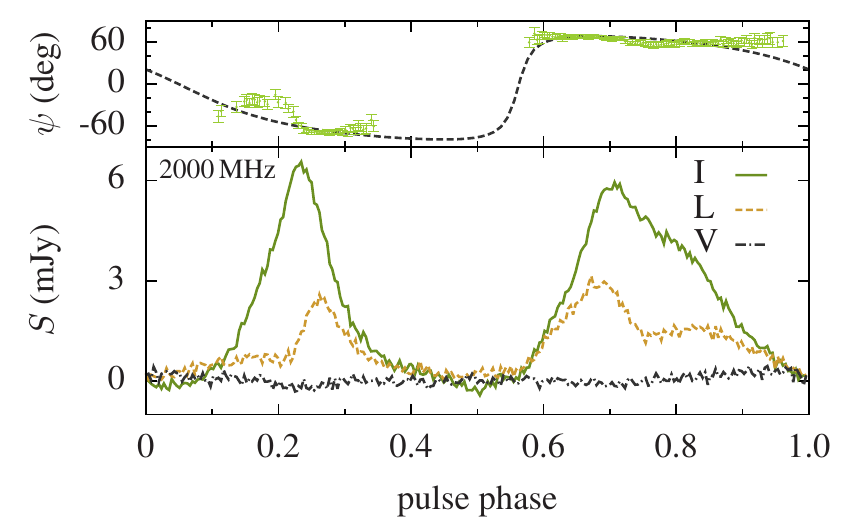}
\or
\includegraphics[width=\columnwidth]{f15a_bw.pdf}
\includegraphics[width=\columnwidth]{f15b_bw.pdf}
\fi
\end{center}
\caption{\label{f:GBT_stokes} Full Stokes polarization-angle profiles
  at 1500\,MHz (top) and 2000\,MHz (bottom) taken at GBT. The
  horizontal axis is rotation phase of the pulsar.  The bottom half of
  each plot shows the radio flux-density $S$ in intensity $I$ (solid),
  linearly polarized component $L=\sqrt{U^2 + Q^2}$ (dashed) and
  circularly polarized component $V$ (dash-dotted).  The top half of
  each plot also shows the derived polarization angle $\psi$ from
  Equation~\eqref{eq:psi_rvm}, corrected with rotation measure RM$ = -230$
  for Faraday rotation arising from the Galactic magnetic field . The
  dashed lines show $\psi_\textrm{RVM}$ for the best-fit rotating
  vector models given by Equation~\eqref{eq:RVM} and
  Table~\ref{tab:RVMparam}.}
\end{figure}

The GBT observations also provided full Stokes polarization parameters
$I$, $Q$, $U$, and $V$ at 1500 and 2000\,MHz, from which the 
polarization angle
\beq
\psi = \frac{1}{2}\arctan\left(\frac{U}{Q}\right)
\label{eq:psi_rvm}
\eeq can be computed as a function of the pulsar rotation phase. These
polarization-angle profiles are shown in Figure~\ref{f:GBT_stokes} as a
function of pulsar rotation phase, along with estimated measurement
uncertainties $\Delta \psi$.  We use PSR/IEEE sign conventions for
$\psi$ and $V$, as defined by
\citet{2010PASA...27..104V10.1071/AS09084} and employed by
\textsc{PSRCHIVE}.

\subsubsection{Emission/Beam Geometry}\label{sec:emissgeo}

The polarization-angle profiles can be used to infer the beam geometry
from the Rotating Vector Model (RVM; \citealp{RVMpaper}).  In the RVM,
the beam geometry is defined by four free parameters: $\alpha$, $\zeta
\equiv \alpha + \beta$ , $\psi_0$ and $\phi_0$.  Here, $\alpha$ is the
angle from the spin vector to the ``visible'' magnetic axis, and
$\zeta$ is the minimum angle from the spin vector to the
pulsar-observer line of sight. These angles are described and
illustrated in Figure~1 of \citet{2001ApJ...553..341E}, whose
conventions we adopt.  The polarization angle at pulsar rotation phase
$\phi_0$ is denoted $\psi_0$.

The polarization angle $\psi_\textrm{RVM}$ as a function
of the pulsar's rotation phase $\phi$ is
\begin{eqnarray}
\label{eq:RVM}
\tan\left(\psi_\textrm{RVM} - \psi_0\right) & &  =  \label{eq:rvm_define} \\
& & - \, \frac{\sin\left(\phi - \phi_0\right) \sin\left(\alpha\right)} 
     {\sin\left(\zeta\right) \cos\left(\alpha\right) -
       \cos\left(\zeta\right)\sin\left(\alpha\right) \cos\left(\phi -
       \phi_0\right)}.
\nonumber
\end{eqnarray}
The sign on the rhs occurs because we follow the ``observer's'' or
``IAU/IEEE'' convention for which the polarization angle $\psi$
increases in the counter-clockwise direction on the sky, as detailed
in \citet{2001ApJ...553..341E}.  This polarization angle convention is
the same as PSR/IEEE \citep{2010PASA...27..104V10.1071/AS09084}.
 
The values of the four RVM parameters were determined by a
least-squares fit.  We began with the measurements of the polarization
angle $\psi$ for $N_{1500} = 158$ different values of the rotation
phase at 1500~MHz, and for $N_{2000} = 143$ different values of the
rotation phase at 2~GHz, as shown in the upper parts of
Figure~\ref{f:GBT_stokes}. At each point of a 4-dimensional cubical grid
(spacing $0^\circ.5$) in $\left(\alpha,\zeta, \psi_0, \phi_0
\right)$-space, we calculated the normalized sum of the
squared-residuals, \beq \chi^2 = \frac{1}{N-4} \sum_{i=1}^N
\frac{\left( \psi(\phi_i) - \psi_\textrm{RVM}(\phi_i) \right)^2}{
  (\Delta \psi_i)^2 }, \nonumber \eeq between the RVM-predicted and
measured polarization angles.  Here $i$ labels the $N = N_{1500}$ or
$N = N_{2000}$ distinct pulsar rotation phases $\phi_i$ for which
$\psi_i = \psi(\phi_i)$ was measured, and $\Delta \psi_i$ is the
experimental measurement uncertainty in $\psi_i$.

\begin{figure}
\begin{center}
\ifcase \bwswitch
\includegraphics[width=\columnwidth]{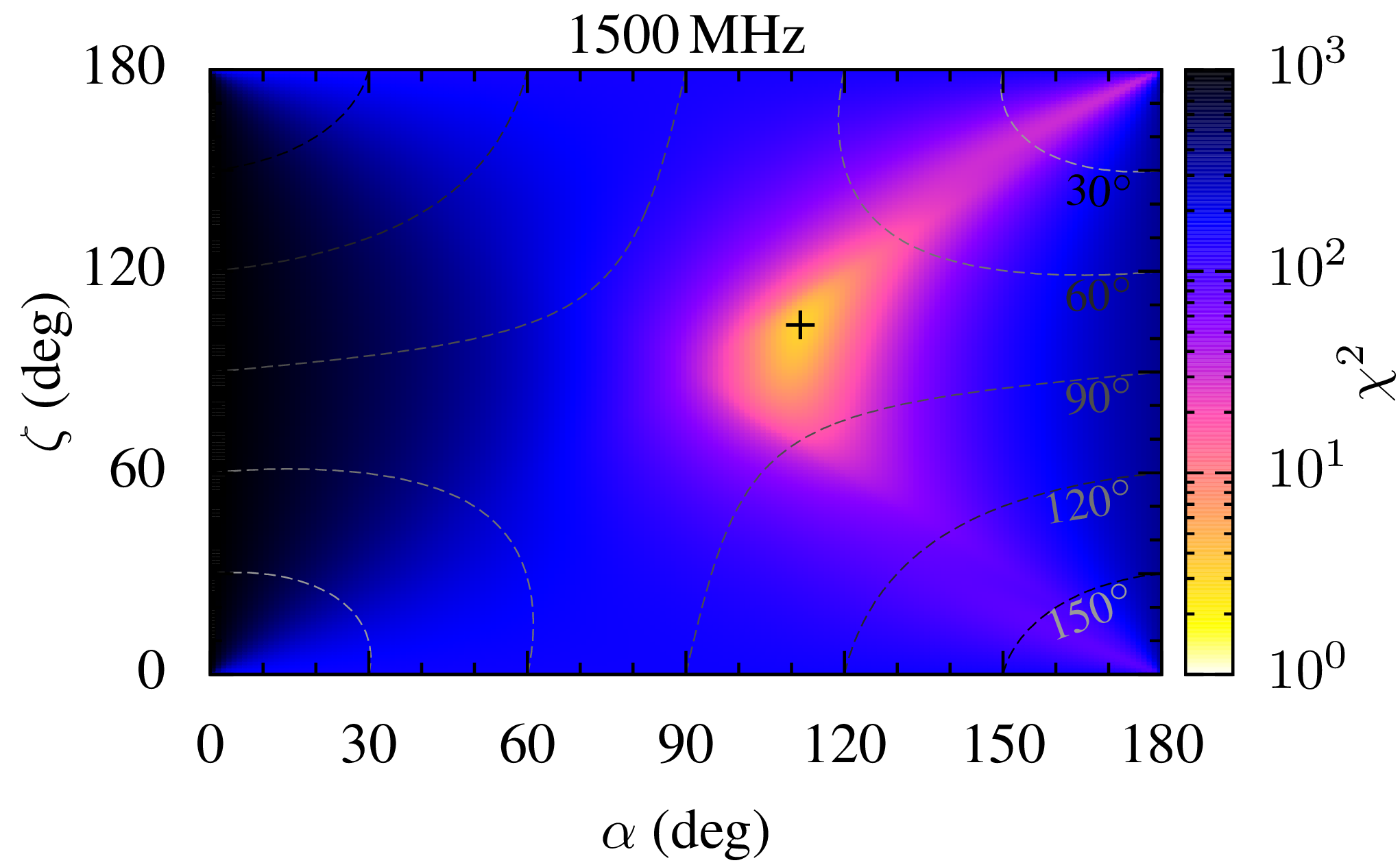}
\includegraphics[width=\columnwidth]{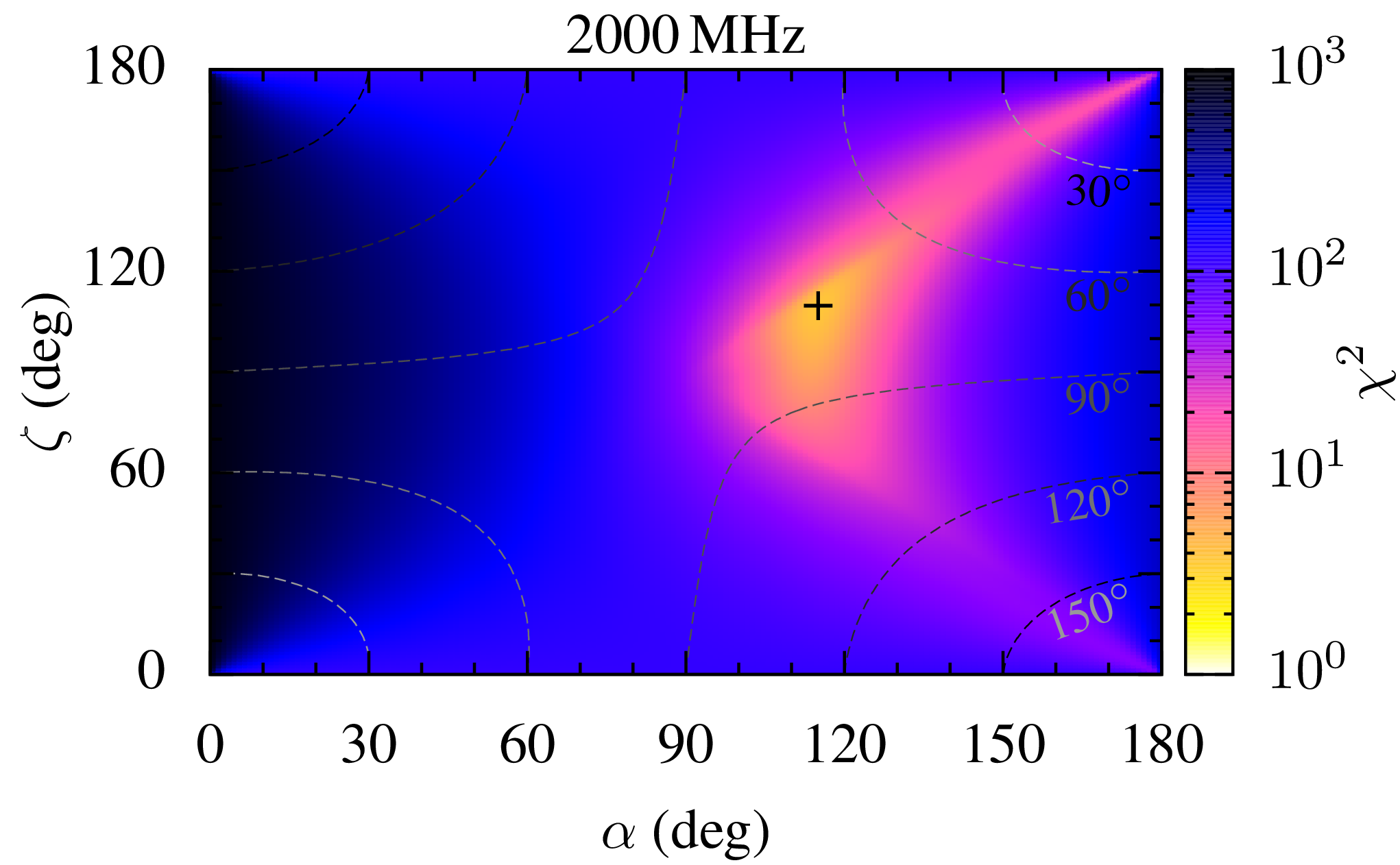}
\or
\includegraphics[width=\columnwidth]{f16a_bw.png}
\includegraphics[width=\columnwidth]{f16b_bw.png}
\fi
\end{center}
\caption{The reduced chi-squared values $\chi^2$ as a function of
  ($\alpha$, $\zeta$), obtained by fitting the measured polarization
  angle to the RVM model Equation~\eqref{eq:rvm_define} as described
  in the text. At each point the $\chi^2$ was minimized over $\phi_0$
  and $\psi_0$. The dashed lines are the contours of constant
  emission-cone half-opening-angle as defined by
  Equation~\eqref{eq:rvm_halfopening}.
\label{f:chi2maps}}
\end{figure}

Because the number of degrees of freedom is $N-4$, $\chi^2$ is a
conventionally-normalized reduced-chi-squared statistic.  Values of
$\chi^2$ near unity indicate that RVM fits the data well (consistent
with Gaussian-distributed errors of standard deviation $\Delta \psi$
in the values of $\psi$). Large values of $\chi^2$ indicate a poor
fit.  Figure~\ref{f:chi2maps} shows the minimum value of $\chi^2$ as a
function of $\left(\alpha,\zeta\right)$; note that the color code has
a logarithmic scale.  The minimum $\chi^2$ values obtained over all
four parameters, and the corresponding best-fit parameter values, are
shown in Table~\ref{tab:RVMparam}.  These best-fit values are shown by
black crosses in Figure~\ref{f:chi2maps}.

\begin{table}
\renewcommand{\arraystretch}{1.5}
\begin{tabular*}{\columnwidth}{@{\extracolsep{\fill}}cccccc}
\hline
Frequency & $\alpha$ & $\beta$  & $\phi_0$ & $\psi_0$ & $\chi^2$ \\
\hline
1500\,MHz & $111.7(5)^\circ$ & $-7(1)^\circ$ & $192(2)^\circ $ & $ 12.7(8)^\circ $  & 3.13 \\
2000\,MHz & $115.1(8)^\circ$ & $-5(1)^\circ$ & $202(3)^\circ $ & $ 5(2)^\circ $ & 3.74 \\
\hline
\end{tabular*}
\caption{The best-fit RVM parameters for PSR~J2007+2722 obtained from
  fitting the model in Equation~\eqref{eq:rvm_define} to the measured
  polarization angle as a function of pulsar rotation phase. $\chi^2$
  is the minimum reduced-chi-squared value, and the numbers in
  parentheses are the estimated 1~$\sigma$ errors.
\label{tab:RVMparam}}
\end{table}

The corresponding best-fit polarization-angle profiles are displayed
by dashed lines in the top panels of Figure~\ref{f:GBT_stokes}.  The
fit is acceptable in the sense that it is not untypical when compared
with other radio pulsars.  Overall, the RVM reproduces the form of the
observed profile, especially at 1500~MHz, but leaves unmodeled
structure below pulse phase 0.2 and above pulse phase 0.9. The largest
deviations are at 2000~MHz below phase 0.25. Nevertheless it is
encouraging that the independent fits at 1500 and 2000~MHz lead to
very similar beam geometry parameters, and surprisingly tight bounds
on their values, as shown in Table~\ref{tab:RVMparam}.

However the fit can not be characterized as good; the deviations
between data and model that are visible in Figure~\ref{f:GBT_stokes}
give rise to reduced-chi-squared $\chi^2$ values that have very low
statistical likelihood of being explained by the polarization-angle
measurement errors.  The failure to fit the RVM very well may arise
because the pulsar does not ever ``shut off'' but is emitting over its
entire rotation.  This can affect the polarimetry; in
Figure~\ref{f:GBT_stokes} one can see regions where the intensity $L$ of
the linearly-polarized component is greater than the total intensity
$I$.  This can not happen in nature; the inconsistency probably
indicates that some aspect of the polarimetry measurement can not be
trusted.  It could well be an artifact of not being able to identify
the uniform level of flux corresponding to zero pulsed emission.
However the lack of a good fit is also consistent with the
interpretation that PSR~J2007+2722 is a DRP:
many recycled pulsars are not well-fit by the basic RVM
\citep{1990ApJ...361..644T,1997ApJ...486.1019N,1998ApJ...501..286X,1999ApJS..123..627S}.

One can infer the opening-angle of the radio emission-cone from the
RVM parameters together with the observed separation between the pulse
peaks.  The emission-cone half-opening-angle $\rho$ is related to
the measured separation $W$ of the pulse peaks by
\beq
\cos\left(\rho\right) = \cos\left(\alpha\right)\cos\left(\zeta\right) +
\sin\left(\alpha\right)\sin\left(\zeta\right) \cos\left(\frac{W}{2}\right).
\label{eq:rvm_halfopening}
\eeq At 1500\,MHz we estimate a peak-to-peak width of the pulse
profile $W_{1500} = 163^\circ$; at 2000\,MHz $W_{2000} =
171^\circ$. For these values of $W$, the dashed lines in
Figure~\ref{f:chi2maps} show contours of constant emission-cone
half-opening-angle $\rho$ as a function of $\alpha$ and $\zeta$.
Using the best-fit $\alpha, \zeta$ values from
Table~\ref{tab:RVMparam} we obtain radio-emission-cone
half-opening-angles $\rho_{1500} = 77^\circ$ and $\rho_{1500} =
78^\circ$ at 1500\.MHz and 2000\,MHz, respectively.

\subsection{Timing Model}
\label{ss:timingmodel}
A timing model for PSR~J2007+2722 has been found using two distinct
data sets, obtained at the Arecibo Observatory and at Jodrell Bank.
The Arecibo data were collected in two short (268~s) survey
observations on 2007 February 11 and 16; the first of these provided
the data used in the \EAH{} discovery.  The Jodrell data were
collected in 75 targeted observations between 2010 July 15 and 2012
November 30, starting soon after the discovery.

The Arecibo data (described earlier) covering a 100\,MHz bandwidth
centered at 1452\,MHz, were used to construct TOAs in four distinct
25\,MHz frequency bands.  A model pulse profile was used to obtain 22
distinct TOAs.

The Jodrell Bank observations used a dual-polarization cryogenic
receiver on the 76-m Lovell telescope, having a system equivalent flux
density of 25\,Jy on cold sky.  Observations typically lasted 20 or 30
minutes. Data were processed using a digital filterbank which covered
a bandwidth of 350\,MHz centered around 1525\,MHz in channels of
0.5\,MHz bandwidth. The data were folded at the nominal topocentric
period of the pulsar for sub-integration times of 10\,s. After
inspection and removal of any RFI, the profiles were de-dispersed and
summed over frequency and time to produce integrated profiles. For
each observation, a single TOA as obtained by cross-correlation of the
profile with a standard template using standard analysis tools from
\textsc{PSRCHIVE}.

The 97 distinct TOAs were analyzed using the TEMPO2 software package
\citep{MNR:MNR10302,MNR:MNR10870}.  In the fitting procedure to
determine the pulsar parameters, a single adjustable offset time
(TEMPO2 ``jump'') was introduced between the two data sets.  This is
needed because different model pulse profiles were used to derive the
Arecibo and Jodrell TOAs, and avoids the need for absolute time
synchronization between the two observatories.

\begin{figure}[h]
\begin{center}
\includegraphics[width=1.0\columnwidth]{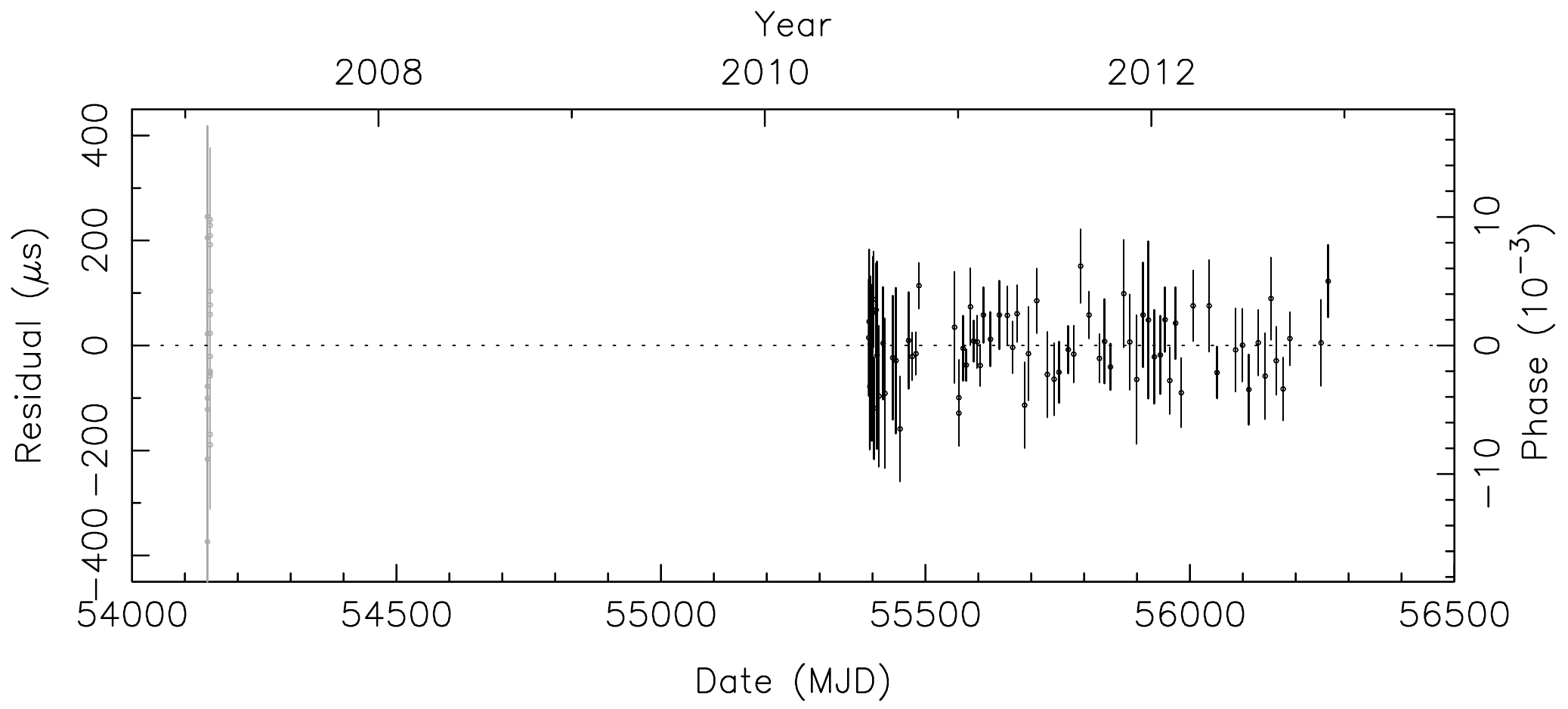}
\end{center}
\caption{Timing residuals obtained by fitting a timing model to TOA
  data from Arecibo Observatory (taken in 2007 February) and TOA data
  from Jodrell Bank (taken between 2010 July and 2012 November). The
  horizontal axis is the date of the TOA observation, and the vertical
  axis is post-fit residuals in seconds.}
\label{f:residuals}
\end{figure}

The parameters of PSR~J2007+2722 obtained from this TEMPO2 analysis
are shown in Table~\ref{tab:timingmodel}; the resulting fitting
residuals are shown in Figure~\ref{f:residuals}. The fit is remarkably
good: the residuals have a weighted rms of 66\,$\mu$s and the
reduced $\chi^2 =1.059$ is very close to unity.  In pulsar astronomy
it is standard practice to re-scale the uncertainties by the
square-root of this value; we have done that here, but it only changes
the estimated one-sigma errors by about 3\%.

The pulsar parameters obtained by timing (sky position, frequency, and
spindown) are reasonably consistent with the announcement paper
\citep{2010Sci...329.1305K} published one month after the
discovery\footnote{To facilitate comparison,
  Table~\ref{tab:timingmodel} specifies the pulsar's parameters at the
  same epoch as \citet{2010Sci...329.1305K}, rather than at the (more
  conventional) midpoint of the observational sample.}.  That paper
gave the sky position (found as described in
Section~\ref{subsec:detskypos}) as R.A. $\rm 20^h07^m15^s.77$, decl. $\rm
27^o 22^{\prime}47^{\prime \prime}.7$ with errors less than
order $1^{\prime \prime}$.  The position found here is consistent with
that. The discovery paper gave the frequency (at MJD 55399) as $f= \rm
40.820677620(6)\,$Hz.  The frequency found here is about one standard
deviation \textit{outside} of that range; this may have been due to
our lack of knowledge about the precise spin-down rate. Finally, the
discovery paper only gave a bound on the spin-down rate, of $|\dot f|<
3 \times 10^{-14}/ \rm s^2$.  The spin-down found here is consistent
with that: $\dot f = -1.6 \times 10^{-15}/ \rm s^2$. This corresponds
to a characteristic age $-f/2 \dot f = \rm 404\,$Myr, an inferred
surface dipole magnetic field strength of $4.9 \times 10^9\,$G, and a
spin-down luminosity $\dot E = 2.6 \times 10^{33}$~erg/s (assuming the
canonical moment-of-inertia $I=10^{45} ~ \text{g} ~ \text{cm}^2$).

\begin{table}
\begin{tabular}{ll}
\multicolumn{2}{c}{Fit and data-set} \\
\hline
Pulsar name\dotfill & JJ2007+2722 \\ 
MJD range\dotfill & 54142.7---56261.4 \\ 
Number of TOAs\dotfill & 97 \\
Rms timing residual ($\mu s$)\dotfill & 65.6 \\
Weighted fit\dotfill &  Y \\ 
Reduced $\chi^2$ value \dotfill & 1.057 \\
\hline
\multicolumn{2}{c}{Measured quantities} \\ 
\hline
Right ascension, $\alpha$\dotfill &  20:07:15.8288(4) \\ 
Declination, $\delta$\dotfill & +27:22:47.914(6) \\ 
Pulse frequency, $\nu$ (s$^{-1}$)\dotfill & 40.820677605083(15) \\ 
First derivative of pulse frequency, $\dot{\nu}$ (s$^{-2}$)\dotfill & $-$1.6015(4)$\times 10^{-15}$ \\ 
Dispersion measure, $DM$ (cm$^{-3}$pc)\dotfill & 127.0(4) \\ 
\hline
\multicolumn{2}{c}{Set quantities} \\ 
\hline
Epoch of frequency determination (MJD)\dotfill & 55399 \\ 
Epoch of position determination (MJD)\dotfill & 55399 \\ 
Epoch of dispersion measure determination (MJD)\dotfill & 55399 \\ 
\hline
\multicolumn{2}{c}{Derived quantities} \\
\hline
$\log_{10}$(Characteristic age, yr) \dotfill & 8.61 \\
$\log_{10}$(Surface magnetic field strength, G) \dotfill & 9.69 \\
$\log_{10}$(Canonical spin-down luminosity, erg/s) \dotfill & 33.4 \\
\hline
\end{tabular}
\caption{The parameters describing PSR J2007+2722 obtained by timing analysis of data spanning about six years using a DE405 solar-system ephemeris model. Figures in parentheses are  the nominal 1$\sigma$ \textsc{tempo2} uncertainties in the least-significant digits quoted. For easy comparison, the Epoch has been chosen to be the same as \citet{2010Sci...329.1305K} rather than at the midpoint of the observational interval.
\label{tab:timingmodel}}
\end{table}

\subsection{Multi-wavelength Electromagnetic Counterparts}

With the final sky position given in Table~\ref{tab:timingmodel}, we
searched for electromagnetic counterparts at different wavelengths.
The pulsar is not in any known globular cluster or near a cataloged
supernova remnant \citep{GreenCatalog}. We then checked infrared,
gamma-ray and X-ray catalogs for counterparts.  {\bf Infrared}: The
nearest sources visible in infrared images (J, H, K-band) obtained
from the Two Micron All Sky Survey \citep{2006AJ....131.1163S}
are more than $13^{\prime \prime}$~distant from the pulsar position.  {\bf
  Gamma-ray}: No counterpart was found in the second \textit{Fermi} LAT Point Source Catalog \citep{2012ApJS..199...31N}.  {\bf
  X-ray}: There are three archival \textit{Chandra} X-ray
observations\footnote{\url{http://heasarc.gsfc.nasa.gov/docs/archive.html}};
from these, no X-ray counterpart could be identified.  We then carried
out more detailed followups starting from the raw gamma-ray and X-ray
data as described below.

Since the launch of \textit{Fermi} in 2008, the on-board LAT
\citep{2009ApJ...697.1071A} has observed pulsations
from more than 120 pulsars\footnote{See
  \url{https://confluence.slac.stanford.edu/display/GLAMCOG/Public+List+of+LAT-Detected+Gamma-Ray+Pulsars/}},
and new blind-search methods similar to those used in this paper are
finding even more \citep{2012ApJ...744..105P,
  2012ApJ...755L..20P,2012Sci...338.1314P}.  The LAT has also
confirmed that many radio-detected, both normal and millisecond,
pulsars are emitting rotation-phase-synchronous gamma-rays
\citep{2009Sci...325..848A,Ray+2012}.  So, we here consider the
possibility of PSR~J2007+2722 also being a gamma-ray pulsar.

Unfortunately the characteristics of PSR~J2007+2722 make it an
unlikely source for gamma-ray emissions or pulsations, when comparing
to the known gamma-ray pulsar population \cite{2013arXiv1305.4385T}.  Its
spin-down power $\dot E = 2.6 \times 10^{33}$~erg/s is near the lower
end of the known gamma-ray pulsar population, and at a distance of
$d=5.4$~kpc, the spin-down flux density $\dot E/d^2$ is smaller than
that of any known gamma-ray emitting pulsar by a factor of a few.  In
addition, PSR~J2007+2722 is in a high-background region close to the
Galactic plane. The {\em Fermi}-LAT Second Source Catalog
\citep{2012ApJS..199...31N} does not contain any source positionally
overlapping with the pulsar's location.

Nevertheless, we searched the LAT data for gamma-ray pulsations
synchronous with the radio-pulse rotation phase.  We extracted the LAT
photons within $2^\circ$ of PSR~J2007+2722's sky position from the
start of data taking in 2008 August up to 2013 January.  We folded
them for different cuts on minimum energy (between 40~MeV and 0.8~GeV)
and different angular cuts (between $0.5^\circ$ and $2^\circ$).  There was no
sign of a signal; the LAT does not detect gamma-ray pulsations from
PSR~J2007+2722. In principle, one could carry out a spectral analysis
of the region and construct a source model for PSR~J2007+2722 to assign
probability weights to the LAT photons as in
\cite{2012ApJ...744..105P}. However, given the extremely low pulsation
significance of the unweighted fold, we concluded this was unlikely to
make much of a difference.

\subsection {X-ray Limits, and the Nature of PSR~J2007+2722}

\begin{figure}
\begin{center}
\includegraphics[width=0.8\columnwidth]{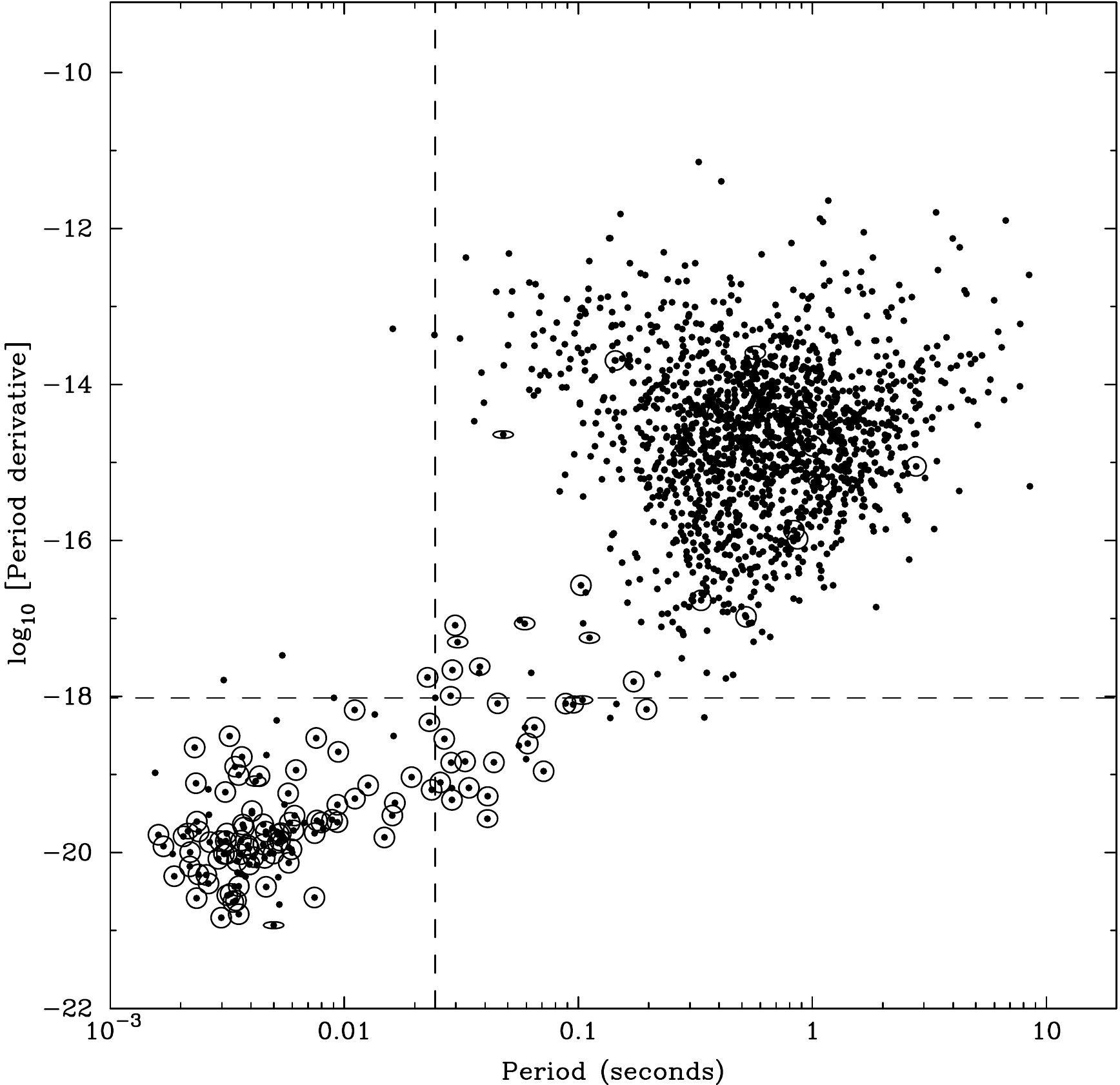}
\end{center}
\caption{The population of known radio pulsars, plotted as a function
  of spin-period (horizontal axis) and rate of change of the
  spin-period with time (vertical axis).  PSR J2007+2722 is at the
  intersection of the dotted lines: a region populated almost
  exclusively by old recycled pulsars in binary systems, indicated by
  circled points.  In contrast to PSR J2007+2722, almost all isolated
  pulsars (uncircled points) are in the region populated by much
  younger non-recycled systems.
\label{fig:ppdot}}
\end{figure}

As shown in Figure~\ref{fig:ppdot}, timing measurements of
PSR~J2007+2722 place it in a region of the ($P, \dot P$)-diagram
normally occupied by old neutron stars in binary systems spun-up due
to accretion torques (i.e. ``recycled''). These pulsars naturally have
shorter periods ($P \simlt 100$~ms) than the younger, isolated
rotation-powered pulsars and are constrained to lie below the spin-up
limit for recycled pulsars $P({\rm ms}) = 1.9 (B/10^9 \ {\rm
  G})^{6/7}$ \citep{1987IAUS..125..393V}, where the magnetic field
restricts the minimal achievable rotation period.

Together with the lack of a stellar companion at any wavelength or
unmodeled systematics in the timing residual to indicating otherwise,
there is no evidence that PSR~J2007+2722 is currently part of a binary
system.  Instead, its moderately short period suggests that it was
partially recycled and is possibly a DRP.
These isolated neutron stars are born in a binary system and become unbound by a
second supernova event involving the companion; they are defined in
\citet{MNR:MNR16970} as isolated radio pulsars in the Galactic disk
with magnetic field strength $|B| < 3 \times 10^{10}$\,G and
spin-frequency $f<50$\,Hz. Their evolutionary origin explains their
location on the region of the ($P,\dot P$)-diagram which is populated
by weak magnetic field pulsars, whose fields have decayed over $\sim
10^8$~yr.  The work by \citet{MNR:MNR16970} describes the 12 DRPs
known at the time of publication; one more (PSR J1821+0155) has
subsequently been discovered \citep{2012arXiv1209.4108R}.
PSR~J2007+2722 would be the 14th and most rapidly spinning member of
this class.

DRPs are an enigma: standard evolutionary models for binary systems
cannot easily explain the observed ratio of isolated recycled pulsars
relative to the number of DNS systems
\citep{MNR:MNR16970}.  The models predict about DNS
system for every ten DRPs, but roughly equal numbers are
observed. Furthermore, there is no independent evidence that all
isolated pulsars overlapping the binary population are actually
derived from binaries.  Indeed, recent observations of manifestly
young pulsars in supernova remnants reveal that neutron stars can be
born with anomalously low surface dipole magnetic fields of order $B
\sim 10^{10}$~G (see \citealt{2013arXiv1301.2717G} for details).
These so-called anti-magnetars occupy an overlapping region in the
($P,\dot P$)-diagram with the DRPs and therefore suggest that their
descendants might be mis-identified as DRPs \citep{gottinprep}. If in
fact PSR~J2007+2722 is a young object instead of a $\sim 10^8$~yr-old
DRP, neutron star cooling curves predict that thermal X-ray surface
emission should be observable for up to 1~Myr
\citep{2009ApJ...707.1131P}, long after its supernova remnant has
dissipated. After this time, the internal temperature drops rapidly
and thermal emission becomes negligible.

\begin{figure}
\begin{center}
\includegraphics[width=0.95\columnwidth,angle=270]{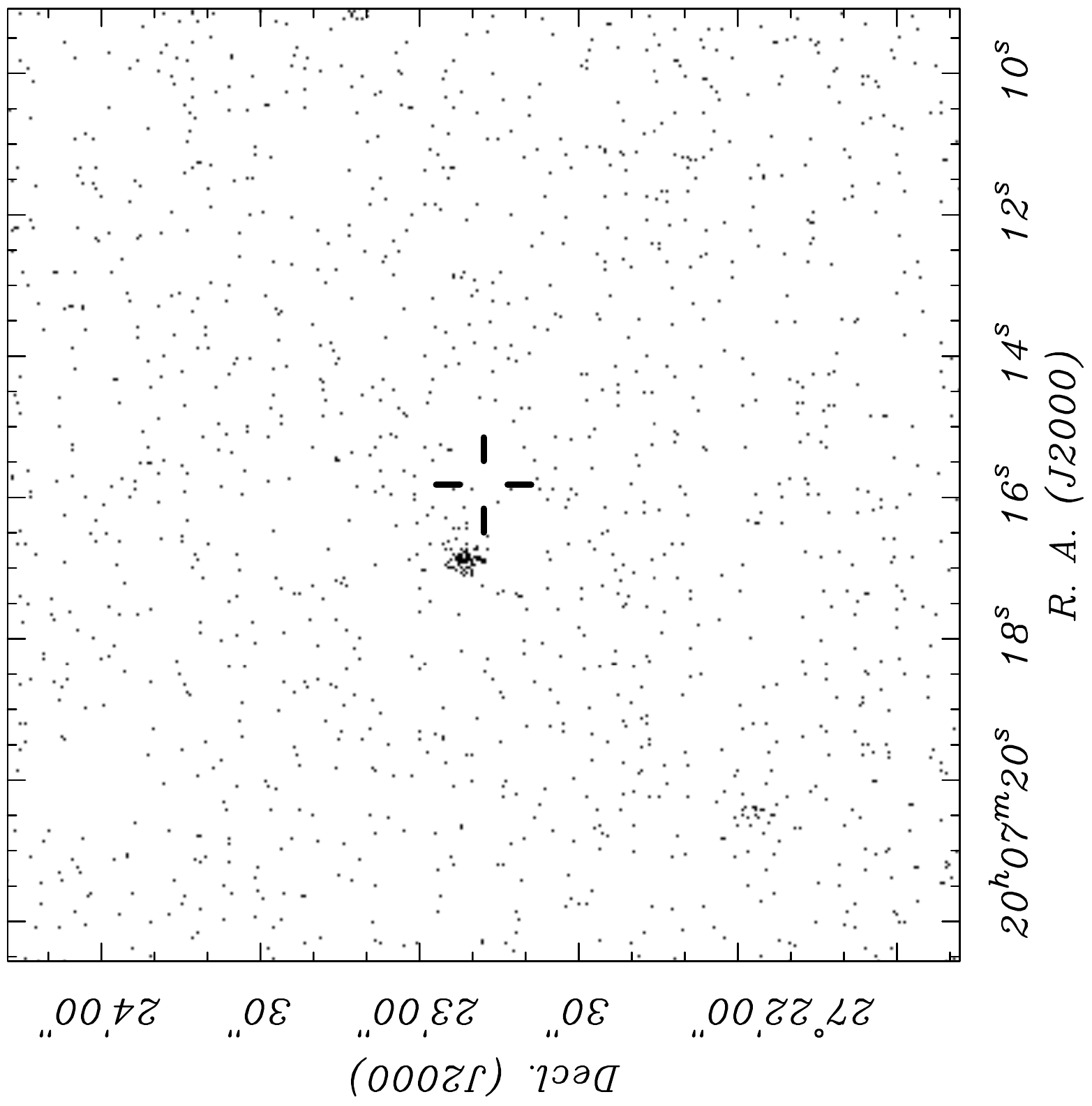}
\end{center}
\caption{\textit{Chandra} ACIS X-ray image ($0.3-2$~keV) of the field
  containing PSR~J2007+2722, whose location is marked by the
  cross. The field-of-view is $3^{\prime}\times3^{\prime}$.  The
  nearest resolved point source is $14\farcs3$ away.
\label{f:Chandra}}
\end{figure}

To investigate the possibility that PSR~J2007+2722 might be a young,
hot object, we examined fortuitous archival X-ray observations
covering the location of the pulsar. A total of 95~ks of good
\textit{Chandra}/ACIS-I data are available as data sets ObsIDs 6438,
7254 and 8492, acquired on 2006 December 10, 2007 January 07 and 29,
respectively \citep{2012AJ....144..101G}.  The expected location of
the pulsar falls $6^{\prime}$ off-axis for each observation, where the
point response function of the telescope is degraded to
$5^{\prime\prime}$ ($99\%$ enclosed energy fraction). Within the
nominal absolute astrometry error of $0\farcs6$ radius no X-ray source
is found that overlaps with the subarcsec pulsar coordinates presented
herein.  As shown in Figure~\ref{f:Chandra}, the closest source is
$14\farcs3$ away from PSR~J2007+2722.

To attempt to place a lower limit on the age of PSR~J2007+2722 we use
the \textit{Chandra} data to determine the minimum detectable flux
expected from a cooling neutron star of radius $R=14$~km at the DM
derived distance of 5.4~kpc. Following the method described in
\citet{gottinprep}, we compute an upper-limit on the number of
expected counts for a non-detection at the $99.73\%$ confidence level
($3\sigma$). Based on the local background rate of
$1.6\times10^{-5}$~cps in the $r=5^{\prime\prime}$ aperture, we
require 6.5 photons from the pulsar in the composite \hbox{ACIS-I}
observation in the $0.3-2$~keV energy band at the off-axis pulsar
location. We convolve an absorbed blackbody spectrum with the
telescope response function generated for these observations and
integrate over the energy band to compute the detected number of
counts as a function of temperature.  The blackbody normalization is
fixed to the ratio of the neutron star radius to its distance and the
column density is set to $N_{\rm H} \approx 4 \times
10^{21}$~cm$^{-2}$, estimated from the DM and by assuming a
rule-of-thumb $N_e/N_{\rm H} \sim 0.1$.  This procedure yields a
temperature of $k{\rm T} \approx 69$~eV and bolometric luminosity of
$L({\rm bol}) \approx 6 \times 10^{32}$~erg~s$^{-1}$ implying a lower
limit on the neutron star cooling age $\simgt 1 - 5 \times
10^5$~yr, depending upon the range of cooling-curve models
\citep{2009ApJ...707.1131P}.  This luminosity is less than 10\% of
what would be expected for a typical young neutron star.

The uncertainty in this upper limit on luminosity is difficult to
estimate. The contribution from the unknown column density depends on
the uncertainty in the Galactic electron density distribution,
estimated as 20\% in \citet{2002astro.ph..7156C}.  A recent calibration
of the ratio $N_e/N_{\rm H}$ shows over an order-of-magnitude scatter
in this relationship \citep{he2013}.  If $N_{\rm H}$ varied by an
order-of-magnitude away from our assumed value, then the lower limits
on the age could be as small as $10^4$~yr.  Moreover, the effects of
any uncertainty on $N_{\rm H}$ are amplified because the derived
temperature falls at the edge of the ACIS-I response function where
the detector sensitivity falls off rapidly.

It appears unlikely that PSR~J2007+2722 is a young pulsar, but current
data cannot prove that it was formed through recycling in a binary
system versus being simply an isolated pulsar born with a low magnetic
field. For a typical rotation-powered pulsar emitting non-thermal
X-rays with power-law spectrum of photon index $\Gamma = 1.5$, the
$2-10$~keV luminosity upper limit for PSR~J2007+2722 is $2.2 \times
10^{31}$ erg~s$^{-1}$.  However, based on its spin-down energy of
$\dot E = 2.58\times 10^{33}$ erg~s$^{-1}$, the predicted X-ray
luminosity in this band is only $L_{\rm x} = 2.7\times 10^{29}$
erg~s$^{-1}$ \citep{2002A&A...387..993P}.  So no definite constraint
is possible.

\section{Discussion and Conclusions}
\label{s:conclusion}

We have presented a detailed description and full timing model for PSR
J2007+2722, the first \EAH{} radio pulsar discovery.  Evidence from
polarization studies, lack of associated remnants, and its location on
the $(P, \dot P)$ diagram, support the hypothesis that it is a
DRP, about 0.4\,Gyr old. However there remains
the possibility that it is a much younger object born with a low
magnetic field.  In this case it is probably at least 100\,kyr old.

PSR J2007+2722 has other unusual properties.  Its (pulsed) radio
spectrum peaks at higher (GHz) frequencies than most pulsars, and has
a relatively flat spectral index above 1~GHz.  The pulse profile is
remarkably wide with emission over almost the entire spin period, and
the beam geometry is well constrained by the RVM.  The beam geometry
is also atypical: the pulsar is almost an orthogonal rotator, the
magnetic field axis passes quite close to the line-of-sight and the
beam opening angle is unusually broad.

We have also given a detailed description of the \EAH{} radio pulsar
search. To date, \EAH{} has found nearly 50 radio pulsars using the
methods described here. Some of these discoveries have already been
published \citep{Knispel1952, 2013arXiv1302.0467K} and others are
forthcoming.  The \EAH{} project continues to analyze data from
GW detectors, from the \textit{Fermi} gamma-ray
satellite, and from radio telescopes. We will continue to search PALFA
data as the survey progresses, and also plan to search Effelsberg data
from the HTRU survey \citep{2011AIPC.1357...52B,2010tsra.confE..66N}.
Because it enables efficient searches over larger volumes of parameter
space, we believe that the \EAH{} can have a significant impact on
pulsar astronomy.

At the end of this decade, Volunteer Distributed Computing might play
an even larger role.  For example to carry out a complete pulsar
survey using data from the upcoming SKA will
requires Exaflop computing resources \citep{2009A&A...493.1161S}. We
expect that this will be pushing ``state of the art'' in computing and
thus will be challenging and expensive.  But based on reasonable
extrapolations about consumer computing hardware, several million
volunteers should be able to provide those compute cycles at very low
cost to the scientific community or funding agencies
(B. Allen et al. 2013, in preparation).

Volunteer Distributed Computing might also provide a novel solution
for SKA data storage (B. Allen et al. 2013, in preparation).
The SKA data
rate is so high (Tb/s) that raw data must be processed and discarded
within a few hours.  In contrast, Volunteer Distributed Computing
might permit \textit{all} SKA data to be stored \textit{forever},
broadening the range of scientific work that could be carried
out. This is possible because both the public Internet capacity and
consumer storage device capacity are anticipated to continue growing
at 40\% annual rates through the end of the decade; it is sufficient
if several million volunteers provide a fraction of that storage.
Existing file-sharing and replication techniques could provide a
statistical guarantee of retrievability and validity.  The key
requirement is that SKA have a Tb/s network connection to the public
Internet, presumably in a major city.

Extrapolating a few years into the future, we expect that laptop and
desktop computers will provide a decreasing fraction of the compute
cycles available from volunteers. A larger fraction will come from
tablets and smart-phones that are being charged.  While less powerful
than conventional machines, they are being sold in very large numbers.

In short, we believe that the approach described here is not a fad,
and will provide a substantial computing resource for astronomy in the
long term.

\section*{ACKNOWLEDGMENTS}

We thank the thousands of Einstein@Home volunteers who made this work
possible, and particularly Chris and Helen Colvin of Ames, Iowa, USA
and Daniel Gebhardt of Mainz, Germany, whose computers processed the
``discovery'' workunits in which PSR~J2007+2722 appeared with the
highest statistical significance.  We also thank David Nice and Duncan
Lorimer for their assistance and encouragement, and Ralph Eatough for
his contributions and helpful comments.

This work was supported by the Max Plank Gesellschaft, by the
Netherlands Foundation for Scientific Research (NWO), and by US
National Science Foundation (NSF) grants 1104902, 1105572, 1148523 and
0555655.

Pulsar research at UBC is supported by an NSERC Discovery Grant and
Discovery Accelerator Supplement, by CANARIE and by the Canada
Foundation for Innovation.

V.M.K. was supported by an NSERC Discovery Grant, the Canadian Institute
for Advanced Research, a Canada Research Chair, FQRNT, and the Lorne
Trottier Chair in Astrophysics.

P.F. gratefully acknowledges financial support by the European
Research Council for the ERC Starting Grant BEACON under contract no.
279702.

P.L. acknowledges the support of IMPRS Bonn/Cologne and NSERC PGS-D.

The Arecibo Observatory is operated by SRI International under a
cooperative agreement with the National Science Foundation
(AST-1100968), and in alliance with Ana G. M{\'e}ndez-Universidad
Metropolitana, and the Universities Space Research Association.

This work makes use of data obtained from the \textit{Chandra} Source
Catalog, provided by the \textit{Chandra} X-Ray Center (CXC) as part
of the \textit{Chandra} Data Archive.

This work makes use of archival data taken with the National Radio
Astronomy Observatory (NRAO) Very Large Array (VLA). NRAO is a
facility of the National Science Foundation (NSF) operated under
cooperative agreement by Associated Universities, Inc.  We also
acknowledge use of the NRAO VLA Sky Survey (NVSS) Catalog
\citep{NVSS}.

This publication makes use of data products from the Two Micron All
Sky Survey \citep{2006AJ....131.1163S}, which is a joint project of
the University of Massachusetts and the Infrared Processing and
Analysis Center/California Institute of Technology, funded by the
National Aeronautics and Space Administration and the National Science
Foundation.

This research has made use of SAOImage DS9, developed by Smithsonian
Astrophysical Observatory.

This research has made use of NASA's Astrophysics Data System.

This work was supported by CFI, CIFAR, FQRNT, MPG, NAIC, NRAO, NSERC, NSF
and STFC.

\bibliographystyle{yahapj}
\bibliography{references}

\end{document}